\ifdefined\pdfoutput\pdfoutput=1\fi
\documentclass[aps,prd,onecolumn,11pt,superscriptaddress,nofootinbib,floatfix,preprintnumbers,longbibliography]{revtex4-2}

\usepackage[T1]{fontenc}
\usepackage[utf8]{inputenc}
\usepackage[english]{babel}
\usepackage{amsmath,amssymb}
\usepackage{graphicx}
\usepackage{array,booktabs}
\usepackage[dvipsnames]{xcolor}
\usepackage{microtype}
\makeatletter\@ifundefined{@removefromreset}{\RequirePackage{remreset}}{}\makeatother
\usepackage[colorlinks=true,linkcolor=RoyalBlue,citecolor=RoyalBlue,urlcolor=RoyalBlue]{hyperref}

\makeatletter
\def\switch@array{\let\@array@sw\@array@sw@LaTeX}
\def\switch@tabular{\let\@array@sw\@array@sw@array}
\makeatother

\makeatletter
\def\section{\@startsection{section}{1}{\z@}{0.8cm \@plus1ex \@minus .2ex}{0.4cm}{\normalfont\normalsize\bfseries\raggedright}}
\def\subsection{\@startsection{subsection}{2}{\z@}{0.6cm \@plus1ex \@minus .2ex}{0.3cm}{\normalfont\normalsize\bfseries\raggedright}}
\def\subsubsection{\@startsection{subsubsection}{3}{\z@}{0.5cm \@plus1ex \@minus .2ex}{0.25cm}{\normalfont\normalsize\itshape\raggedright}}
\makeatother
\newcommand{\colhead}[1]{\begin{tabular}[b]{@{}c@{}}#1\end{tabular}}
\newcommand{\colheadl}[1]{\begin{tabular}[b]{@{}l@{}}#1\end{tabular}}

\makeatletter
\renewcommand{\thesection}{\arabic{section}}
\renewcommand{\thesubsection}{\thesection.\arabic{subsection}}
\renewcommand{\thesubsubsection}{\thesubsection.\arabic{subsubsection}}
\renewcommand{\p@subsection}{}
\renewcommand{\thetable}{\arabic{table}}
\renewcommand{\p@subsubsection}{}
\makeatother

\usepackage{dcolumn}
\usepackage{bm}
\usepackage{amsfonts}

\usepackage{bm}

\usepackage{verbatim}

\usepackage{xspace}

\usepackage{mathtools}
\usepackage{multirow,rotating}
\usepackage{stackrel}
\usepackage{cancel}
\usepackage{pifont}
\usepackage{dcolumn}

\usepackage[compat=1.1.0]{tikz-feynman}
\usepackage{tikz-cd}

\newcommand{\be}{\begin{equation}}
\newcommand{\ee}{\end{equation}}
\newcommand{\bea}{\begin{eqnarray}}
\newcommand{\eea}{\end{eqnarray}}

\newcommand{\mathsym}[1]{{}}

\newcommand{\anu}{\overline{\nu}}

\newcommand{\etaa}{\eta_{a,\theta}}   

\begin{document}

\preprint{Published in Universe \textbf{12}, 320 (2026)}
\preprint{\href{https://doi.org/10.3390/universe12110320}{doi:10.3390/universe12110320}}

\title{Left-Chiral Neutrino Secret Interactions Probed by High-Energy Neutrinos Surviving the Cosmic Neutrino Background}

\author{Maria Petropavlova}
\email{mariia.petropavlova@cvut.cz}
\affiliation{Institute of Experimental and Applied Physics, Czech Technical University in Prague, Husova 240/5, 110\,00 Prague 1, Czech Republic}
\affiliation{Faculty of Mathematics and Physics, Charles University, Ke Karlovu 3, 121\,16 Prague 2, Czech Republic}

\begin{abstract}
This work explores the potential of astrophysical neutrinos to probe neutrino secret interactions ($\nu$SI) through their propagation in the Cosmic Neutrino Background (C$\nu$B). Using representative energies and distances for SN 1987A, NGC 1068, TXS 0506+056 and PKS 0735+178, together with distance hypotheses for KM3-230213A, we derive single-event survival benchmarks for the mediator mass and coupling. We focus on scenarios that do not populate right-handed neutrino states in the early Universe and therefore avoid the strongest cosmological constraints: a massive spin-one mediator with a left-chiral coupling to Dirac neutrinos and a scalar mediator coupled to Majorana neutrinos. We also consider scalar exchange with Dirac neutrinos, a common benchmark in propagation studies, and demonstrate that on the standard relic background its resonance is helicity suppressed, with a strength controlled by the relic-neutrino mass. The analysis covers a broad range of mediator masses, including the light-mediator, resonant and heavy-mediator regions, and accounts for the effects of the relic-neutrino masses and of cosmological evolution along the propagation path. We illustrate the main features of each scenario in a common-mass approximation, with analytic expressions for the non-relativistic and ultra-relativistic relic background, and then extend the analysis to realistic non-degenerate neutrino mass spectra. Owing to its energy, the KM3-230213A event shifts the resonant sensitivity to heavier mediators, while distant-source hypotheses increase the propagation sensitivity in this mass range.
\end{abstract}

\maketitle
% Fix reference numbering to the order of the published version
\nocite{Berryman_2023,Carena_2003,Flowers:1976kb,Berbig:2020wve,Blinov_2019,Venzor_2023,Abreu_2023,Cyr_Racine_2014,Esteban_2021,Das_2022,Camarena_2023,Kolb:1987qy,Bustamante:2020mep,Doring:2023vmk,Chang_2023,Wu_2024,Bakhti_2019,Berryman_2018,Gelmini:1980re,Forastieri_2019,Lessa_2007,Barik:2024kwv,Kelly:2018tyg,Escudero_2020,Heeck:2018nzc,Farzan2016,OKADA2020135845,Kelly2020,Kreisch_2020,Deppisch:2020sqh,Deppisch_2020,Blum_2018,de_Gouv_a_2020,BelleII,Cortina_Gil_2021,Brdar_2020,Dev:2024twk,Pasquini:2015fjv,Laha:2013xua,Grohs_2020,Huang_2018,Escudero:2018mvt,Sabti:2019mhn,Luo:2020sho,Luo:2020fdt,Abazajian:2019oqj,Kamada_2015,Ioka_2014,Ng_2014,Ibe_2014,Ambrosone:2024zrf,He:2025bex,Troitsky:2023nli,PhysRevD.107.042005,Wang:2025qap,Bally2020,DiFranzo:2015qea,rozhkov2024,KM3NeT:2025npi,Aglietta:1987mc,Alekseev:1987qf,Bionta:1987qt,Hirata:1987hu,Panagia:2003rt,Scholberg_2012,Markham:2026ngc,2022IceCube,Urry:1995mg,IceCube:2018dnn,Paiano:2018qeq,2018,VERITAS:2023eso,Dik_2023,Petkov:2022fnz,Omeliukh:2024kgk,Sahakyan_2022,Falomo2021,Nilsson2012,KM3NeT:2026blazars,KM3NeT:2026galactic,KM3NeT:2025bxl,KM3NeT:2025vut,Long_2014,Lesgourgues:2006nd,Abbasi_2025,Kuznetsov:2025ehl,Planck:2018vyg,Blum:2014ewa,Herrera:2023nww,Arguelles:2026rjr,Farzan_2008,Elbers:2025DESI,Murase:2019xqi,Cline:2022qld,IceCube:2013dkx,Dror:2017nsg}

\section{Introduction}

In the Standard Model (SM),  massless neutrinos interact with each other at tree level only through $Z$-boson exchange. In~extensions of the SM, massive neutrinos can also interact with each other via the Higgs boson and/or electromagnetically via loop corrections. Yet direct observation of these interactions remains beyond current experimental capabilities. 
For example, in~the case of SM neutrino self-interactions mediated by $Z$-boson, despite cross-sections comparable to electron-neutrino scattering (in ultra-relativistic regime), detection of such a process is complicated by the simultaneous detection of two outgoing neutrinos~\cite{Berryman_2023, Carena_2003, Flowers:1976kb}. 
Due to their weak and neutral nature, neutrinos are the least understood SM particles,  yet they offer a pathway to beyond Standard Model (BSM) physics.

Beyond Standard Model scenarios offer various solutions to problems that require additional mediators interacting with neutrinos. ``New'' or ``secret'' neutrino self-interactions ($\nu$SI)
can produce larger scattering rates and impact phenomena like neutrino mass generation, solutions to cosmological tensions (e.g., $H_0$, $S_8$), and~dark matter production~\cite{Berryman_2023, Berbig:2020wve, Blinov_2019, Venzor_2023}.

The discovery of high energy neutrinos, first detected by IceCube, apart from their cosmological and astrophysical significance, has opened new opportunities to probe $\nu$SI. 
The high energies (HE) and ultra-high energies (UHE) unattainable in laboratory experiments\footnote{Neutrino %
 experiment FASER confirmed the detection of neutrinos with energies significantly above 200 GeV~\cite{Abreu_2023}, marking the highest-energy neutrinos ever detected from a human-made source.  However, this is still orders of magnitude lower than astrophysical neutrino energies, which can reach hundreds of PeV.} or in other astrophysical environments, and~great distances between the source and detector, can uncover new information about neutrino mixing, masses, and~potential new forms of interactions. Such a new interaction can affect the passage of astrophysical neutrinos through C$\nu$B.

When testing BSM $\nu$SI, the~relevant parameter space depends on the mediator type, the~neutrino nature, and~the flavor structure of the interaction. In~this work, we restrict the propagation analysis to a flavor-universal and flavor-diagonal interaction, so that the interaction is characterized by the mediator mass and the corresponding coupling strength. The~phenomenology also depends on whether neutrinos are Dirac or Majorana particles, since this determines which interaction structures are allowed and affects the resulting scattering processes. In~more general scenarios, the~mediator may also interact with other particles, such as dark matter.
Constraints on $\nu$SI couplings can be derived from a variety of sources, including cosmological~\cite{Cyr_Racine_2014, %
Esteban_2021, Das_2022, Camarena_2023}, astrophysical~\cite{Kolb:1987qy, 
Bustamante:2020mep, Doring:2023vmk, Chang_2023}, solar~\cite{Wu_2024} and laboratory measurements~\cite{Bakhti_2019}, which probe different mass ranges of $\nu$SI mediators and interaction scales of $\nu$SI for various~models. 

Building on these constraints, in~the scenario of minimal coupling, neutrinos may couple to a massive scalar~\cite{Berryman_2018}, a~pseudoscalar~\cite{Gelmini:1980re, Forastieri_2019, Lessa_2007}, a~vector~\cite{Kolb:1987qy, Flowers:1976kb, Bakhti_2019, Barik:2024kwv}, or~an axial-vector particle~\cite{Kelly:2018tyg}. Spin-0 mediators arise naturally in models of neutrino mass generation and in scenarios with spontaneously broken global lepton-number symmetries, with~the Majoron providing a well-known example~\cite{Escudero_2020}. A~massive spin-one mediator provides a natural benchmark for neutrino self-interactions, since new neutral gauge bosons arise in many extensions of the Standard Model based on additional $U(1)'$ symmetries.  Popular realizations, such as $U(1)_{L_\mu-L_\tau}$, $U(1)_{B-L}$, or~sterile-sector $U(1)'$ broken by a singlet, generate $\nu$SI interactions at tree level~\cite{Heeck:2018nzc, Farzan2016, OKADA2020135845, Kelly2020, Berbig:2020wve}.   
Both spin-one and spin-zero mediated $\nu$SI can modify neutrino free-streaming and imprint characteristic phase and amplitude shifts in the cosmic microwave background (CMB) \cite{Forastieri_2019, Blinov_2019, Kreisch_2020, Berbig:2020wve}.  

A new mediator coupled to neutrinos can also bring the right-handed states of Dirac neutrinos into thermal equilibrium in the early Universe, which is strongly constrained by cosmology (Section~\ref{sec:constraints}). We therefore focus on interactions that involve only left-chiral neutrino fields. In~this study, we examine a massive spin-one boson $V^{\mu}$ interacting through a left-chiral $V{-}A$ coupling with Dirac neutrinos, $(V{-}A,D)$, and~a scalar mediator coupled to Majorana neutrinos, $(S,M)$. For~Majorana neutrinos no additional right-handed states exist, and~the scalar interaction involves only the left-chiral field. In~both cases the standard helicity-polarized C$\nu$B populations, specified in Section~\ref{S:Interaction_rate}, are used consistently. For~comparison, we also evaluate scalar exchange with Dirac neutrinos, $(S,D)$, a~common benchmark in propagation studies. On~the same standard background, it shows how the benchmarks depend on the neutrino nature for the same mediator and illustrates the helicity suppression of the~resonance.

\section{Neutrino Secret Interactions: Models and Existing~Constraints}
\label{sec:level1}
\unskip

\subsection{Models}
\label{sec:setup}

We consider a flavor-universal and flavor-diagonal interaction, $g_{\alpha\beta}=g\,\delta_{\alpha\beta}$, so that each scenario is characterized by the mediator mass $M$ and the coupling $g$. The~mediator is assumed not to couple to charged particles, or~to couple to them negligibly. We label the models by the mediator and the neutrino nature. The~main scenarios are $(V{-}A,D)$ and $(S,M)$. For~Dirac neutrinos, the~spin-one mediator $V^\mu$ couples through %
{\be
{\cal L}_{V-A}^{D} = g_{\alpha\beta}\anu_{\alpha}\gamma_{\mu}P_L \nu_{\beta} V^{\mu},\;\; P_L=\frac{1-\gamma^5}{2},\;\;
(\alpha,\beta = e,\mu, \tau).
\label{Lagrangian}\ee
}
The %
 pure-vector coupling to Dirac neutrinos, with~$\gamma_\mu P_L$ replaced by $\gamma_\mu$, is denoted by $(V,D)$ and is used only as a reference (Appendix~\ref{app:vecref}).
For a scalar mediator $\phi$ and Majorana neutrinos, we use
\begin{equation}
\mathcal{L}_{S}^{M}
=
-\frac{g}{2}\,\phi\,\bar{\nu}\nu .
\end{equation}
The factor $1/2$ accounts for the identical Majorana fields. Scalar exchange with Dirac neutrinos, $(S,D)$, is used as a comparison case on the standard relic background (Section~\ref{sec:HPvsSA}),
\begin{equation}
\mathcal{L}_{S}^{D}
=
-g\,\phi\,\bar{\nu}\nu .
\end{equation}
\subsection{Existing~Constraints}
\label{sec:constraints}

\subsubsection{Laboratory %
}
Laboratory experiments can access self-interaction scales up to ${\cal O}$(100~GeV). Constraints arise from double-beta decay with mediator emission~\cite{Deppisch:2020sqh, Deppisch_2020, Blum_2018, de_Gouv_a_2020}, from~collider and fixed-target searches for missing energy~\cite{de_Gouv_a_2020, Bakhti_2019, Berryman_2018, BelleII, Cortina_Gil_2021}, and~from $\tau$, meson, invisible $Z$ and Higgs decays with mediator emission~\cite{BelleII, Brdar_2020, Bakhti_2019, Dev:2024twk, de_Gouv_a_2020, Pasquini:2015fjv, Laha:2013xua}. These bounds constrain individual flavor components of the coupling: $g_{ee}$ through double-beta decay, $g_{e\alpha}$ and $g_{\mu\alpha}$ through meson decays, and~$g_{\tau\tau}$ through $\tau$ and invisible $Z$ decays. A~flavor-universal coupling must satisfy all of them, which makes it more strongly constrained than couplings to $\nu_\tau$ alone~\cite{Blinov_2019, Brdar_2020}. The~bounds also depend on the interaction structure. Double-beta decay with mediator emission requires lepton-number violation and constrains Majorana scalars. For~a spin-one mediator coupled only to neutrinos, meson and $W$ decays with mediator emission strongly constrain light mediators~\cite{Dror:2017nsg}. %

\subsubsection{Cosmology}
\label{sec:cosmo}

Cosmological observables probe $\nu$SI through BBN and light-element abundances~\cite{Grohs_2020,Huang_2018}, the~radiation density parameterized by $N_{\rm eff}$ \cite{Escudero:2018mvt,Sabti:2019mhn}, CMB anisotropies~\cite{Cyr_Racine_2014,Kreisch_2020}, and~the matter power spectrum~\cite{Camarena_2023}. Such interactions have also been studied in connection with the $H_0$ tension~\cite{Venzor_2023}. In~a standard thermal history, a~mediator lighter than a few MeV is brought into equilibrium with neutrinos unless its coupling is extremely small, and~BBN and the CMB then require $M\gtrsim$ few MeV, largely independently of the neutrino nature and flavor structure~\cite{Blinov_2019,Escudero:2018mvt,Sabti:2019mhn}. For~heavier mediators, the~CMB constrains the effective interaction strength $g^2/M^2$ through its effect on neutrino free-streaming~\cite{Cyr_Racine_2014,Kreisch_2020}. The~models differ in whether the interaction populates right-handed states. For~$(V{-}A,D)$, the~mediator couples only to left-chiral fields, and~right-handed states are produced only through chirality flips suppressed by the neutrino mass~\cite{Blinov_2019}. For~$(S,M)$, no additional states exist. For~$(S,D)$ and $(V,D)$, the~interaction connects left- and right-chiral fields and generically brings the right-handed states into equilibrium, which is constrained by $N_{\rm eff}$ \cite{Luo:2020sho,Luo:2020fdt,Abazajian:2019oqj}. These bounds are derived under the thermal-equilibrium and cosmological assumptions specified in Refs.~\cite{Escudero:2018mvt,Sabti:2019mhn}. Their applicability to modified thermal histories requires a separate~assessment.

\subsubsection{Astrophysics}

Supernovae, blazars, active galactic nuclei (AGN), and~other known astrophysical sources of neutrinos test characteristic self-interaction scales up to ${\cal O}$(10 GeV).
The presence of new self-interactions can modify the observed spectrum of high-energy neutrinos at IceCube~\cite{Kamada_2015, Ioka_2014, Ng_2014, Ibe_2014}, KM3NeT~\cite{Ambrosone:2024zrf, He:2025bex} and Baikal-GVD~\cite{Troitsky:2023nli, PhysRevD.107.042005}. In~particular, HE/UHE neutrino propagation is sensitive to resonant scattering on the C$\nu$B, which can produce absorption features in the diffuse spectrum. The~diffuse supernova neutrino background can also be attenuated by scattering on the relativistic C$\nu$B~\cite{Wang:2025qap}, and~self-interactions converting solar neutrinos into antineutrinos are constrained by inverse-beta-decay searches~\cite{Wu_2024,Bally2020}. In~this work, we build upon the single-event analysis of Ref.~\cite{Kolb:1987qy} and on later propagation studies~\cite{Bustamante:2020mep, Ng_2014, DiFranzo:2015qea, Kelly:2018tyg, Doring:2023vmk}, and~test $\nu$SI with high-energy neutrinos scattering on the C$\nu$B.

\section{Neutrino Astronomy: High-Energy Neutrino~Sources}
\label{S:sources}
Cosmic neutrino populations span many orders of magnitude in energy, from~relic-neutrino momenta of order $10^{-4}~\mathrm{eV}$ to cosmogenic neutrinos that may reach energies of order $10^{18}~\mathrm{eV}$.  Their detection is, however, restricted to several characteristic energy ranges because of the weakness of neutrino interactions and the presence of terrestrial backgrounds. These include solar neutrinos at $\mathcal{O}(0.1-10)$~MeV, supernova neutrinos at $\mathcal{O}(1-100)$~MeV, and~high- and ultra-high-energy astrophysical neutrinos from TeV to hundreds of PeV~\cite{rozhkov2024, KM3NeT:2025npi}.

The detection of a diffuse high-energy astrophysical neutrino flux  marked the beginning of a new era in neutrino astronomy. High-energy neutrino spectra have been reconstructed from multiple analyses, primarily from IceCube data. A~key milestone was the independent confirmation of high-energy neutrinos by Baikal-GVD, whose spectrum is consistent with IceCube measurements~\cite{PhysRevD.107.042005, rozhkov2024}.  Astrophysical HE and UHE neutrinos have been conclusively detected by large-volume neutrino telescopes, including IceCube, Baikal-GVD, and~ KM3NeT.

In this analysis, we focus on point-like sources of high-energy neutrinos.  
While the $\nu$SI benchmarks derived in this work are based on neutrino propagation, they depend on the source identification and distance, which are inferred from multi-messenger data. Several remarkable high-energy and ultra-high-energy neutrino sources, identified by IceCube, Baikal-GVD, KM3NeT, and~other experiments, will be used in this study, which are listed below and in Table~\ref{tab:2} with light-travel distances and redshifts, and~a representative (characteristic) neutrino energy used in our analytic single-event survival benchmarks. For~single-event associations, we take the reported reconstructed energy as representative of that event and for multi-event excesses, we adopt a characteristic energy scale and do not perform an event-by-event likelihood~analysis.

\paragraph{{SN 1987A%
}}
SN 1987A provides the historical benchmark for propagation bounds on $\nu$SI~\cite{Kolb:1987qy}.
The core-collapse supernova SN 1987A was detected on February 23, 1987 by LSD~\cite{Aglietta:1987mc}, BUST~\cite{Alekseev:1987qf}, IMB~\cite{Bionta:1987qt}  and Kamiokande II~\cite{Hirata:1987hu} from a distance of $51.4\pm1.2$~kpc~\cite{Panagia:2003rt}, and~provided strong observational support for core-collapse supernova models, in~which in the final evolutionary stage of massive stars, nuclear fuel depletion leads to gravitational core collapse, releasing immense energy. Most of this energy is carried away by neutrinos, which eventually escape from the core and propagate outward with little attenuation~\cite{rozhkov2024, Scholberg_2012}.
For SN 1987A we use a characteristic energy scale rather than a full likelihood~analysis.

\paragraph{{NGC 1068}}
NGC 1068 is a barred spiral galaxy hosting AGN at its center and is one of the closest and best-studied Seyfert II galaxies.
It stands out as one of the most prominent nearby sources of high-energy neutrinos likely produced in the extreme environment surrounding its AGN.  We adopt the Cepheid-based redshift-independent distance $D=10.72\pm0.52$~Mpc~\cite{Markham:2026ngc}. At~this distance, cosmological corrections are negligible, and~we use it as the propagation~distance.

Recent IceCube data, showing an excess of  $79^{+22}_{-20}$  muon-neutrino events, confirmed with significance $4.2 \sigma$ neutrino emission from NGC 1068, with~the dominant contribution from neutrinos in the 1.5 TeV to 15 TeV energy range~\cite{2022IceCube}. The~IceCube detection is reported as a statistical excess over background across a broad energy range. We do not attempt event-by-event signal-background classification and instead use a characteristic energy scale for the analytic optical-depth~estimate.

\paragraph{{TXS 0506+056}}
Blazars are a subclass of AGNs with a relativistic jet pointing close to the observer’s line of sight~\cite{Urry:1995mg}.  
A high-energy neutrino event detected by IceCube on 22 September 2017 was coincident in direction and time with a gamma-ray flare from the blazar TXS 0506+056~\cite{IceCube:2018dnn}.
This coincidence provided the first compelling evidence linking blazars to the high-energy astrophysical neutrino flux, marking TXS 0506+056 as the first identifiable source. 
The redshift of TXS 0506+056 is $z = 0.3365 \pm 0.0010$ \cite{Paiano:2018qeq}. Furthermore, in~2014-2015 during the 5-month period IceCube found a signal excess of $13 \pm 5$ muon-neutrino events at the coordinates of TXS 0506+056~\cite{2018}. For~TXS 0506+056 we use the 2017 alert event energy. The~2014-2015 IceCube excess is not used in the present~estimate.

\paragraph{{PKS 0735+178}}
In December 2021, multiple high-energy neutrino events were detected in temporal and spatial coincidence with the record gamma-ray flare of the blazar PKS 0735+178. Neutrino telescopes IceCube~\cite{VERITAS:2023eso}, Baikal-GVD~\cite{Dik_2023, Troitsky:2023nli},  and~BUST~\cite{Petkov:2022fnz}, all reported events: IceCube detected a 172 TeV track-like event (IceCube-211208A), Baikal-GVD observed a 43~TeV cascade event (GVD20211208CA), and~BUST registered a muon neutrino with energy above 1 GeV four days prior. PKS 0735+178 is a candidate association supported by reports from several neutrino telescopes~\cite{Troitsky:2023nli, Omeliukh:2024kgk, Sahakyan_2022}.  Reported estimates of the redshift of PKS 0735+178 include $z \sim 0.65$ \cite{Falomo2021} or $z = 0.45\pm 0.06$ \cite{Nilsson2012}. In~this work we adopt the commonly used value $z=0.65$ as a representative choice.
For the survival benchmarks shown in this manuscript, we use the IceCube-211208A reconstructed energy as the representative input. The~Baikal-GVD and BUST reports are cited as supporting multi-messenger context but are not incorporated into the present survival~estimate.

\paragraph{{KM3-230213A neutrino event}}
The most energetic neutrino event ever observed, near-horizontal KM3-230213A, was detected by the KM3NeT neutrino telescope on 13 February 2023. The~KM3NeT Collaboration estimates the median neutrino energy as 220 PeV. UHE$\nu$ could originate from various astrophysical sources such as AGNs, Gamma-Ray Bursts (GRBs), Tidal Disruption Events (TDEs), or~be cosmogenic, i.e.,~produced in interactions of UHE cosmic rays (UHECR) with background photons~\cite{KM3NeT:2025npi,KM3NeT:2026blazars}.  

The origin of the event remains unclear, though~we can use four hypotheses of its origin, following the~\cite{KM3NeT:2025npi} selection, to~adopt the possible distance: galactic, local
Universe, transient, and~extragalactic origin~\cite{KM3NeT:2026galactic}. According to~\cite{KM3NeT:2025npi}, no transient source was identified. As~for extragalactic sources, the~search for counterparts has identified several blazar candidates, but~further investigations are needed to establish any definitive associations~\cite{KM3NeT:2025bxl}. A~cosmogenic origin also remains plausible under specific UHECR assumptions~\cite{KM3NeT:2025vut}.

For through-going muon track events, such as IceCube-170922A and KM3-230213A, the~reported neutrino energy has substantial uncertainty and can depend on assumptions about the parent neutrino spectrum and event reconstruction methodology. Our analysis adopts a representative energy and does not account for this~uncertainty.
 \begin{table}[htbp]
\centering
    \caption{HE and UHE neutrino sources and events with representative energies,
    light-travel distances, and~adopted redshifts.}
    \label{tab:2}
\setlength{\tabcolsep}{12pt}
\begin{tabular}{lccc}
        \toprule
       \multicolumn{1}{c}{\raisebox{1.4ex}{\textbf{Source or Event}}}
        & \colhead{\textbf{Representative}\\\textbf{Energy}}
        & \colhead{\textbf{Light-Travel}\\\textbf{Distance}}
        & \colhead{\textbf{Adopted}\\\textbf{Redshift} \boldmath{$z$}} \\
        \midrule

        SN~1987A
        & $10~\mathrm{MeV}$
        & $51.4~\mathrm{kpc}$
        & --- \\

        NGC~1068
        & $10~\mathrm{TeV}$
        & $10.72~\mathrm{Mpc}$
        & --- \\

        TXS~0506+056
        & $290~\mathrm{TeV}$
        & $1.2~\mathrm{Gpc}$
        & $0.3365$ \\

        PKS~0735+178
        & $172~\mathrm{TeV}$
        & $1.9~\mathrm{Gpc}$
        & $0.65$ \\

        \midrule
        KM3-230213A
        & $220~\mathrm{PeV}$
        & ---
        & --- \\

        \bottomrule
    \end{tabular}
\end{table}
\unskip

\section{High-Energy Neutrino Scattering on C\boldmath{$\nu$}B }
\label{S:Interaction_rate}
A flux of high-energy neutrinos traveling from their astrophysical sources to detectors on Earth may be sensitive to C$\nu$B. Potential scattering of HE neutrinos on C$\nu$B, mediated by $\nu$SI, can modify the energy spectrum and flux of the neutrinos observed at Earth. The~freely propagating C$\nu$B is an incoherent ensemble of neutrino mass eigenstates. We therefore describe the scattering of a fixed incident high-energy mass eigenstate on the background mass eigenstates $i$. For~Dirac neutrinos, the~coupling matrix remains proportional to the identity under the transformation to the mass basis, $g_{ij}=(U^\dagger gU)_{ij} = g\delta_{ij}$. For~the Majorana scalar it transforms as $(U^TgU)_{ij}$, which is not diagonal for nonzero CP phases. For~$(S,M)$ we therefore take the coupling to be universal and diagonal in the mass basis, $g_{ij}=g\delta_{ij}$, which differs from flavor universality. Thus, all three C$\nu$B mass eigenstates interact with the same coupling, although~their individual scattering rates may differ because their kinematics depend on the target masses $m_i$.

For Dirac neutrinos, the~C$\nu$B contains both relic neutrinos and relic antineutrinos, and~for a fixed incident antineutrino mass eigenstate $\bar{\nu}_a$ both populations must be included. The~relic neutrinos decoupled while ultrarelativistic and were produced through left-chiral interactions. Free streaming preserves helicity, so the C$\nu$B is helicity polarized: only the states of the left-chiral field are populated, namely negative-helicity neutrinos $\nu_{i,-}$ and positive-helicity antineutrinos $\anu_{i,+}$ \cite{Long_2014}. We assume the incident antineutrino has positive helicity, $\anu_{a,+}$. The~allowed processes with these fixed initial helicities depend on whether the incident and target states are identical, $i=a$, or~different, $i\neq a$:
\begin{align*}
&\bar{\nu}_{a,+}+\bar{\nu}_{i,+}\rightarrow\bar{\nu}_a+\bar{\nu}_a, && i=a,\ \ t+u\ \text{channels},\\
&\bar{\nu}_{a,+}+\nu_{i,-}\rightarrow\bar{\nu}_a+\nu_a, && i=a,\ \ s+t\ \text{channels},\\
&\bar{\nu}_{a,+}+\nu_{a,-}\rightarrow\bar{\nu}_j+\nu_j, && j\neq a,\ \ s\ \text{channel},\\
&\bar{\nu}_{a,+}+\bar{\nu}_{i,+}\rightarrow\bar{\nu}_a+\bar{\nu}_i, && i\neq a,\ \ t\ \text{channel},\\
&\bar{\nu}_{a,+}+\nu_{i,-}\rightarrow\bar{\nu}_a+\nu_i, && i\neq a,\ \ t\ \text{channel},\\
&\bar{\nu}_{a,+}+\nu_{i,-}\rightarrow\phi+\phi, && i=a,\ \ t+u\ \text{channels}.
\end{align*}
For Majorana neutrinos, particle and antiparticle coincide and both target helicities are populated. For~a fixed incident mass eigenstate $\nu_{a,+}$, the~allowed processes are
\begin{align*}
&\nu_{a,+}+\nu_{a,\pm}\rightarrow\nu_a+\nu_a, && i=a,\ \ s+t+u\ \text{channels},\\
&\nu_{a,+}+\nu_{a,\pm}\rightarrow\nu_j+\nu_j, && j\neq a,\ \ s\ \text{channel},\\
&\nu_{a,+}+\nu_{i,\pm}\rightarrow\nu_a+\nu_i, && i\neq a,\ \ t\ \text{channel},\\
&\nu_{a,+}+\nu_{a,\pm}\rightarrow\phi+\phi, && i=a,\ \ t+u\ \text{channels}.
\end{align*}
The scalar vertex is diagonal in the mass basis, so the $u$-channel is absent for $i\neq a$. In~the UR limit, the~$s$-channel requires equal initial helicities, whereas mediator-pair production requires opposite ones, so both processes~contribute.

Figure~\ref{diagrams} illustrates the channel structure used in the calculation. For~identical states, including Majorana neutrinos, the~$s$-, $t$- and $u$-channel amplitudes enter with relative signs fixed by Fermi statistics, so that the total amplitude is antisymmetric under the exchange of identical fermions. For~different states, only the allowed $s$- or $t$-channel contributions are~retained.

\begin{figure}[htbp]
\centering
    \begin{minipage}[ht]{0.47\linewidth}
        \begin{tikzpicture}
         \begin{feynman}
            \vertex (a1);
            \vertex [right=1cm of a1](a2){\(\nu_{C\nu B}\)};
            \vertex [left=1.1cm of a1] (a){\(\nu_{C\nu B}\)};
            \vertex [below=1.5cm of a1] (b1);
            \vertex [below=1.5cm of a2] (b2){\(\anu\)};
            \vertex [left=1.1cm of b1] (b){\(\anu\)};
            
    \diagram* { [layered layout, horizontal=a to a2] {
      (a)  --  (a1) -- (a2), 
      (b)  --  (b1) -- (b2), 
      (a1)--[boson, edge label'={\(\phi\)}] (b1), }
    };
            \end{feynman}
        \end{tikzpicture}
    \end{minipage}
    \hfill
    \hfill
    \begin{minipage}[ht]{0.47\linewidth}
        \begin{tikzpicture}
         \begin{feynman}
             \vertex (a1);
            \vertex [right=1cm of a1](a2);
            \vertex [above left=1cm of a1] (a){\(\nu_{C\nu B}\)};
            \vertex [above right=1cm of a2] (a3){\(\nu_{C\nu B}\)};
            \vertex [below left=1cm of a1] (b1){\(\anu  \)};
            \vertex [below right=1cm of a2] (b2){\(\anu \)};
            
    \diagram* { [layered layout, horizontal=a to a3] {
        (a)-- (a1)-- [boson, edge label'={\(\phi\)}](a2)--(a3),
        (a1)-- (b1),  
        (a2)-- (b2),}
    };
            \end{feynman}
        \end{tikzpicture}
    \end{minipage}
\hfill
    \begin{minipage}[ht]{0.47\linewidth}
        \begin{tikzpicture}
         \begin{feynman}
            \vertex (a1);
            \vertex [right=1cm of a1](a2){\(\anu_{C\nu B}\)};
            \vertex [left=1.1cm of a1] (a){\(\anu_{C\nu B}\)};
            \vertex [below=1.5cm of a1] (b1);
            \vertex [below=1.5cm of a2] (b2){\(\anu \)};
            \vertex [left=1.1cm of b1] (b){\(\anu \)};
            
    \diagram* { [layered layout, horizontal=a to a2] {
      (a)  -- (a1) --(a2), 
      (b)  -- (b1) -- (b2), 
      (a1)--[boson, edge label'={\(\phi\)}] (b1), }
    };
            \end{feynman}
        \end{tikzpicture}
    \end{minipage}
    \hfill
    \hfill
   \begin{minipage}[ht]{0.47\linewidth}
        \begin{tikzpicture}
         \begin{feynman}
            \vertex (a1);
            \vertex [right=1cm of a1](a2){\(\anu_{C\nu B}\)};
            \vertex [left=1.1cm of a1] (a){\(\anu_{C\nu B}\)};
            \vertex [below=1.5cm of a1] (b1);
            \vertex [below=1.5cm of a2] (b2){\(\anu \)};
            \vertex [left=1.1cm of b1] (b){\(\anu \)};
            
    \diagram* { [layered layout, horizontal=a to a2] {
      (a)  --  (a1) -- (b2), 
      (b)  --  (b1) -- (a2), 
      (a1)--[boson, edge label'={\(\phi\)}] (b1), }
    };
            \end{feynman}
        \end{tikzpicture}
\end{minipage}  
\caption{The %
 tree-level Feynman diagrams illustrating the same-state and different-state neutrino scattering channels. The~diagrams are shown for Dirac neutrinos. For~Majorana neutrinos the topologies are the same, while the direction of the fermion flow is not fixed and is chosen independently for each~diagram.}
\label{diagrams}
\end{figure}
\unskip

\subsection{Cross-Section}

The processes listed above provide the building blocks for the total interaction rate. We calculate their differential cross sections for fixed (helicity-polarized) initial helicities, corresponding to the populated relic states specified above. The~differential cross sections for the $(V{-}A,D)$, $(S,M)$ and $(S,D)$ models are given in Tables%
~\ref{tab:VA_FH_diff}, \ref{tab:ScalarMajorana} and \ref{tab:SD_FH_diff}, respectively, and~the corresponding total cross sections in Tables~\ref{tab:VA_FH_tot}, \ref{tab:SM_av_tot} and \ref{tab:SD_FH_tot}. In~the same tables (Appendices~\ref{app:vectorD}--\ref{app:scalarM}), we provide pure-vector cross sections as well as spin-averaged ones, which correspond to equally populated relic helicities for the parity-even pure-vector and scalar interactions. For~massless fermions with fixed helicities, the~$V{-}A$ amplitudes of the same-flavor processes coincide with the corresponding helicity amplitudes of M{\text \o}ller and Bhabha scattering with the photon propagator replaced by the massive one, providing a useful~cross-check.

Neutrino-mass-suppressed contributions to the matrix elements and cross sections can generally be neglected within the present analysis, since the relevant center-of-mass energies are much larger than the relic-neutrino masses. The~exception is the chirality-flip $s$-channel in $(S,D)$, where the mass term is the leading contribution. The~neutrino masses, however, remain important through the target kinematics, in~particular through $s\simeq 2Em_i$ for a non-relativistic~target.

\subsection{Interaction~Rate}

The interaction rate is obtained by averaging the corresponding scattering cross section over the momentum distribution of the C$\nu$B targets, %
\be
\mathcal R_{i}(E)=\int \frac{d^3\textbf{p}_i}{(2\pi)^3}f_i(\textbf{p}_i)\;v_{M\text{\o}ller}\;\sigma_i(s),\label{Gamma}
\ee
where $\textbf{p}_i$ and $f_i(\textbf{p}_i)$ are the momentum and phase-space distribution of the background mass eigenstate $i$, and~$v_{M\text{\o}ller}$ is the M\text{\o}ller velocity. We denote interaction rates by $\mathcal R$ and reserve $\Gamma_\phi$ for the mediator width. Here $\mathcal R_i$ includes scattering on all populated target states of the mass eigenstate $i$, summed over all allowed final~states.

The present-day C$\nu$B may contain both non-relativistic and ultra-relativistic neutrino mass eigenstates, depending on the absolute neutrino mass scale. The~measured neutrino mass splittings imply that at least two mass eigenstates are non-relativistic today, while the lightest state may remain relativistic. We therefore treat the NR and UR regimes as two limiting cases of the same background mass eigenstate $i$:
{\begin{align}
&\text{NR:}\;\;
\displaystyle\frac{|\textbf{p}_{i}|}{E_i}\rightarrow 0, \quad
s\rightarrow m_i^2+2m_iE\simeq2m_iE, \quad
v_{M\text{\o}ller}= 1,\\
&\text{UR:}\;\;
\displaystyle\frac{|\textbf{p}_{i}|}{E_i}\rightarrow 1,\quad s\rightarrow 2EE_i(1-c_{\theta}), \quad v_{M\text{\o}ller}=\frac{s}{2 E E_i},
\end{align}
}
with corresponding interaction rates
\bea
&&\mathcal R_i^{\rm NR}= n_i\;\sigma_i(2Em_i),\qquad n_i=\int\frac{d^3\textbf{p}_i}{(2\pi)^3}f_i(\textbf{p}_i),
\\
&&\mathcal R_i^{\rm UR}=  \frac{1}{4\pi^2}\int_0^\infty d|\textbf{p}_i|\,|\textbf{p}_i|^2 f_i(|\textbf{p}_i|)
\int_{-1}^{1}dc_\theta\,
(1-c_\theta)\,\sigma_i(c_{\theta}, E_i).
\eea
Here $E$ is the incident HE$\nu$ energy, $m_i$ is the mass of the background eigenstate, $E_i$ is its C$\nu$B energy, $c_{\theta}=\cos\theta$, and~$\theta$ is the angle between the incident and the background neutrino.
The present-day number density of each populated relic state is taken to be $n_i=56~\mathrm{cm}^{-3}$, with~the Fermi--Dirac momentum distribution $f_i(p)=[1+e^{|\textbf{p}_i|/T_{C\nu B}}]^{-1}$ and a present-day effective temperature of $1.68\times10^{-4}$ eV~\cite{Lesgourgues:2006nd}. The~UR rate functions use the Fermi--Dirac expansion of Appendix~\ref{app:conventions}. Maxwell--Boltzmann expressions provide the analytic building blocks, with~the computational prescriptions summarized in Table~\ref{tab:UR_prescriptions}.

For a fixed incident mass eigenstate $a$, the~interaction rate due to all three C$\nu$B target mass eigenstates is
\be
\mathcal R_a
=
\sum_{i=1}^{3}\mathcal R_{ai},
\label{total_CnuB_rate}
\ee
where $\mathcal R_{ai}$ is the rate of Equation~\eqref{Gamma} on the target eigenstate $i$. In~Equation~\eqref{Gamma} and in the degenerate case below, the~index $a$ is~suppressed.

To illustrate the characteristic behaviour of the interaction models and isolate the kinematic differences between the NR and UR regimes, we first consider a degenerate neutrino mass spectrum, $m_1\simeq m_2\simeq m_3\equiv m$. In~this illustrative case, all three C$\nu$B mass eigenstates are simultaneously either NR or UR. For~Dirac neutrinos with the helicity-polarized C$\nu$B and a fixed incident state, the~combined rate can be written in terms of the channel classes
\be
\mathcal R^{\rm deg}_D  = \mathcal R_{t+u}  + \mathcal R_{s+t}  + 2\mathcal R_{t++} + 2\mathcal R_{t+-}  + 2\mathcal R_s + \mathcal R_{\rm pair}.
\label{degenerate_total_rate}
\ee
The first two terms describe scattering on the same-state background populations $\bar{\nu}_{a,+}$ and $\nu_{a,-}$, respectively. The~factor $2$ in the following terms counts the two target states with $i\neq a$ for $\mathcal R_{t++}$ and $\mathcal R_{t+-}$, the~$t$-channel rates on $\bar{\nu}_{i,+}$ and $\nu_{i,-}$, and~the two final species $j\neq a$ for $\mathcal R_s$, the~$s$-channel rate per final species, and~$\mathcal R_{\rm pair}$ is the mediator-pair production on $\nu_{a,-}$. For~a scalar mediator $\mathcal R_{t++}=\mathcal R_{t+-}$.

For Majorana neutrinos, the~combined rate is
\be
\mathcal R^{\rm deg}_{M}=\mathcal R_{s+t+u}+2\mathcal R_s+2\mathcal R_t+\mathcal R_{\rm pair},
\label{degenerate_total_rate_M}
\ee
where each term includes both target helicities. The~first term describes scattering on the same mass eigenstate, and~the factor $2$ in the next two terms counts the final species $j\neq a$ and the target states $i\neq a$, respectively.

\subsection{Resonant~Scattering}
\label{sec:resonant}

We separate the thermally averaged rate into the resonant $s$-channel pole and an off-resonant remainder, $\mathcal R=\mathcal R_{\rm off}+\mathcal R_{\rm res}$. Near~the pole the propagator is regularized by the mediator width, $\Gamma_\phi\propto g^2M$, so the integrated pole scales as $g^2$ in the narrow-width approximation (NWA) \cite{Doring:2023vmk}, whereas the off-resonant remainder scales as $g^4$. The~off-resonant $s$-channel is obtained by subtracting the NWA pole from the finite-width thermal integral before taking its zero-width limit. Equivalently, it is the finite part of the thermal integral with the bare propagator (Appendix~\ref{app:pole}). The~$s$--$t$ interference has no inverse-width enhancement and contributes to the~remainder.

Regardless of the background regime, thermal motion of the relic neutrinos broadens the resonance~\cite{Doring:2023vmk,Wang:2025qap}. The~broadening is weakest for NR targets, $\delta M/M\sim T_{C\nu B}/m_i$, and~becomes of order unity in the UR~regime.

In the approximation of strictly massless UR targets with a Fermi--Dirac momentum distribution, the~resonant rate for a fixed incident mass eigenstate, summed over open final species, is
\begin{equation}
\mathcal R_{\rm res}
=\kappa\,\frac{g^2}{2\pi}
\frac{T_{C\nu B}M^2}{(2E)^2}
\ln\left[1+\exp\left(-\frac{M^2}{4ET_{C\nu B}}\right)\right],
\label{eq:res_massless}
\end{equation}
which is maximal at $M^2\simeq4.6\,ET_{C\nu B}$. Here $\kappa=1$ for $(V{-}A,D)$ and $\kappa=\tfrac12$ for $(S,M)$. The~massless UR approximation serves as a transparent and simple analytic~reference.

For the main scenarios, $(V{-}A,D)$ and $(S,M)$, retaining a finite relic mass suppresses the low-$M$ side of the resonance for $M^2\ll2Em_i$. Away from this suppression, the~pole retains a similar overall profile to the massless UR result. The~scalar Dirac comparison case, $(S,D)$, behaves differently, as~its pole is helicity suppressed and proportional to $m_i^2$. The~corresponding pole rates are given in Appendix~\ref{app:URmass}.

For NR targets, the~resonance is rather narrow and centred at $M^2=2Em_i$. For~a fixed incident energy and a relic at rest, the~resonance would be a Breit--Wigner of width $\Gamma_\phi\propto g^2M$, and~the rate at the pole would become independent of $g$. The~relative thermal spread is $\delta s/s\sim p_i/m_i$, with~$s\simeq2E(E_i-p_i\cos\theta)$. This spread exceeds $\Gamma_\phi/M$ by many orders of magnitude for the couplings considered. The~pole is therefore evaluated in the NWA with this thermal broadening, which regularizes the resonance and gives it a finite depth. For~sources at cosmological distances, the~resonance is further spread by the redshift evolution along the path (Section~\ref{sec:redshift}). In~the semi-relativistic regime, $m_i\sim T_{C\nu B}$, neither limiting form applies, and~the pole rates are evaluated numerically from the full expressions. The~massive-target kernels for arbitrary $m_i/T_{C\nu B}$ are given in Appendix~\ref{app:NR}.

The partial mediator widths per neutrino species, which enter the resonant rates, are
\begin{equation}
\Gamma_{\phi}^{V{-}A}\simeq\frac{g^2 M}{24\pi},
\qquad
\Gamma_{\phi}^{S,D}\simeq\frac{g^2 M}{8\pi},
\qquad
\Gamma_{\phi}^{S,M}\simeq\frac{g^2 M}{16\pi}.
\end{equation}
The spin-one width corresponds to the left-chiral normalization of Equation~\eqref{Lagrangian}. The~total width sums over the same open final species as the $s$-channel~rate.

\section{Survival Criterion and~Propagation}
\label{S:survival_limits}

Despite the wide range of possible astrophysical and cosmogenic production channels for HE and UHE neutrinos, theoretical expectations and observational limits indicate that their combined flux at Earth is very low~\cite{Abbasi_2025, Kuznetsov:2025ehl}. The~detection of even a single neutrino from a distant source is therefore sensitive to interactions that would strongly suppress its probability of reaching Earth. The~attenuation accumulated along the propagation path is characterized by the optical depth $\tau$, which gives the expected number of interactions along the line of sight. The~probability that the neutrino reaches the observer without undergoing an interaction is $P_{\rm surv}=e^{-\tau}$.

\subsection{Optical Depth and Cosmological~Evolution}
\label{sec:redshift}

For nearby sources, cosmological evolution can be neglected. The~interaction rate then defines the mean free path $\lambda$,
\be
\lambda^{-1}
\simeq
\mathcal R,
\label{total_CnuB_mean_free_path}
\ee
where $\mathcal R$ is the total interaction rate, and~the optical depth takes the intuitive form
\be
\tau
\simeq
D\,/\lambda, \label{nearby_optical_depth}
\ee
where $D$ is the Euclidean distance to the~source.

For sources at cosmological distances, the~neutrino energy decreases due to cosmological redshift, while the density and temperature of the C$\nu$B also decrease as the Universe expands. This modifies the resonance condition, the~scattering cross sections and the target density along the propagation path. For~a neutrino emitted at redshift $z_s$, the~optical depth~becomes
\be
\tau
=
\int_0^{z_s}
\frac{dz}{(1+z)H(z)}\,
\mathcal R\bigl(E(1+z),z\bigr),
\label{eq:tau_z}
\ee
where $E$ is the neutrino energy observed at Earth and $H(z)=H_0\sqrt{\Omega_m(1+z)^3+\Omega_\Lambda}$, with~the radiation contribution neglected. We use $H_0=67.4~\mathrm{km\,s^{-1}Mpc^{-1}}$ and $\Omega_m=1-\Omega_\Lambda=0.315$ \cite{Planck:2018vyg}. The~integral of the propagation measure alone gives the light-travel distance,
\be D = \int_0^{z_s} \frac{dz}{(1+z)H(z)},\ee
which reduces to the Euclidean distance for nearby sources. The~redshift dependence of the scattering rate follows from the substitutions~\cite{DiFranzo:2015qea, Blum:2014ewa, Doring:2023vmk}
\be
\text{NR:}\;\; E \rightarrow E (1+z), \;\; T_{C\nu B} \rightarrow T_{C\nu B} (1+z), \;\; n_{C\nu B} \rightarrow n_{C\nu B} (1+z)^3,
\ee
\be
\text{UR:}\;\; E \rightarrow E (1+z), \;\; T_{C\nu B} \rightarrow T_{C\nu B} (1+z), \;\; E_{i} \rightarrow E_{i} (1+z).
\ee

For the nearby sources SN~1987A and NGC~1068, the~optical depth is given by \mbox{Equation~\eqref{nearby_optical_depth}}. For~TXS~0506+056, PKS~0735+178 and the distant-source hypotheses for KM3-230213A, Equation~\eqref{eq:tau_z} is evaluated~numerically.

\subsection{Single-Event Survival~Condition}

In the absence of a source-flux likelihood analysis, we adopt the
single-event survival condition
\be
\tau
\lesssim 1.
\label{total_single_event_condition}
\ee
The boundary $\tau=1$ corresponds to one expected interaction along the path~\cite{Kolb:1987qy}. A~larger threshold would correspond to an exponentially smaller survival probability, but~it would shift the survival boundary only slightly, since the coupling on the boundary scales as $\tau^{1/4}$ away from the resonance and as $\tau^{1/2}$ where the pole dominates. As~our results are not associated with a statistical confidence level, we adopt $\tau=1$ as a simple common survival benchmark, which permits a consistent comparison of the interaction scenarios considered below. At~the same time, the~detection of one event alone does not establish this condition without information about the unattenuated source flux. Such information may be obtained from electromagnetic observations of the source and from theoretical modelling of its intrinsic neutrino emission~\cite{Herrera:2023nww,Arguelles:2026rjr}, which is beyond the scope of this~work.

The coupling enters the rate in two ways. Away from the resonance, the~interaction strength enters through the scattering cross section, $\sigma\propto g^4$, so that $\mathcal R\propto g^4$. Near~the resonance, a~pole treated within the NWA scales as $g^2$ (Section~\ref{sec:resonant}). For~one incident state the rate therefore has the form $\mathcal R = Ag^4+Bg^2$, where $Bg^2=\mathcal R_{\rm res}$ is the pole contribution and $A$ contains the off-resonant contributions retained in the computational prescription. For~a fixed background, the~condition $\tau = D\mathcal R=1$ is a quadratic equation in $g^2$, with~solution
\be
g^2=\frac{2}{D\left(B+\sqrt{B^2+4A/D}\right)}.
\label{eq:g_general}
\ee
It has two limiting cases. Away from the resonance the pole is negligible, $B\to0$, and \mbox{Equation~\eqref{eq:g_general}} reduces to the off-resonance limit
\be
g\leq
\left(\frac{1}{\tau\vert_{g=1}}
\right)^{1/4}=(DA)^{-1/4},
\label{coupling_limit_from_optical_depth}
\ee
where $\tau|_{g=1}$ is the optical depth evaluated at unit coupling~\cite{Kolb:1987qy}. For~nearby sources this bound takes the particularly transparent form
\be
g \leq \left(\frac{\lambda|_{g=1}}{D}\right)^{1/4}.
\ee
When the pole dominates, $A\to0$, it reduces to $g^2 = 1/(DB)$.

\section{Survival Benchmarks for a Degenerate C\boldmath{$\nu$}B}
\label{sec:degenerate}

\begin{figure}[htbp]
\centering
\includegraphics[width=0.48\linewidth]{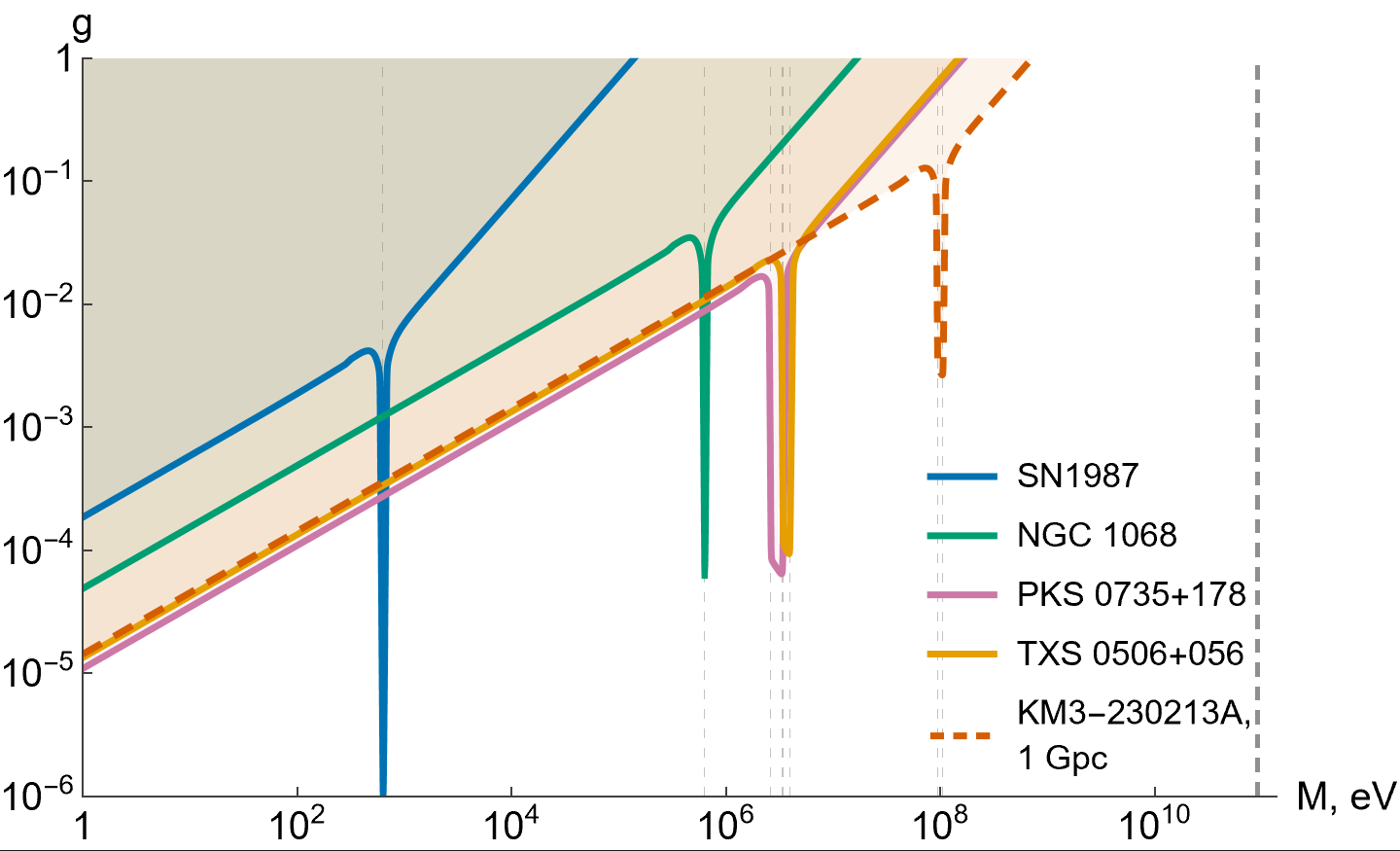}\hfill
\includegraphics[width=0.48\linewidth]{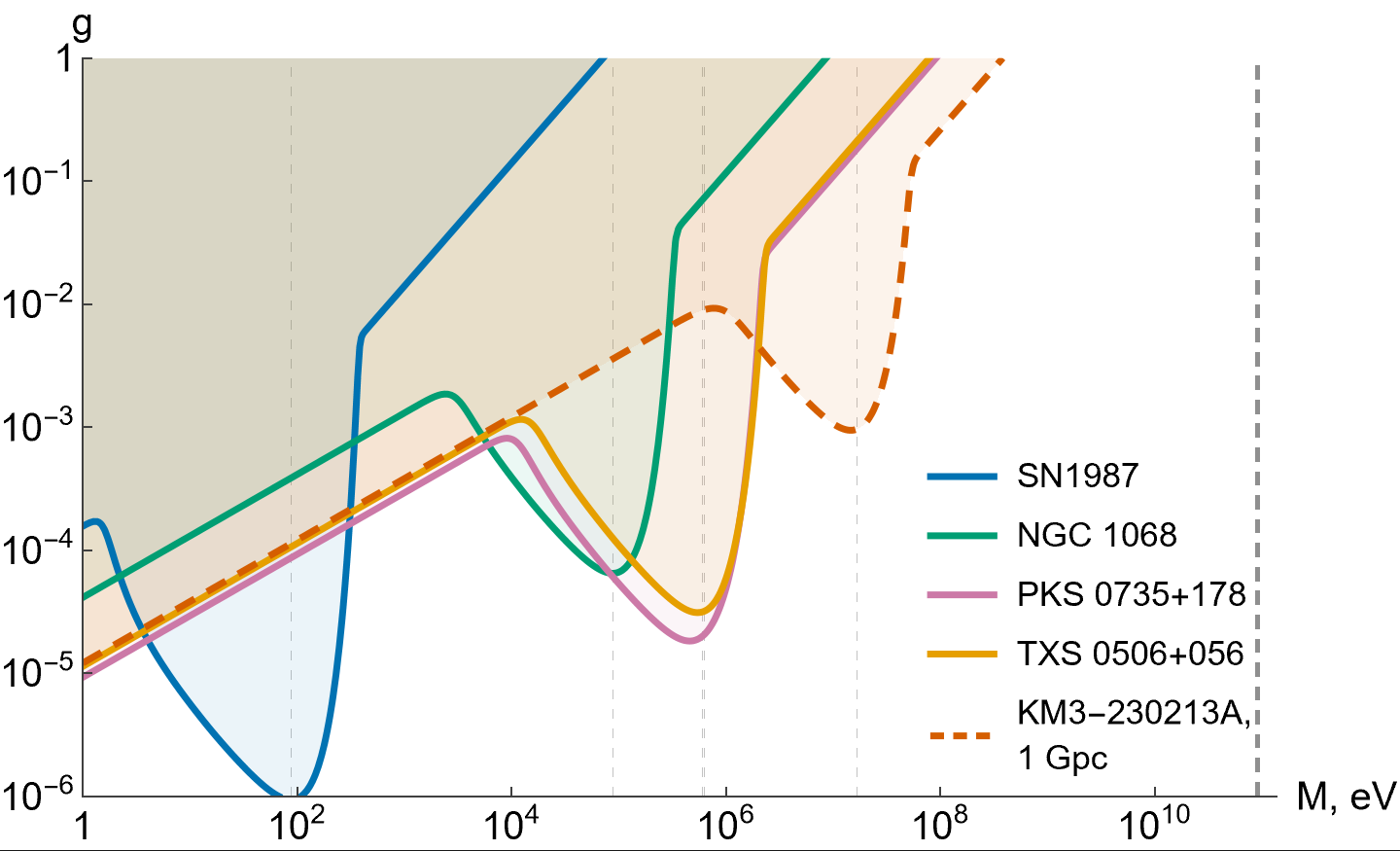}
\caption{Single-event survival benchmarks on $g$ as a function of $M$ for $(V{-}A,D)$ with a helicity-polarized C$\nu$B and a degenerate relic mass spectrum. Left: NR benchmark with $m=2\times10^{-2}~\mathrm{eV}$. Right: UR benchmark with $m=10^{-5}~\mathrm{eV}$ and the finite-mass pole prescription. The~curves use the source inputs in Table~\ref{tab:2}. For~KM3-230213A, a~light-travel distance of $1~\mathrm{Gpc}$ is assumed and the curve represents a conditional sensitivity. The~shaded regions correspond to $\tau>1$. The~thick vertical dashed line marks the $Z$-boson~mass.}
\label{fig:nrur}
\end{figure}

\begin{figure}[htbp]
\centering
\centering
\includegraphics[width=0.48\linewidth]{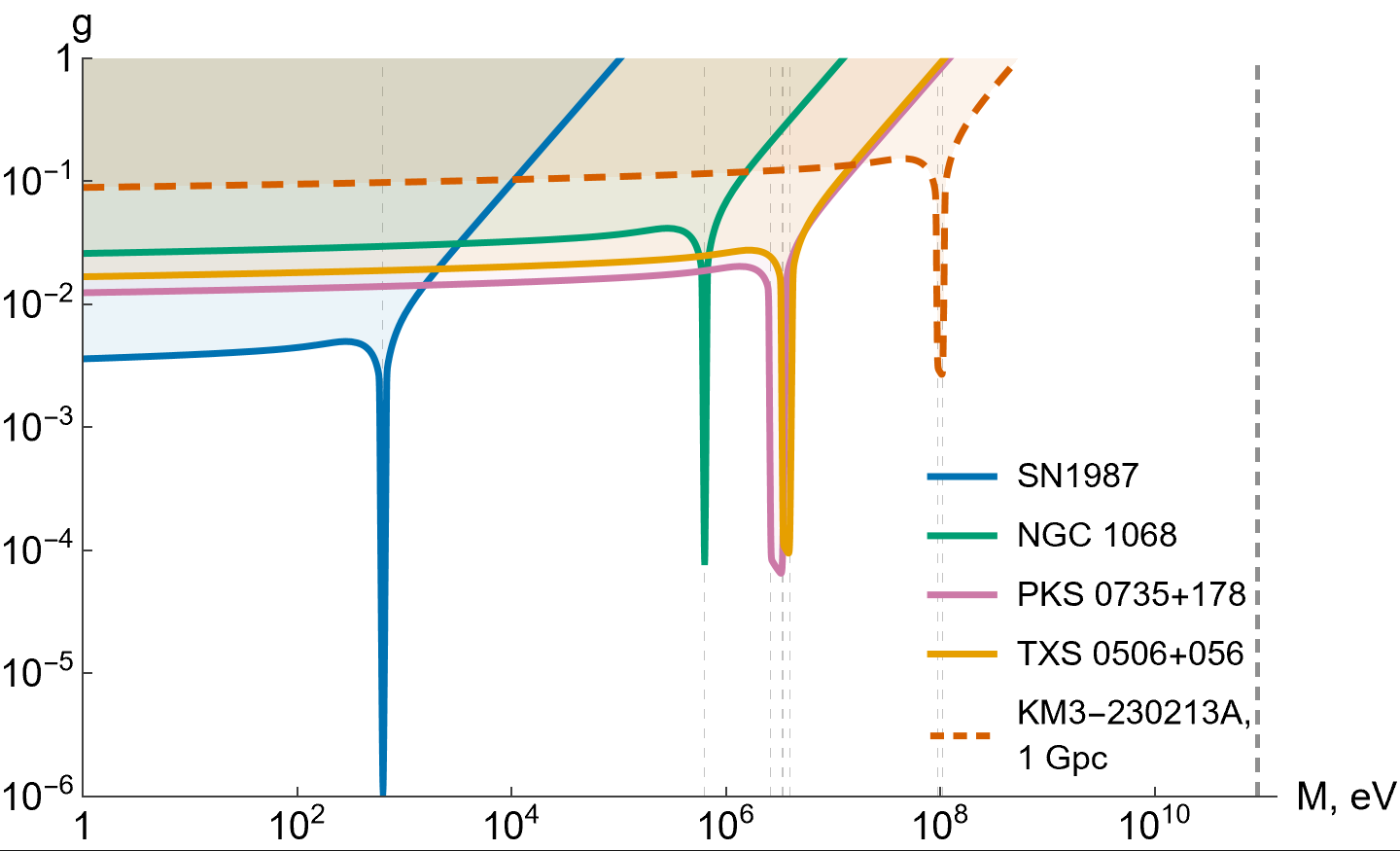}\hfill
\includegraphics[width=0.48\linewidth]{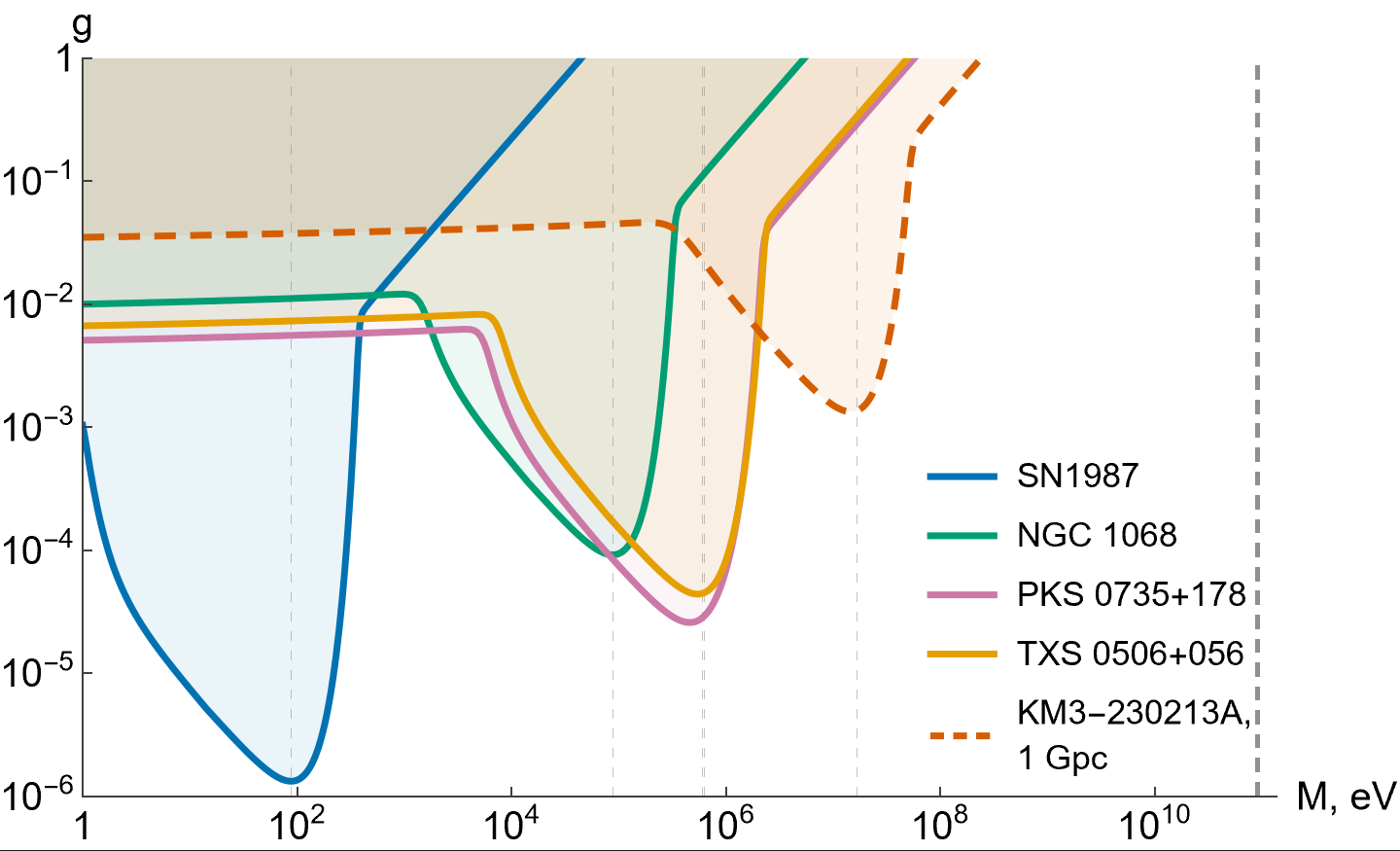}
\caption{Single-event survival benchmarks on $g$ as a function of $M$ for $(S,M)$ with a degenerate relic mass spectrum, including both populated target helicities. Left: NR benchmark with $m=2\times10^{-2}~\mathrm{eV}$. Right: UR benchmark with $m=10^{-5}~\mathrm{eV}$ and the finite-mass pole prescription. Sources, distances and shading are as in Figure~\ref{fig:nrur}. The~thick vertical dashed line marks the $Z$-boson~mass.}
\label{fig:nrur_majorana}
\end{figure}

\begin{figure}[htbp]
\centering
\includegraphics[width=0.48\linewidth]{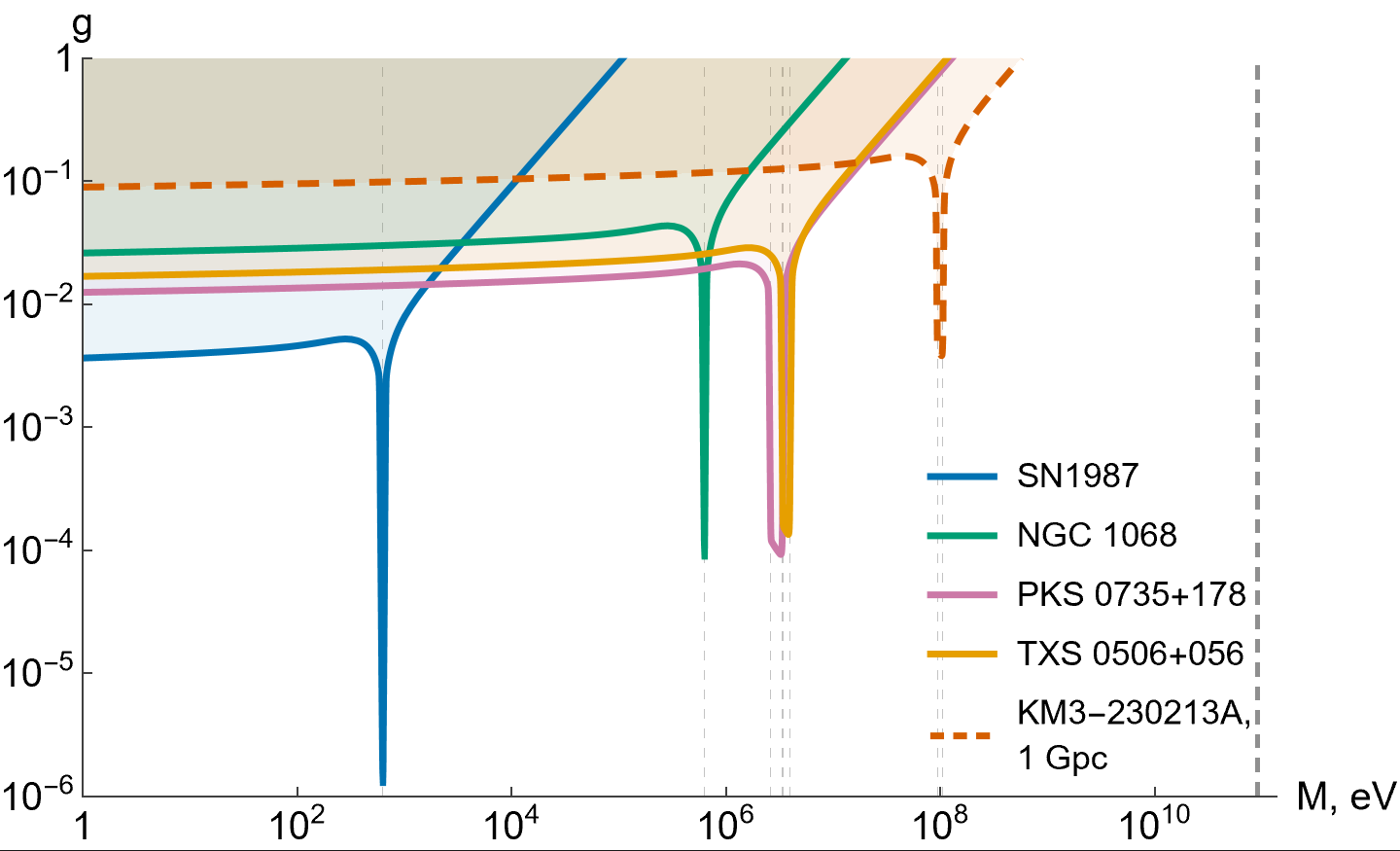}\hfill
\includegraphics[width=0.48\linewidth]{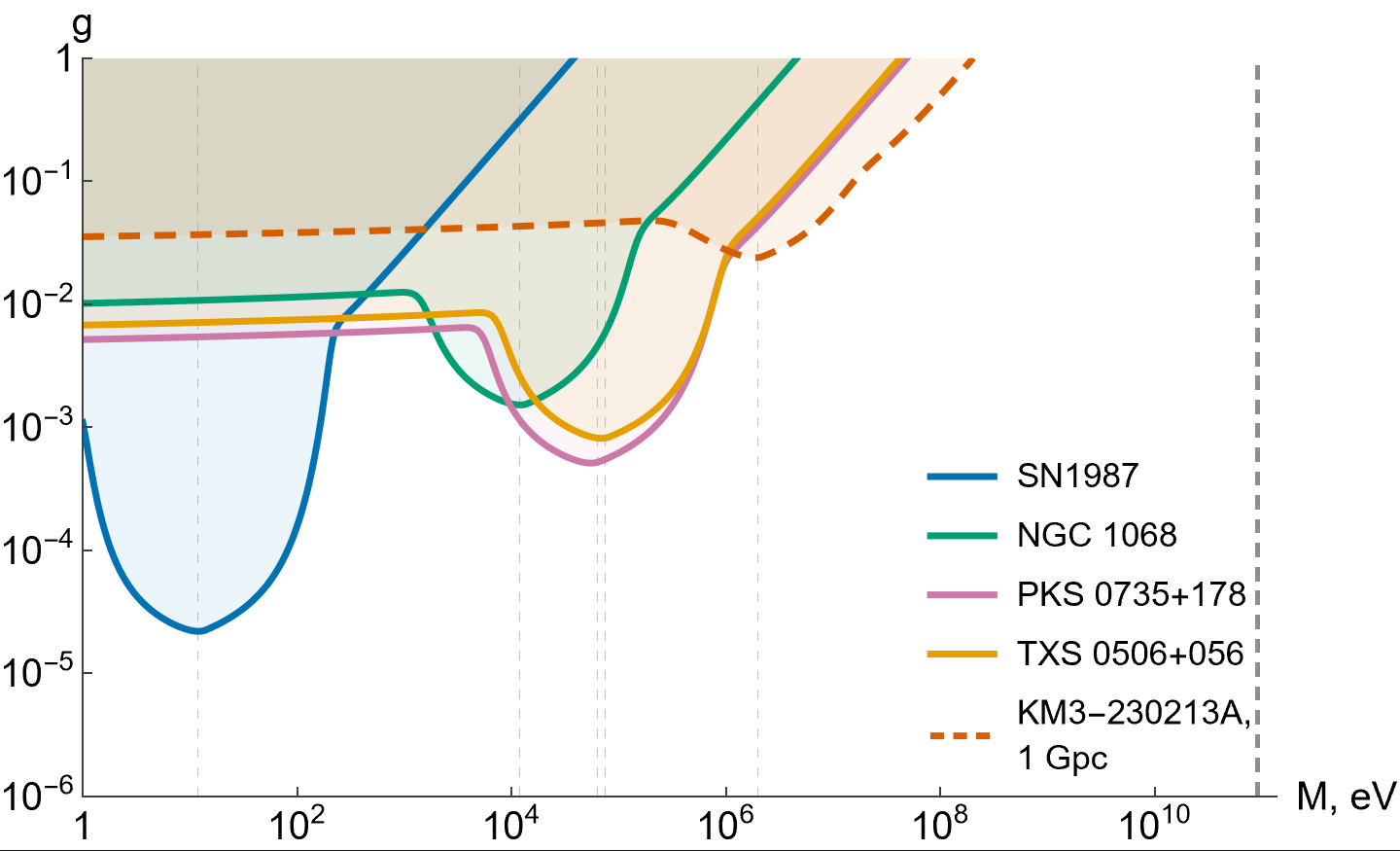}
\caption{Single-event survival benchmarks on $g$ as a function of $M$ for the comparison case $(S,D)$, evaluated on the standard helicity-polarized C$\nu$B with a degenerate relic mass spectrum (Section~\ref{sec:HPvsSA}). Left: NR benchmark with $m=2\times10^{-2}~\mathrm{eV}$. Right: UR benchmark with $m=10^{-5}~\mathrm{eV}$, for~which the helicity-suppressed pole is set by the finite relic mass. Sources, distances and shading are as in Figure~\ref{fig:nrur}. The~thick vertical dashed line marks the $Z$-boson~mass.}
\label{fig:nrur_scalar}
\end{figure}

We derive single-event survival benchmarks using SN~1987A, TXS~0506+056, PKS~0735+178, NGC~1068 and the KM3-230213A event. The~source associations and representative energies and distances are specified in Section~\ref{S:sources} and Table~\ref{tab:2}. Figures~\ref{fig:nrur}, \ref{fig:nrur_majorana} and \ref{fig:nrur_scalar} show the $(V{-}A,D)$, $(S,M)$ and $(S,D)$ results, respectively, with~NR panels on the left and UR panels on the right. Both scenarios use the standard relic populations: a helicity-polarized C$\nu$B for Dirac neutrinos, and~both helicities of each mass eigenstate for Majorana~neutrinos.

The degenerate spectra are illustrative limiting cases. We use $m=2\times10^{-2}~\mathrm{eV}$ for the NR benchmark and $m=10^{-5}~\mathrm{eV}$ for the UR benchmark, unless~another mass is explicitly indicated. The~UR curves retain the finite-mass poles of Equations%
~\eqref{eq:Bres_VA}--\eqref{eq:W0_UR} and \eqref{eq:res_massive}--\eqref{eq:res_massive_SD}. The~redshift evolution of TXS~0506+056, PKS~0735+178 and KM3-230213A is included through Equation~\eqref{eq:tau_z}.

\vspace{-6pt}

\vspace{-6pt}

\subsection{Spin-One Versus Scalar~Exchange}

\begin{figure}[htbp]
\centering
\includegraphics[width=0.48\linewidth]{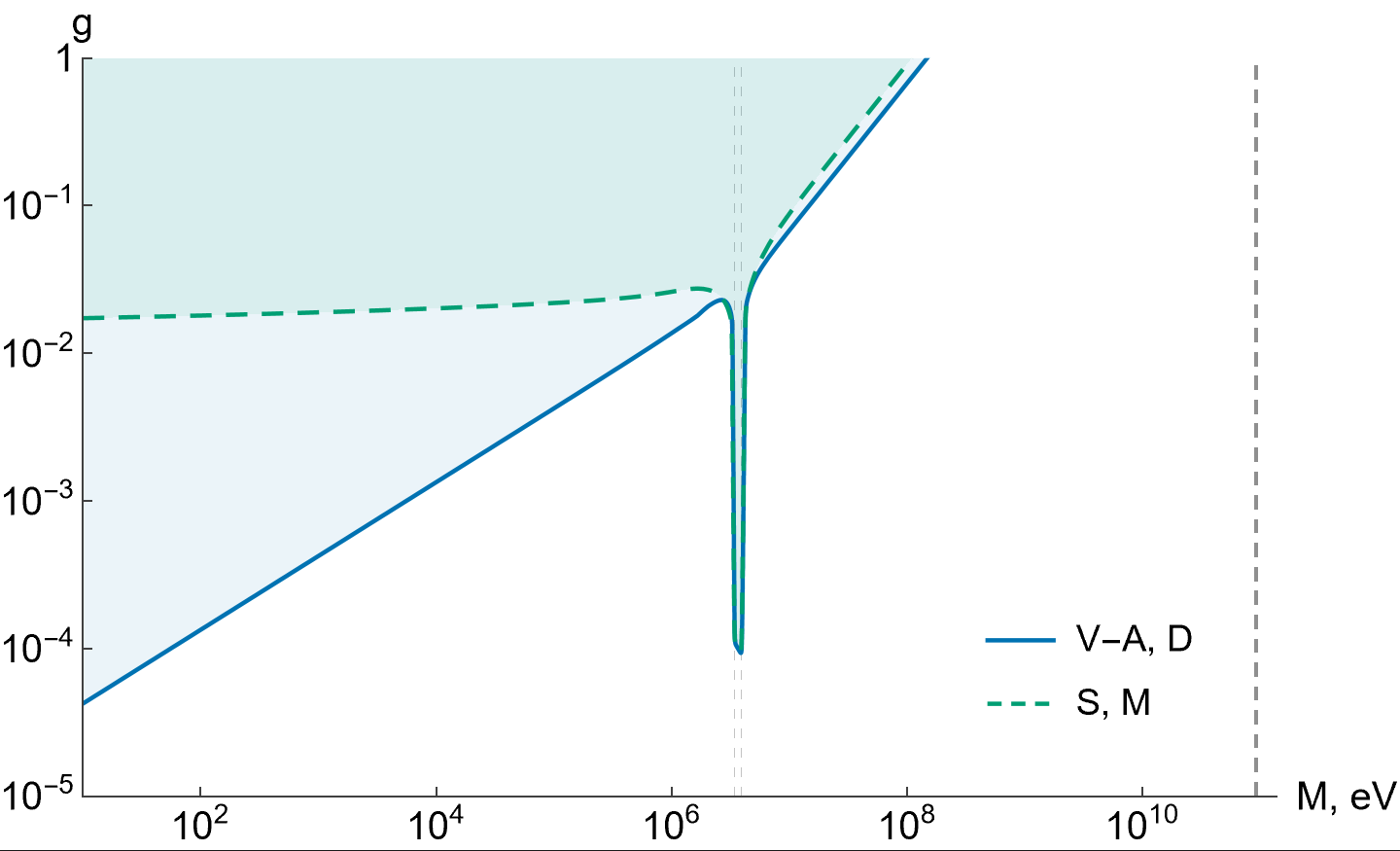}\hfill
\includegraphics[width=0.48\linewidth]{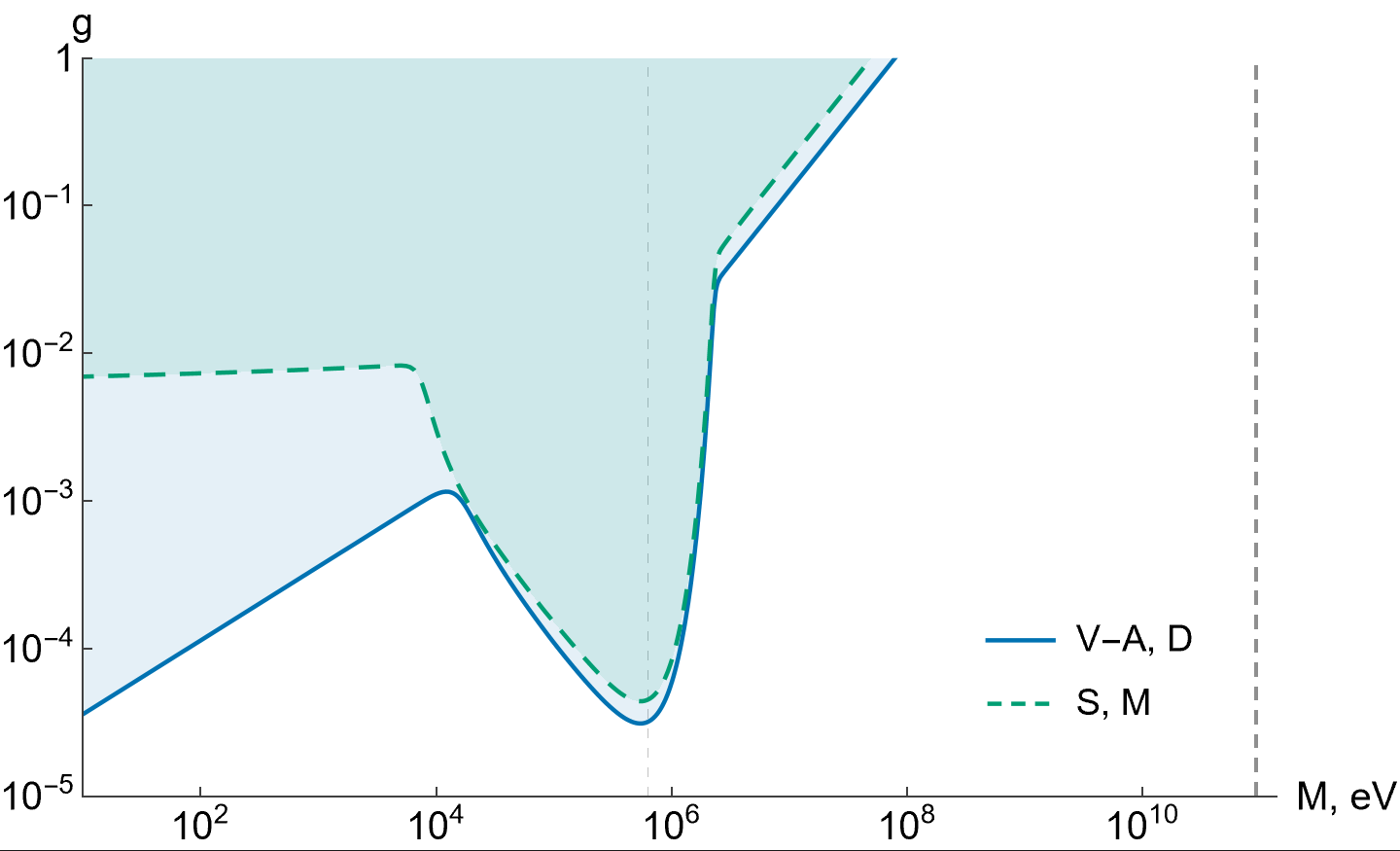}
\caption{{Comparison %
 of $(V{-}A,D)$ (solid) and $(S,M)$ (dashed) single-event survival benchmarks for TXS~0506+056 and a degenerate C$\nu$B. Left: NR regime with $m=2\times10^{-2}~\mathrm{eV}$, with~vertical lines marking $M=\sqrt{2Em}$ and $\sqrt{2E(1+z_s)m}$. Right: UR regime with $m=10^{-5}~\mathrm{eV}$. The~thick vertical dashed line marks the $Z$-boson mass. The shaded regions indicate $\tau>1$.}}
\label{fig:scalar}
\end{figure}

Figure~\ref{fig:scalar} compares the main spin-one and spin-zero scenarios, $(V{-}A,D)$ and $(S,M)$. In~the pre-resonance region, the~forward-enhanced spin-one $t$-channel, $\sigma_t\simeq g^4/(4\pi M^2)$ for $M^2\ll s$, gives substantially stronger survival benchmarks than scalar exchange, for~which $\sigma_t\propto g^4/s$. The~observable significance of this enhancement depends on the energy acceptance (Section~\ref{sec:cutoff}). Most of the difference is therefore confined to lower mediator masses. In~the UR regime, the~broad resonance also reflects the different pole coefficients discussed in Section~\ref{sec:resonant}.
\unskip

\subsection{Scalar Mediator: Majorana Versus Dirac~Neutrinos}
\label{sec:HPvsSA}

\begin{figure}[htbp]
\centering
\includegraphics[width=0.48\linewidth]{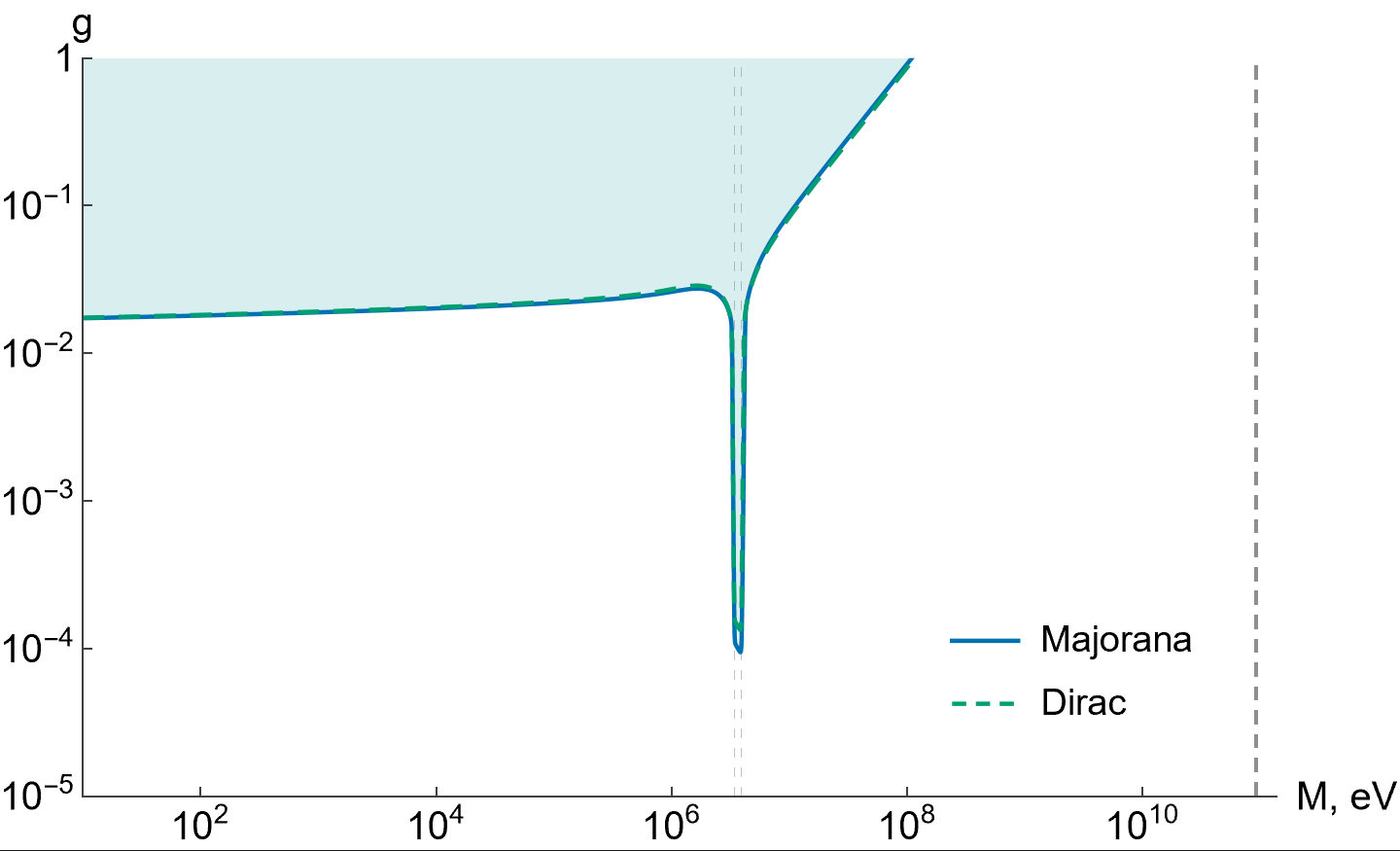}\hfill
\includegraphics[width=0.48\linewidth]{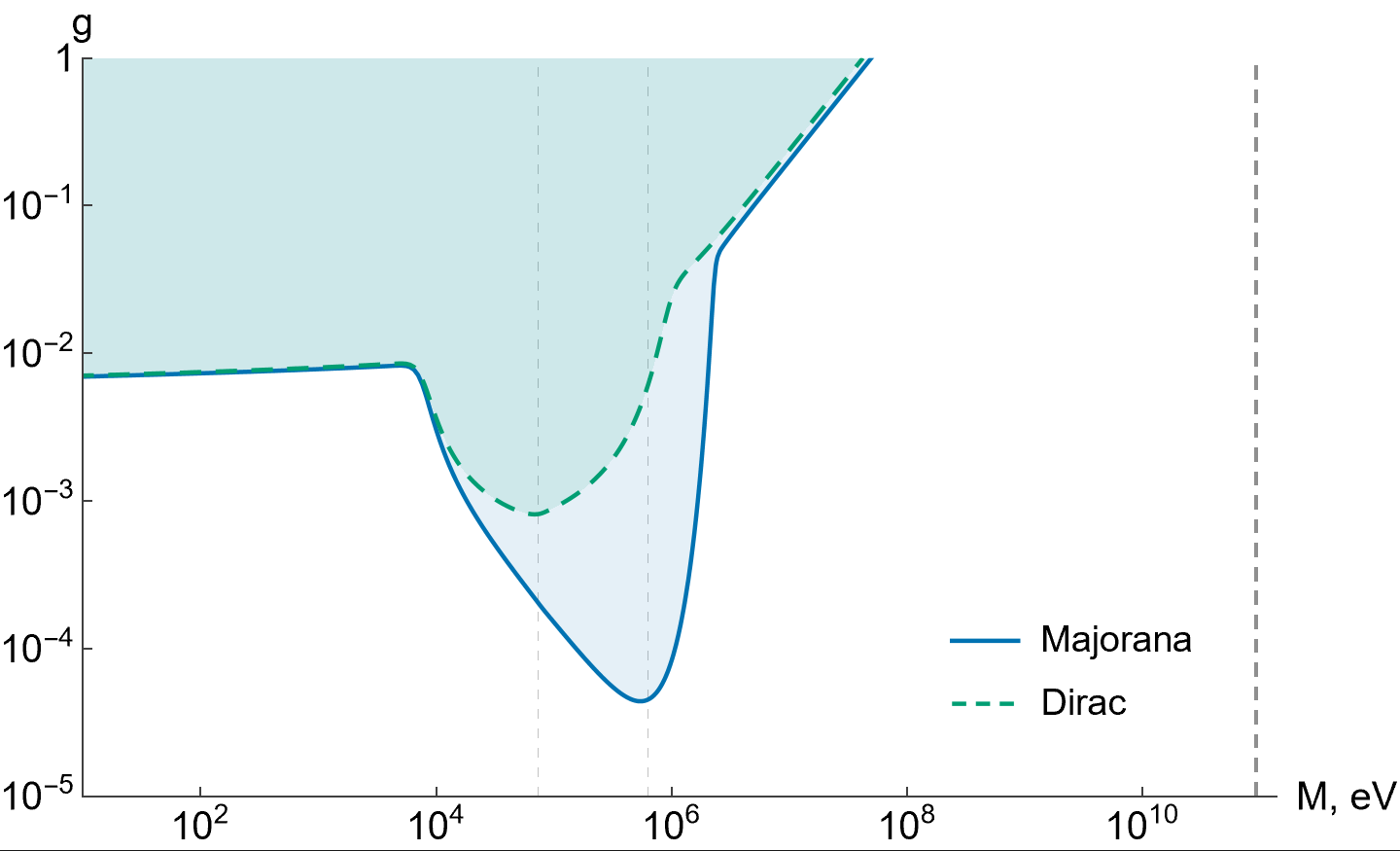}
\caption{Scalar-mediated %
 single-event survival benchmarks for Majorana neutrinos (solid) and Dirac neutrinos (dashed), using TXS~0506+056 and a degenerate C$\nu$B. Left: NR regime with $m=2\times10^{-2}~\mathrm{eV}$. Right: UR regime with $m=10^{-5}~\mathrm{eV}$. Dirac rates use the populated relic helicities, while Majorana rates include both target helicities. The~thick vertical dashed line marks the $Z$-boson~mass. The shaded regions indicate $\tau>1$.}
\label{fig:majorana}
\end{figure}

Scalar exchange with Dirac neutrinos couples left- and right-chiral fields. In~a standard thermal history, the~couplings considered here bring the right-handed states into equilibrium in the early Universe, and~the $(S,D)$ parameter space shown here is in strong tension with cosmological constraints on $N_{\rm eff}$ \cite{Luo:2020sho,Luo:2020fdt,Abazajian:2019oqj,Blinov_2019}. The~$(S,D)$ results therefore do not constrain this model in a standard cosmology. We include this scenario, together with the spin-averaged case discussed in Appendix~\ref{app:refs}, as~a comparison with $(S,M)$ that shows how the benchmarks depend on the neutrino nature for the same mediator, and~because scalar exchange is a common benchmark in propagation studies~\cite{Ng_2014,Ioka_2014,Kelly:2018tyg,Bustamante:2020mep}. Figure~\ref{fig:majorana} compares the scalar Majorana and Dirac benchmarks for TXS~0506+056. The~two cases have different channel combinations and multiplicities (Section~\ref{S:Interaction_rate}). They behave similarly in the light- and heavy-mediator regions and at the NR resonance, but~differ significantly at the UR~resonance.

\begin{figure}[htbp]
\centering
\includegraphics[width=0.48\linewidth]{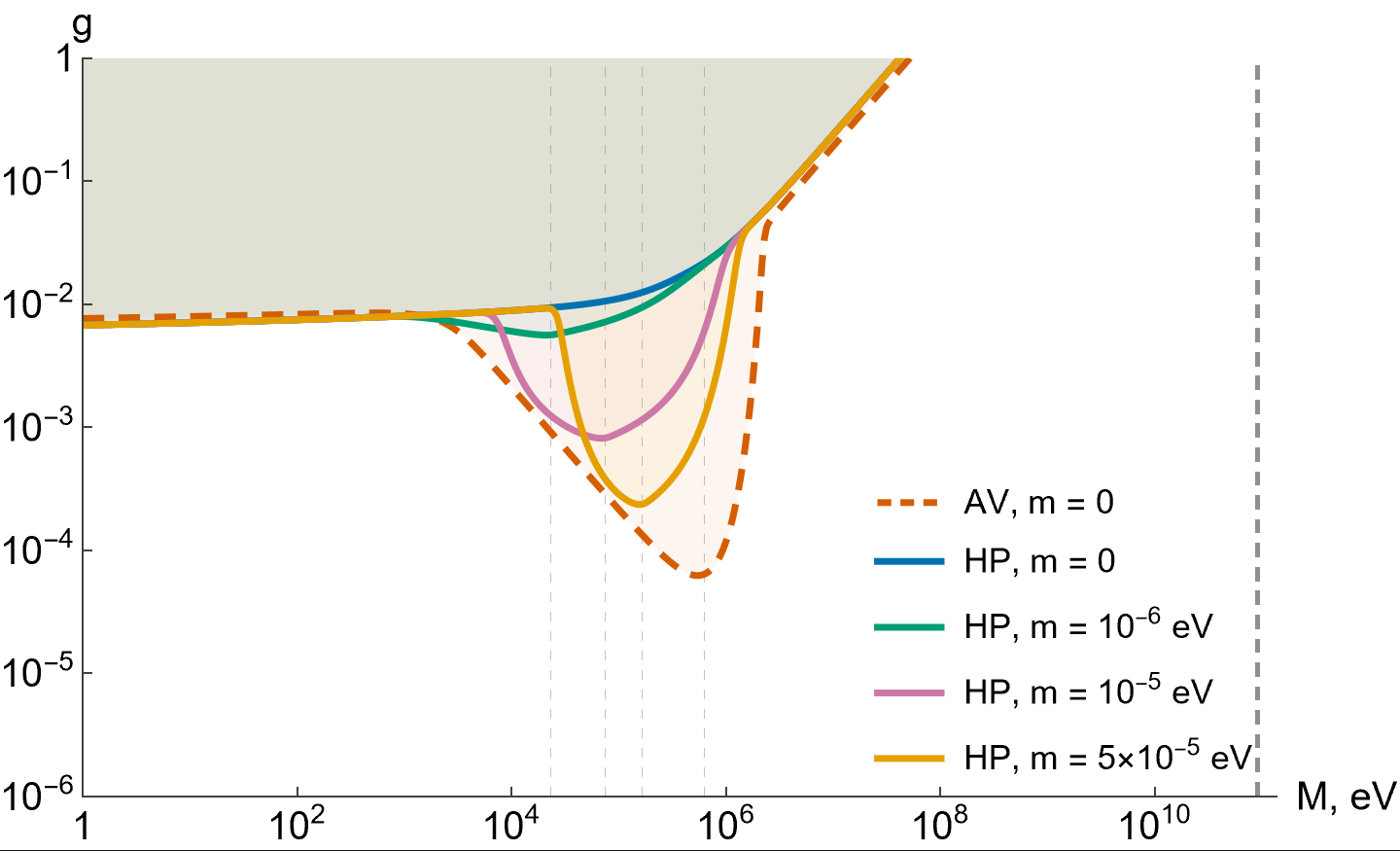}\hfill
\includegraphics[width=0.48\linewidth]{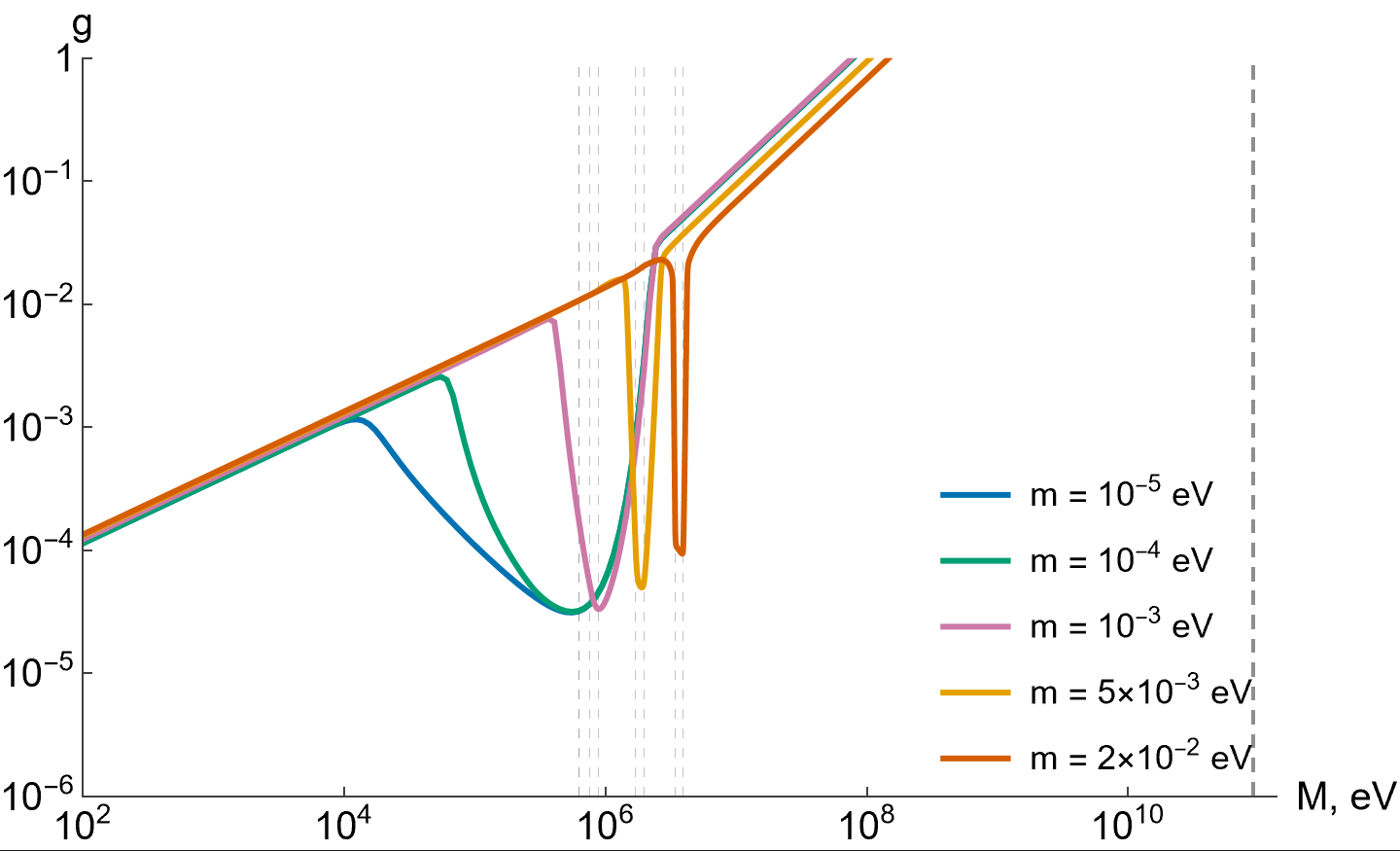}
\caption{Dependence %
 of the single-event survival benchmarks on the common relic mass for TXS~0506+056. Left: $(S,D)$ benchmarks for a degenerate UR C$\nu$B. The~helicity-polarized (HP) curves use $m=0$, $10^{-6}$, $10^{-5}$ and $5\times10^{-5}~\mathrm{eV}$. The~dashed curve is the massless spin-averaged (AV) case, which corresponds to equally populated target helicities (Appendix~\ref{app:refs}). The~HP pole vanishes in the massless limit. Right: $(V{-}A,D)$ benchmarks evaluated numerically with the massive-target kernels for relic masses spanning the UR, transition and NR regimes (Appendix~\ref{app:NR}). Increasing $m/T_{C\nu B}$ moves the resonant feature toward $M=\sqrt{2Em}$ and reduces its relative thermal width. The~thick vertical dashed line marks the $Z$-boson~mass. The shaded regions indicate $\tau>1$.}
\label{fig:HPSA}%
\end{figure}

This difference arises because, for~$(S,D)$, the~$s$-channel for the populated initial pair $\bar\nu_{a,+}\nu_{a,-}$ is helicity suppressed. The~suppression is controlled by the helicity-kinematic factor $\etaa$ of Equation~\eqref{eq:eta_E}, which scales as $m_a^2/E_i^2$ for UR targets and approaches $1+c_\theta$ for NR targets, so that the suppression disappears for $m_a\gtrsim T_{C\nu B}$. For~NR targets, $\langle\etaa\rangle=1$, so the helicity suppression does not persist in the angularly averaged $s$-channel rate, and~the helicity-polarized and spin-averaged prescriptions nearly coincide (Appendix~\ref{app:refs}). The~suppression is therefore most visible for the UR resonance, which also shows the low-$M$ cutoff of the finite-mass resonance (Section~\ref{sec:resonant}). Figure~\ref{fig:HPSA} (left) illustrates the resulting mass dependence for TXS~0506+056: the helicity-polarized pole vanishes for $m=0$ and becomes more pronounced as the mass~increases.
\unskip

\subsection{Ultra-High-Energy Neutrino Event with Undefined~Source}

The extreme energy of KM3-230213A (Section~\ref{S:sources}) extends the mediator masses accessible through resonant scattering. As~the origin of the event remains unclear, we take four representative light-travel distances that span from galactic to extragalactic scales, namely $10~\mathrm{kpc}$, $10~\mathrm{Mpc}$, $1~\mathrm{Gpc}$ and $4.12~\mathrm{Gpc}$. The~corresponding redshifts are approximately $2.3\times10^{-6}$, $2.3\times10^{-3}$, $0.27$ and $12$, respectively. The~largest distance marks the upper end of this range rather than a realistic source scenario. Since these distance hypotheses are not established source associations, the~resulting curves are conditional sensitivities rather than survival benchmarks, and~depend on the assumed distance and representative~energy.

\begin{figure}[htbp]
\centering
\includegraphics[width=0.48\linewidth]{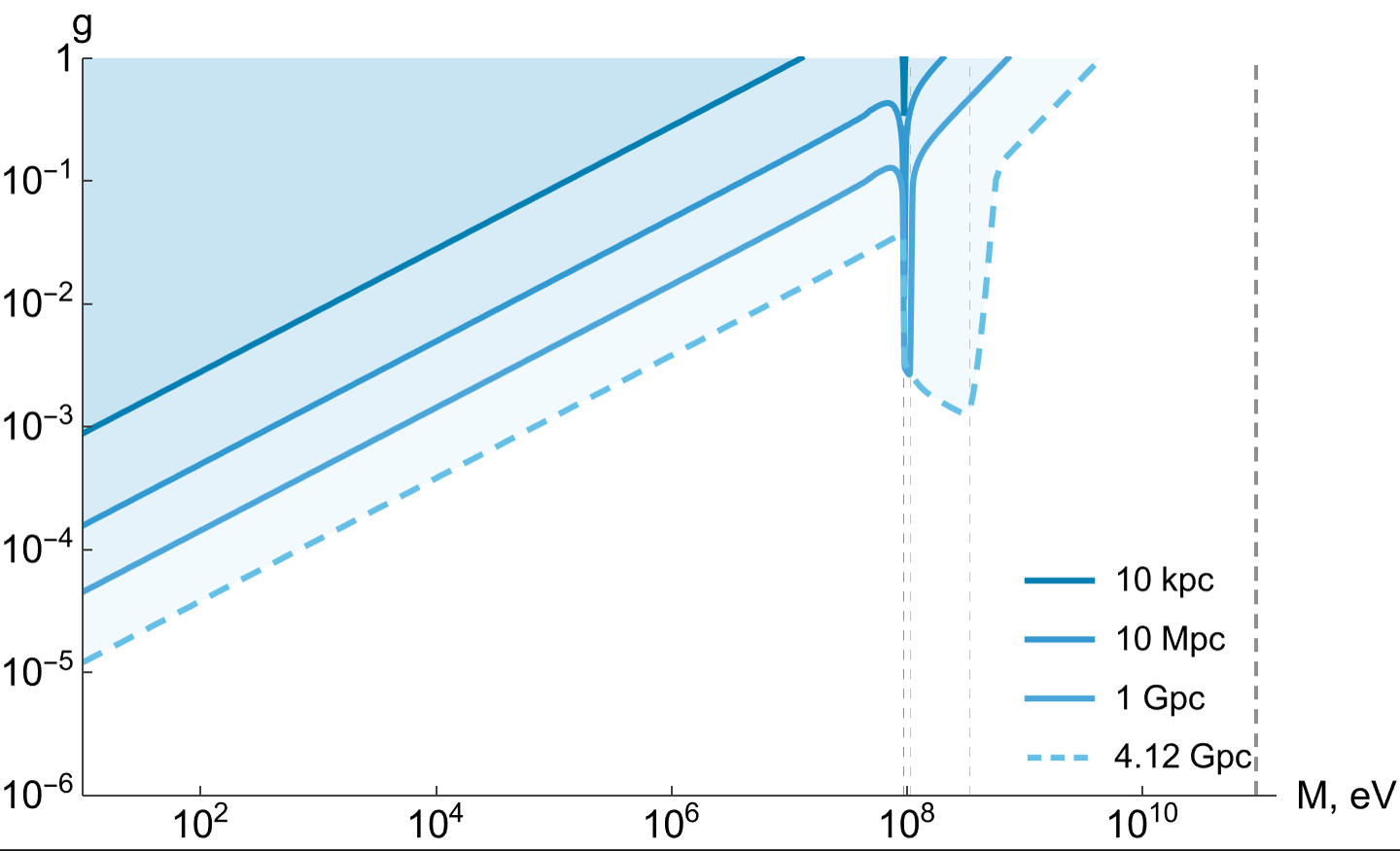}\hfill
\includegraphics[width=0.48\linewidth]{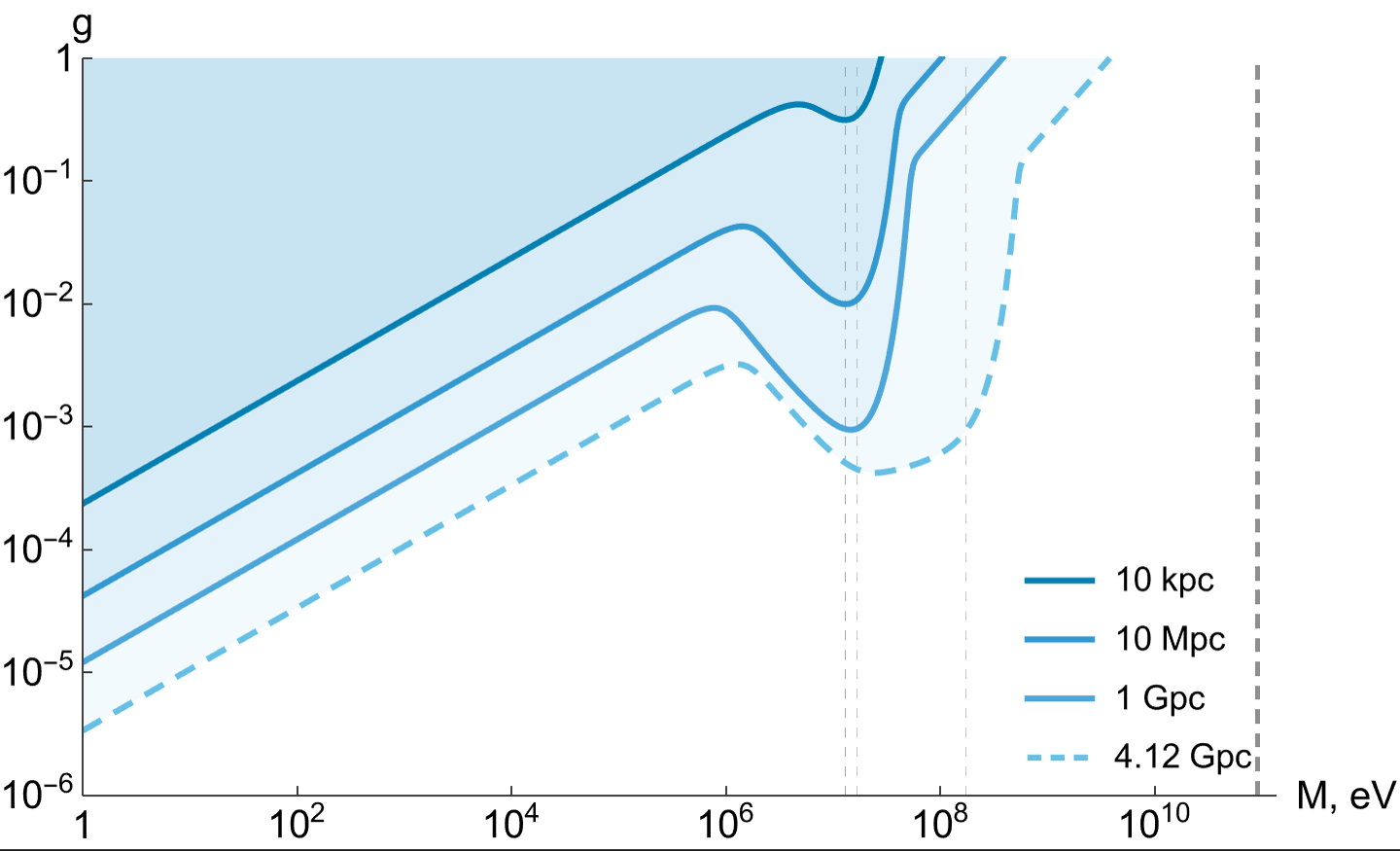}
\caption{Conditional single-event survival sensitivity for KM3-230213A in the $(V{-}A,D)$ model with a helicity-polarized, degenerate C$\nu$B. Left: NR regime with $m=2\times10^{-2}~\mathrm{eV}$. Right: UR regime with $m=10^{-5}~\mathrm{eV}$. The~curves use $E=220~\mathrm{PeV}$ and the light-travel distances indicated in the legend. The~$1~\mathrm{Gpc}$ and $4.12~\mathrm{Gpc}$ curves include cosmological evolution. The~shaded regions correspond to $\tau>1$ for the assumed distance. The~thick vertical dashed line marks the $Z$-boson~mass.}
\label{fig:KM3}
\end{figure}

\begin{figure}[htbp]
\centering
\includegraphics[width=0.48\linewidth]{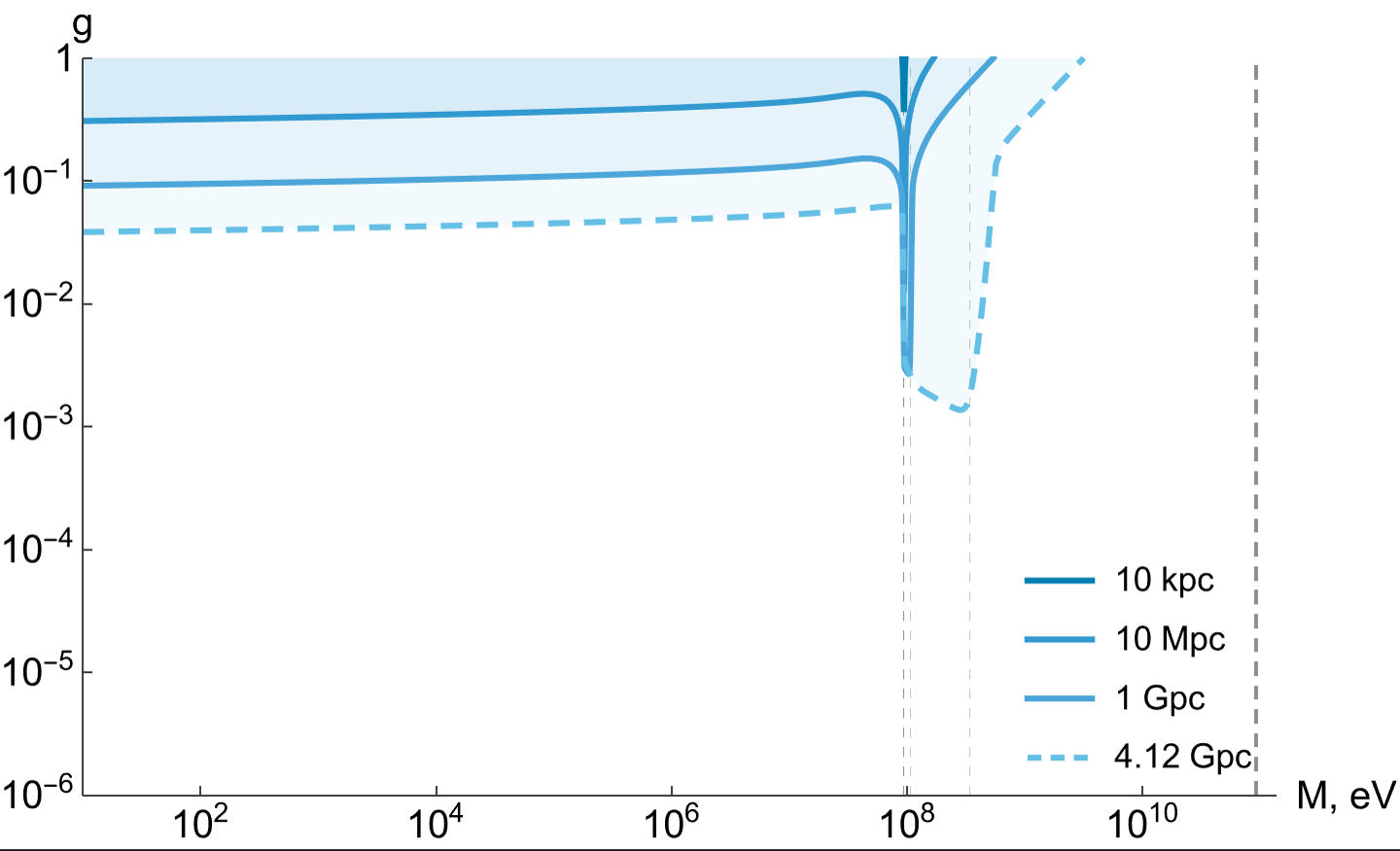}\hfill
\includegraphics[width=0.48\linewidth]{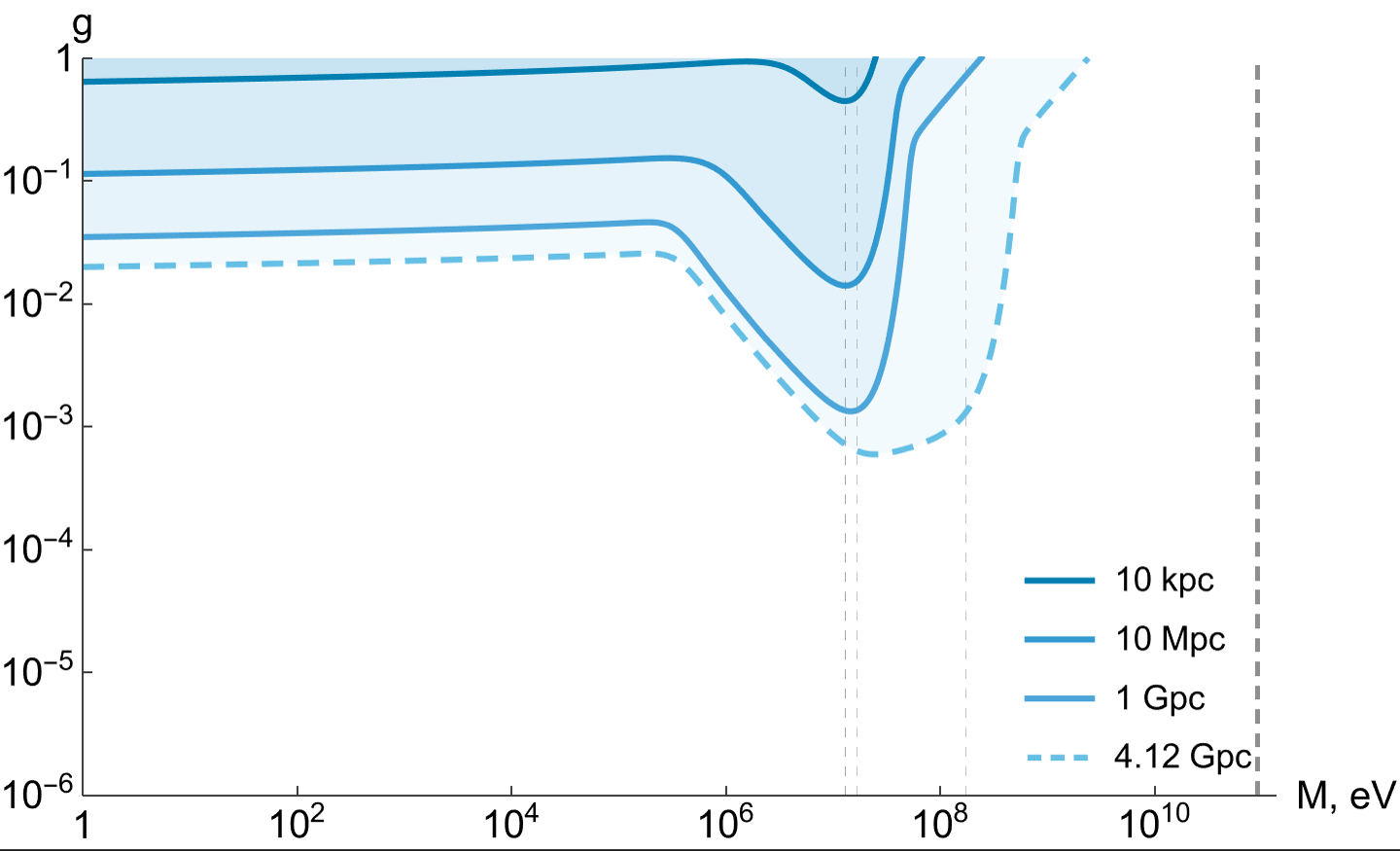}
\caption{Conditional single-event survival sensitivity for KM3-230213A in the $(S,M)$ model with a degenerate C$\nu$B and both populated target helicities. Left: NR regime with $m=2\times10^{-2}~\mathrm{eV}$. Right: UR regime with $m=10^{-5}~\mathrm{eV}$. The~curves use $E=220~\mathrm{PeV}$ and the light-travel distances indicated in the legend. The~$1~\mathrm{Gpc}$ and $4.12~\mathrm{Gpc}$ curves include cosmological evolution. Shading is as in Figure~\ref{fig:KM3}. The~thick vertical dashed line marks the $Z$-boson~mass.}
\label{fig:KM3_Scalar}
\end{figure}

Figures~\ref{fig:KM3} and \ref{fig:KM3_Scalar} show the $(V{-}A,D)$ and $(S,M)$ results, respectively. The~NR and UR panels use $m=2\times10^{-2}~\mathrm{eV}$ and $m=10^{-5}~\mathrm{eV}$, respectively, for~the common relic mass. Compared with the nearby distance hypotheses, the~resonance condition moves along the path as the neutrino energy redshifts, so that the resonance spreads between $M=\sqrt{2Em_i}$ and $\sqrt{2E(1+z_s)m_i}$. The~$4.12~\mathrm{Gpc}$ curves include the redshift integration up to $z\simeq12$. For~such redshifts, the~shift of the resonance exceeds its intrinsic thermal width, so that the resonant region cannot be described by evaluating the rate at a single effective redshift. For~this distance hypothesis, the~relic temperature reaches $T_{C\nu B}(1+z)\approx2\times10^{-3}~\mathrm{eV}$ at the source, so that the NR benchmark mass becomes semi-relativistic, $m/T_{C\nu B}\approx9$, along the early part of the path, where the exact massive-target kernel is used (Appendix~\ref{app:NR}). As~follows from Equation~\eqref{nearby_optical_depth}, larger assumed propagation distances generally increase the sensitivity. Within~the present approach, an~origin of KM3-230213A beyond $1~\mathrm{Gpc}$ would extend the propagation sensitivity to a region not reached by the other sources considered~here.

\vspace{-6pt}
\unskip

\section{Survival Benchmarks for a Non-Degenerate C\boldmath{$\nu$}B}
\label{sec:physicalcnb}
\unskip

\subsection{Mixed Survival~Probability}

In the degenerate-mass limit, the~three incident mass eigenstates are kinematically equivalent and their relative weights do not affect the survival probability. For~non-degenerate masses, we describe the incident neutrino as an incoherent mixture of mass eigenstates~\cite{Farzan_2008} and evaluate the optical depth separately for each incident state,
\begin{equation}
\tau_a=\sum_{i=1}^{3}\tau_{ai}.
\label{optical_depth_per_a}
\end{equation}
The probability of propagation without an interaction is then
\begin{equation}
P_{\rm surv}=\sum_{a=1}^{3}w_a e^{-\tau_a},
\qquad \sum_{a=1}^{3}w_a=1,
\label{eq:Psurv}
\end{equation}
where $w_a$ is the incident mass-eigenstate fraction. We use $w_a=1/3$ as a representative benchmark and impose $P_{\rm surv}\geq e^{-1}$, or~equivalently $\tau_{\rm obs}\equiv-\ln P_{\rm surv}\leq1$. This criterion describes the unscattered component of the assumed mixture and does not include regeneration or detector response. Each $\tau_a$ has the form of Section~\ref{S:survival_limits}, integrated along the propagation path where cosmological evolution is relevant, and~the boundary $P_{\rm surv}(g,M)=e^{-1}$ is found~numerically.

Each incident mass eigenstate has its own $s$-channel resonance, because~resonant annihilation contributes to its optical depth only through targets of the same mass eigenstate. These individual resonances need not produce three deep features in the mixed survival benchmark. Near~an isolated resonance, strong attenuation of one component leaves the other two available to~survive.

For three equally weighted incident components with equal off-resonant optical depths, complete absorption of the resonant component gives the largest possible reduction of the coupling relative to its off-resonant value $g_{\rm off}$. Since the off-resonant optical depth scales as the fourth power of the coupling, this reduction is $(g_{\rm off}-g)/g_{\rm off}\simeq0.122\simeq12\%$.

Increasing the resonant optical depth further cannot deepen the feature, because~that component has already been removed from the surviving mixture. Averaging optical depths before exponentiation does not retain this limitation and can therefore produce artificially deep features at the individual~resonances.

A larger resonant enhancement requires significant attenuation of more than one component at the same mediator mass. This occurs where nearby resonant contributions overlap. At~the couplings relevant to the mixed survival benchmark, the~thermally broadened contributions can remain significant away from their individual peaks, producing the stronger feature between the $m_1$ and $m_2$ resonances in the 3 NR~spectrum.

For a non-degenerate spectrum, the~three target states can belong to different kinematic regimes. We therefore evaluate each contribution $\mathcal R_{ai}(m_i)$ separately, using the appropriate UR, NR or semi-relativistic expression, and~obtain the total rate $\mathcal R_a$ from Equation~\eqref{total_CnuB_rate}.

We consider normal ordering and use two representative neutrino mass spectra: three non-relativistic mass eigenstates (3 NR) and one ultra-relativistic and two non-relativistic mass eigenstates (2 NR + 1 UR). The~corresponding total interaction rates are
{\begin{align}
&\left.\mathcal R_{a}\right|_{m_{1,2,3}\gg T_{C\nu B}}=\mathcal R_{a1}^{\rm NR}+\mathcal R_{a2}^{\rm NR}+\mathcal R_{a3}^{\rm NR},\\
&\left.\mathcal R_{a}\right|_{m_1\ll T_{C\nu B},\,m_{2,3}\gg T_{C\nu B}}=\mathcal R_{a1}^{\rm UR}+\mathcal R_{a2}^{\rm NR}+\mathcal R_{a3}^{\rm NR}.
\end{align}
}
Generally, for~the lightest eigenstate, the~relic mass may lie between the UR and NR regimes. Figure~\ref{fig:HPSA} (right) shows how the relic mass shapes the $(V{-}A,D)$ resonance, which evolves from the broad UR feature to the narrow NR one as the mass increases. For~intermediate masses, $m_i\sim T_{C\nu B}$, the~pole is computed with the exact massive-target kernels (\mbox{Appendix~\ref{app:NR}}). For~comparison, we also show a naive averaged prescription, $\overline{\tau}=\frac13\sum_a\tau_a=1$, which marks the positions of the individual resonances. The~two prescriptions coincide when all $\tau_a$ are equal, as~in the degenerate~limit.

\subsection{$V{-}A$-Mediated Dirac~Neutrinos}
For the $(V{-}A,D)$ analysis, we adopt the following two benchmark mass spectra.
In the 3 NR scenario, we take
\be
m_1=4\times10^{-3}~\mathrm{eV}, \quad
m_2=9.539\times10^{-3}~\mathrm{eV}, \quad
m_3=5.046\times10^{-2}~\mathrm{eV}, \nonumber\ee
giving $\sum_i m_i\simeq0.064~\mathrm{eV}$, motivated by the current cosmological sensitivity to the neutrino mass sum~\cite{Elbers:2025DESI}.
In the 2 NR + 1 UR scenario,
\be m_1=10^{-5}~\mathrm{eV}, \quad
m_2=8.66\times10^{-3}~\mathrm{eV}, \quad
m_3=5.03\times10^{-2}~\mathrm{eV}, \nonumber\ee
where the lightest mass eigenstate is ultra-relativistic, $m_1\ll T_{C\nu B}$,
giving $\sum_i m_i\simeq0.05897~\mathrm{eV}$.

\begin{figure}[htbp]
\centering
    \begin{minipage}[ht]{0.45\linewidth}
\includegraphics[width=0.99\textwidth]{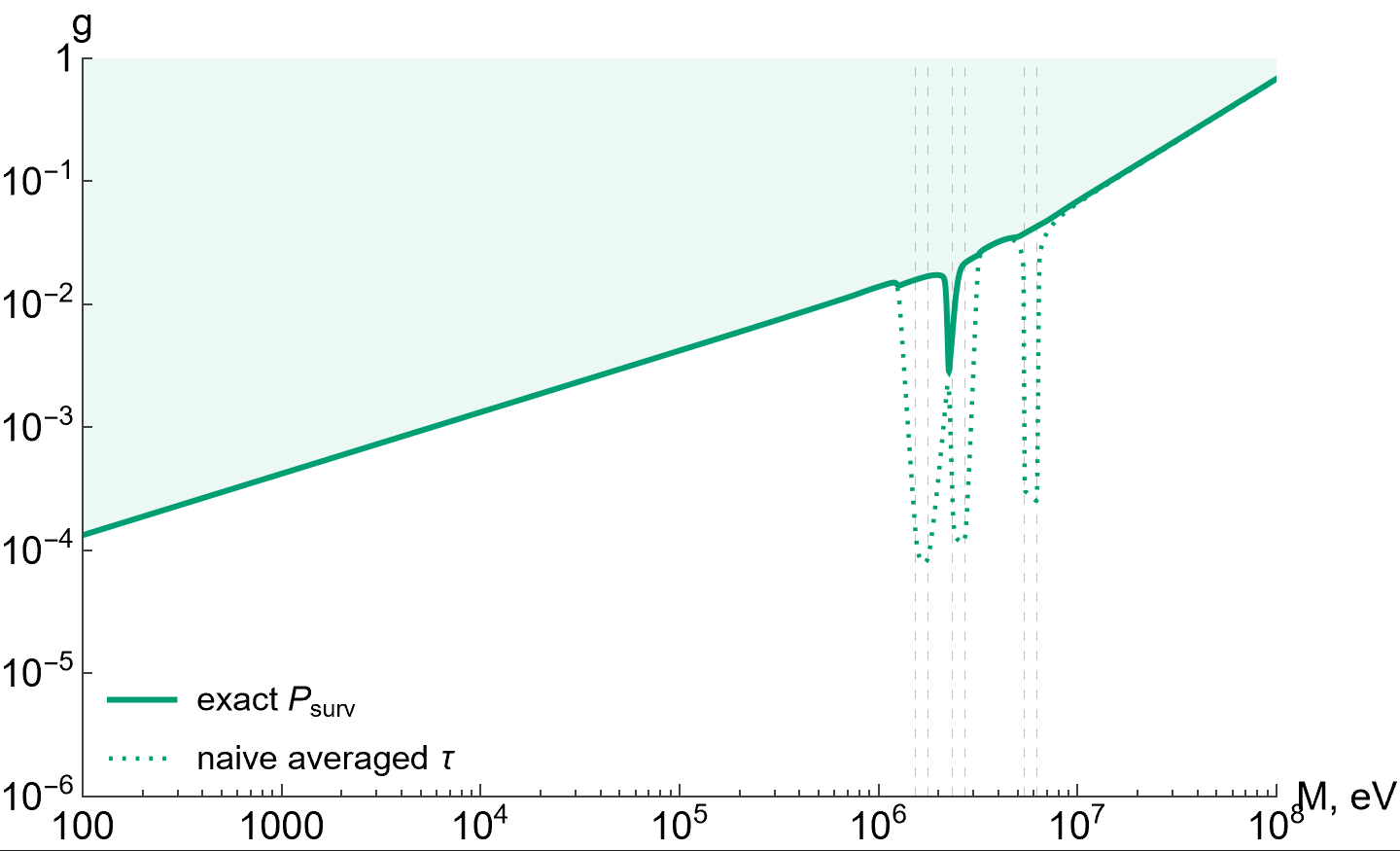}
\end{minipage}
\hfill
    \begin{minipage}[ht]{0.45\linewidth}
\includegraphics[width=0.99\textwidth]{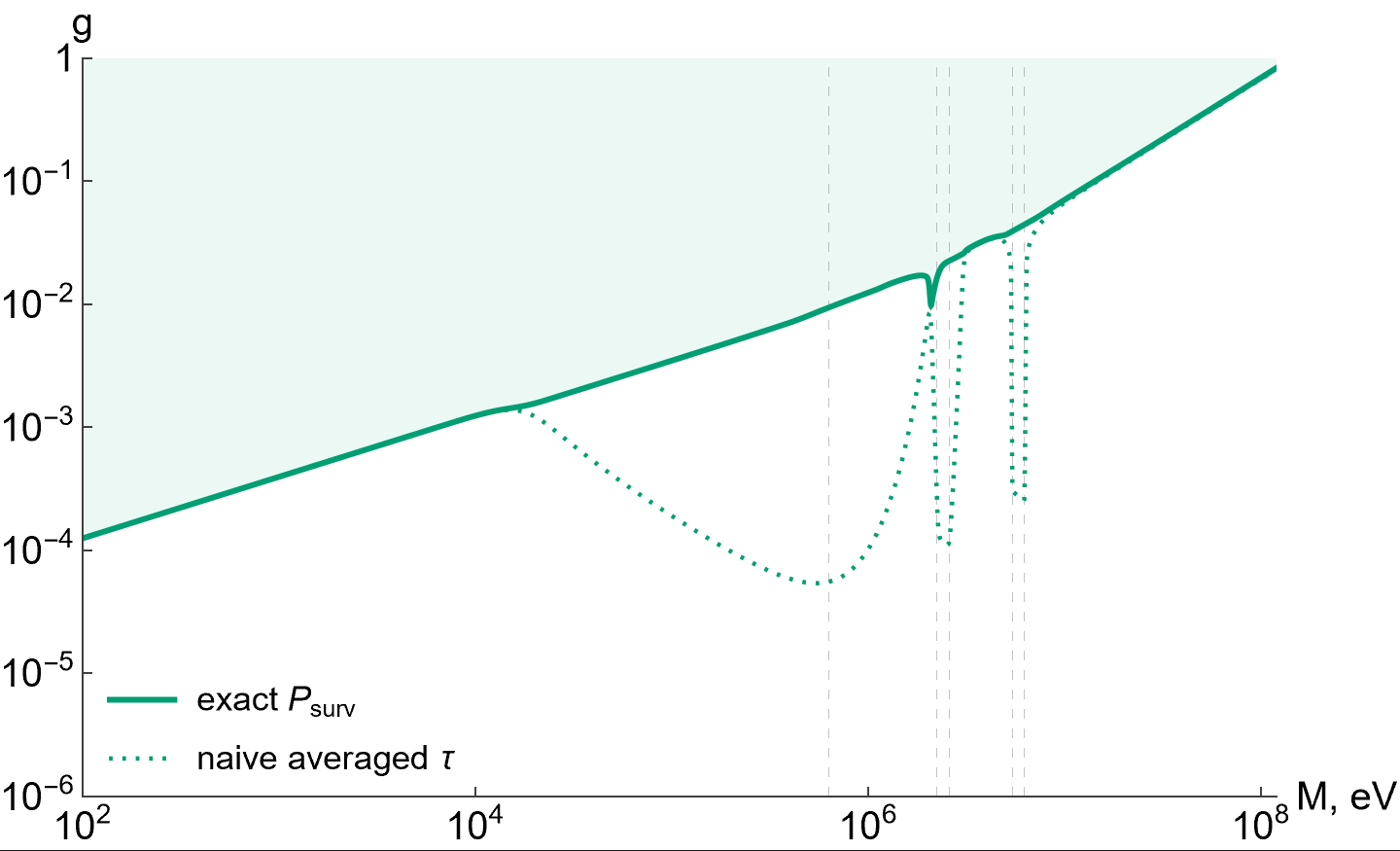}
\end{minipage}
\caption{Single-event %
 survival benchmarks on the coupling constant $g$ as a function of the $V{-}A$ mediator mass $M$ for two representative C$\nu$B neutrino mass spectra in normal ordering, using TXS 0506+056 as the representative source. Left: 3 NR mass eigenstates. Right: 2 NR + 1 UR mass eigenstates. Solid curves show the benchmarks from the mixed survival probability, $P_{\rm surv}=e^{-1}$ (Equation~\eqref{eq:Psurv}), and~dotted curves show the averaged optical depth $\overline{\tau}=1$. The~vertical lines mark $M=\sqrt{2Em_i}$ and $\sqrt{2E(1+z_s)m_i}$ for the NR states. The~benchmarks are shown for Dirac neutrinos with a helicity-polarized C$\nu$B. The shaded regions indicate $P_{\rm surv}<e^{-1}$. \label{fig:physical_vector}}
\end{figure}

The corresponding survival benchmarks for TXS 0506+056 are shown in Figure~\ref{fig:physical_vector}. Away from the resonances, the~exact and averaged prescriptions coincide. Near~each isolated resonance, the~exact boundary changes only slightly, whereas the averaged optical depth produces deep, narrow features. In~the 3 NR spectrum, the~feature of the exact boundary between the $m_1$ and $m_2$ resonances arises from the overlap of the tails of their thermally broadened resonances, which suppresses two incident states~simultaneously.
\unskip

\subsection{Scalar-Mediated Majorana Neutrinos and Comparison with Previous~Work}

\begin{figure}[htbp]
\centering
    \begin{minipage}[ht]{0.45\linewidth}
\includegraphics[width=0.99\textwidth]{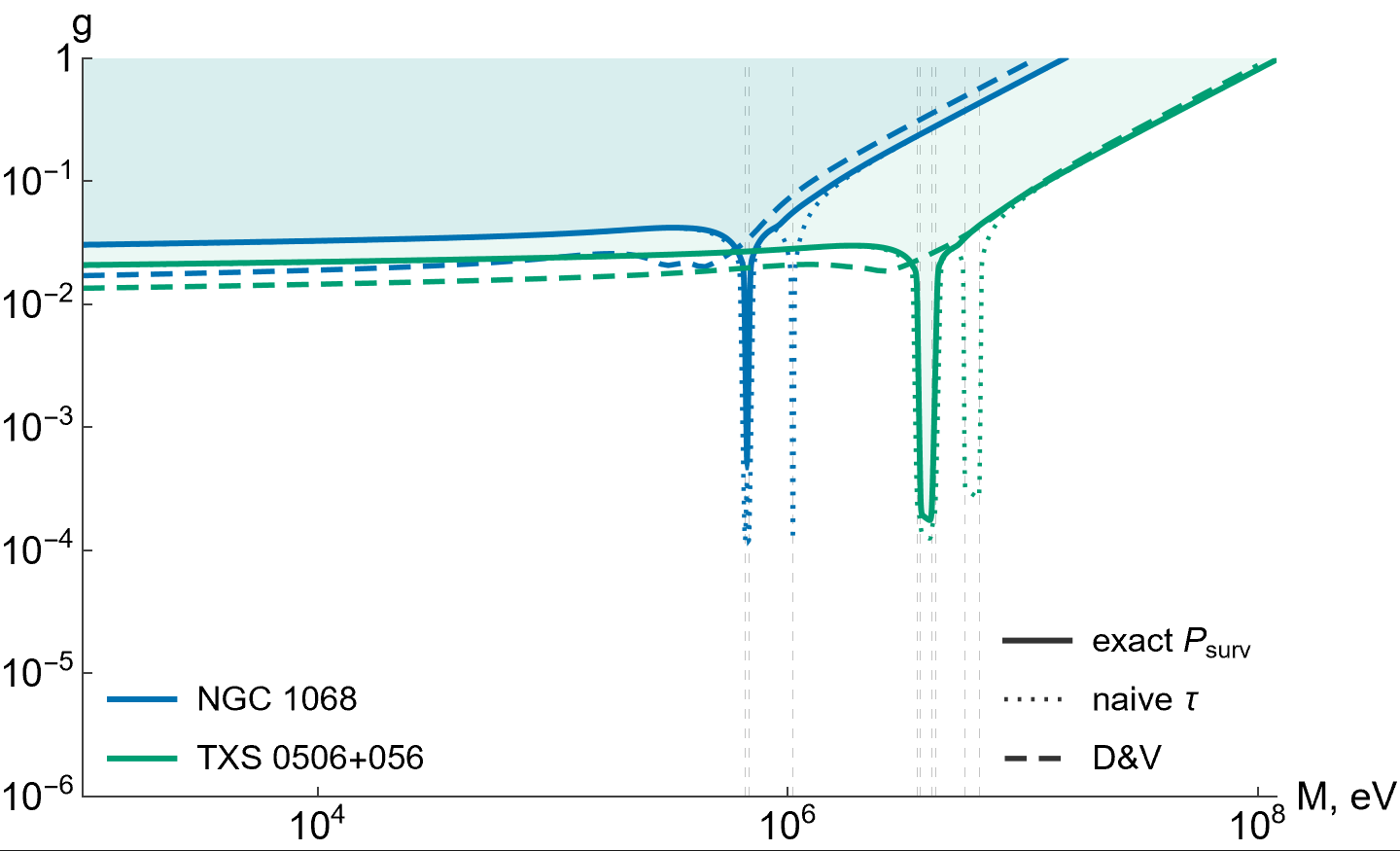}
\end{minipage}
\begin{minipage}[ht]{0.05\linewidth}
\par
\end{minipage}
\begin{minipage}[ht]{0.45\linewidth}
\includegraphics[width=0.99\textwidth]{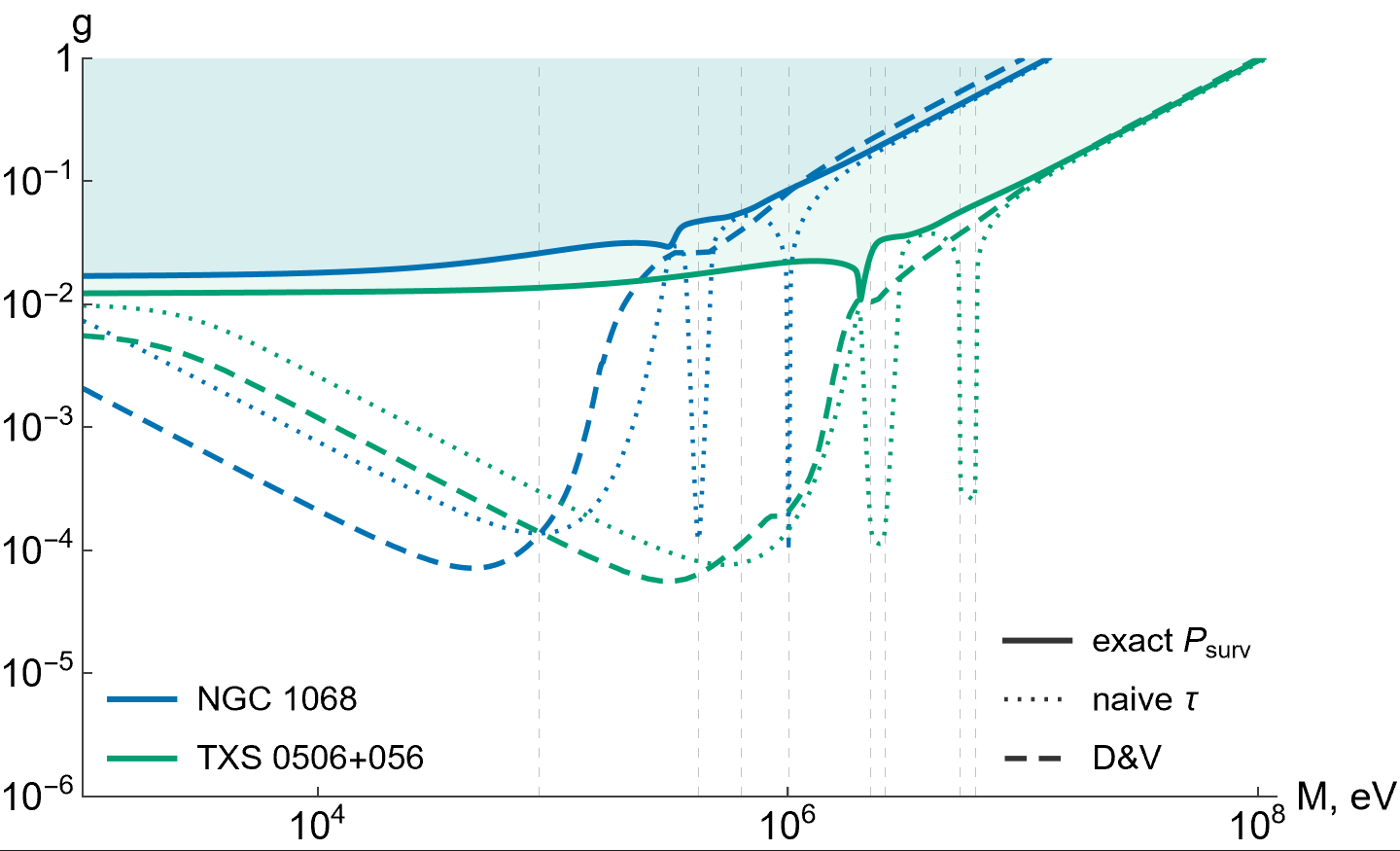}
\end{minipage}
\caption{Comparison %
 of the scalar-mediated Majorana-neutrino benchmarks obtained in the present analysis with Ref.~\cite{Doring:2023vmk} (D\&V), using the same representative neutrino-mass choices and, for~NGC~1068, the~distance of $14.4$~Mpc adopted there. Left: normal ordering with $\sum_i m_i=0.1~\mathrm{eV}$, for~which all three C$\nu$B mass eigenstates are NR. Right: an effectively massless lightest eigenstate, with~one UR and two NR C$\nu$B states. Solid curves show the present single-event survival benchmarks from the mixed survival probability $P_{\rm surv}=e^{-1}$, dotted curves the averaged optical depth, and~dashed curves the results of Ref.~\cite{Doring:2023vmk}. The~source choices are indicated in the~legend. The shaded regions indicate $P_{\rm surv}<e^{-1}$. \label{fig:comparison}}
\end{figure}

We compare the scalar Majorana benchmarks with Ref.~\cite{Doring:2023vmk}, using its representative mass spectra and, for~NGC~1068, its distance of $14.4$~Mpc. The~main difference is the criterion used to obtain the curves. Ref.~\cite{Doring:2023vmk} requires the expected number of events, obtained from the attenuated point-source flux weighted by the detector effective area over the source energy interval, to~be at least half of its unattenuated value, whereas we impose a single-event survival condition at a representative energy. Neither criterion defines a statistical confidence level. The~curves in Figure~\ref{fig:comparison} therefore need not coincide. For~the spectrum with an effectively massless lightest state, the~feature near the $m_2$ resonance arises from the overlap of the broad UR resonance of the lightest state with the $m_2$ resonance.

%\unskip
%\vspace{-8pt}
\section{Discussion and~Limitations}

We discuss the interpretation of the survival contours, compare the scattering prescriptions with earlier work, and~relate the results to existing~constraints.

\subsection{Forward Scattering and Observable~Attenuation}
\label{sec:cutoff}

In the present analysis, the~full mediator-mass dependence is retained in the scattering rate. For~a nonzero mediator mass, the~$t$-channel contribution remains finite, with~$\sigma_t\simeq g^4/(4\pi M^2)$ for $M^2\ll s$. In~contrast, the~massless-mediator limit leads to a divergent total cross section and requires an additional cutoff prescription, as~introduced in Ref.~\cite{Kolb:1987qy} (see also Ref.~~\cite{Kelly:2018tyg}).

Since the finite mediator mass already regularizes the cross section, the~role of an additional cutoff is no longer mathematical but phenomenological. The~enhancement in the light-mediator region is dominated by near-forward scatterings, characterized by momentum transfers $|t|\sim M^2$. In~such interactions, the~incident neutrino transfers only a fraction of order $M^2/s$ of its energy. The~associated deflection and time delay need not exceed the angular and temporal acceptance of the observation~\cite{Murase:2019xqi}. Consequently, even a large total scattering probability does not necessarily imply a comparable depletion of the detectable flux. The~observable attenuation depends on the energy resolution and acceptance of the detector, together with its angular and temporal acceptance. A~quantitative prediction for the accepted flux, including the cumulative energy loss from multiple near-forward scatterings and the regeneration of neutrinos at lower energies, would require transport modelling with energy and angular redistribution~\cite{Murase:2019xqi,Cline:2022qld}, which is beyond the scope of this~work.

Following the physically motivated prescription of Ref.~\cite{Kolb:1987qy}, we adopt the cutoff parameter $\epsilon$, restricting the $2\to2$ scattering phase space to $-s(1-\epsilon)\leq t\leq-\epsilon s$.\footnote{An equivalent approach uses an angular cutoff, $-(1-\epsilon')\le\cos\theta\le1-\epsilon'$ \cite{Kelly:2018tyg}, with~$\epsilon'=2\epsilon$. For~identical final-state neutrinos the observed particle is the leading one, so a cut on its energy loss is symmetric in $t$ and $u$. In~Ref.~\cite{Kolb:1987qy}, the~symmetric form was motivated by the regeneration of an energetic neutrino by the recoiling background particle. For~distinguishable final states the cut may be one-sided.} In Ref.~\cite{Kolb:1987qy}, this cutoff was introduced to account for the fact that scatterings with small energy transfer do not necessarily remove neutrinos from the detectable flux, and~$\epsilon=0.1$ was adopted as a representative threshold for substantial energy loss. In~the present analysis, the~finite mediator mass already regularizes the scattering cross section, so the cutoff serves only to select which scatterings contribute to the effective optical depth. For~$\epsilon\to0$, all scatterings contribute, and~$\tau=1$ corresponds to one expected scattering irrespective of the transferred energy. For~finite $\epsilon$, only scatterings satisfying the imposed energy-transfer criterion~contribute.

The appropriate value of $\epsilon$ is not universal and depends on the neutrino observation. Even for contained events, the~deposited energy is reconstructed with a resolution of about $15\%$ \cite{IceCube:2013dkx}. For~through-going tracks, such as IceCube-170922A~\cite{IceCube:2018dnn} and KM3-230213A~\cite{KM3NeT:2025npi}, the~inferred neutrino energy is uncertain by a factor of a few, and~the reported 90\% confidence intervals span more than an order of magnitude. These uncertainties illustrate why small fractional energy losses cannot generally be distinguished on an event-by-event basis, although~the effect on the observed flux also depends on the source spectrum and event selection. We therefore regard $\epsilon\simeq0.1$ as a representative phenomenological benchmark, rather than a universal detection~threshold.

\begin{figure}[htbp]
\centering
    \begin{minipage}[ht]{0.45\linewidth}
\includegraphics[width=0.99\textwidth]
{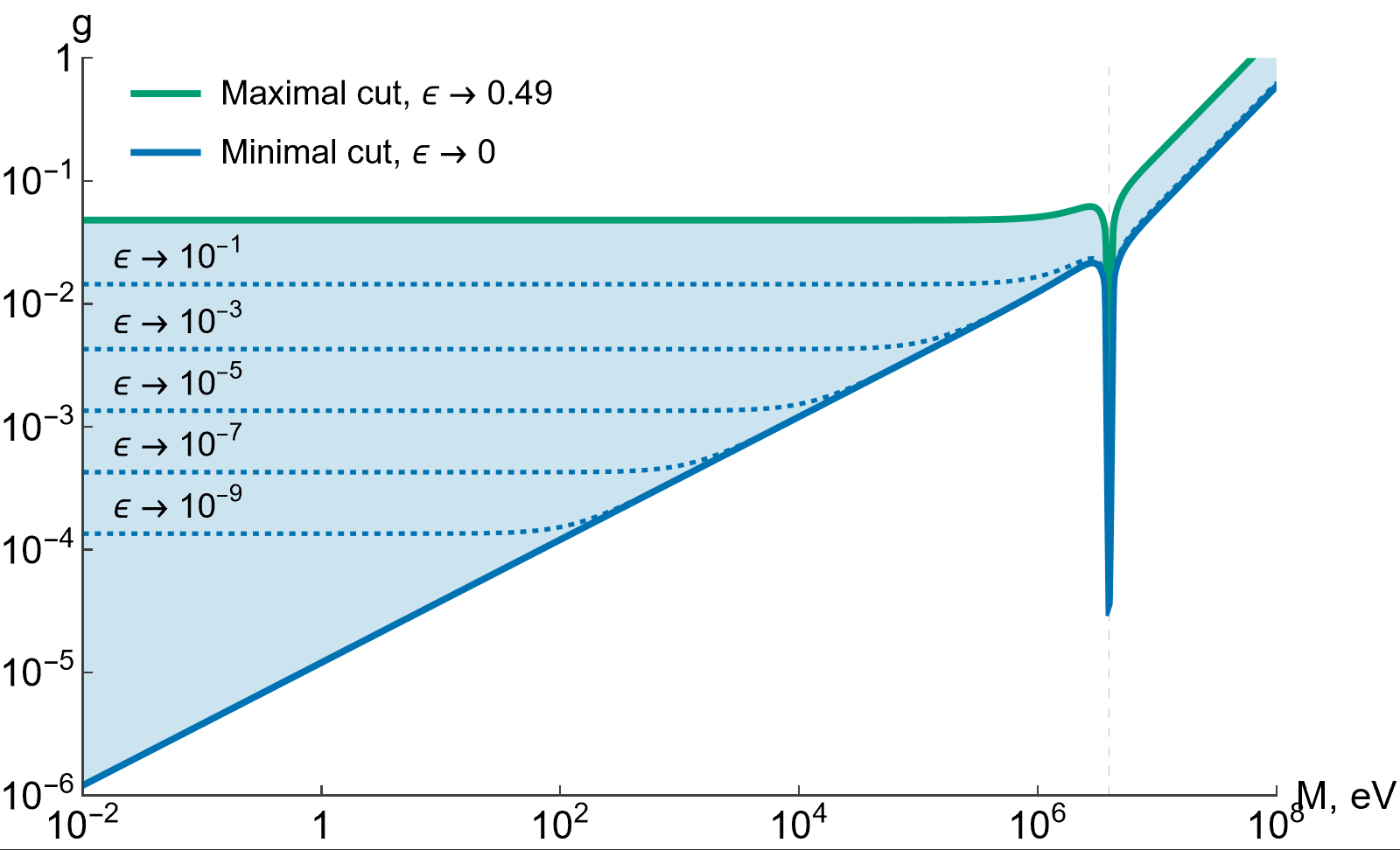}
\end{minipage}
\begin{minipage}[ht]{0.05\linewidth}
\par
\end{minipage}
\begin{minipage}[ht]{0.45\linewidth}
\includegraphics[width=0.99\textwidth]
{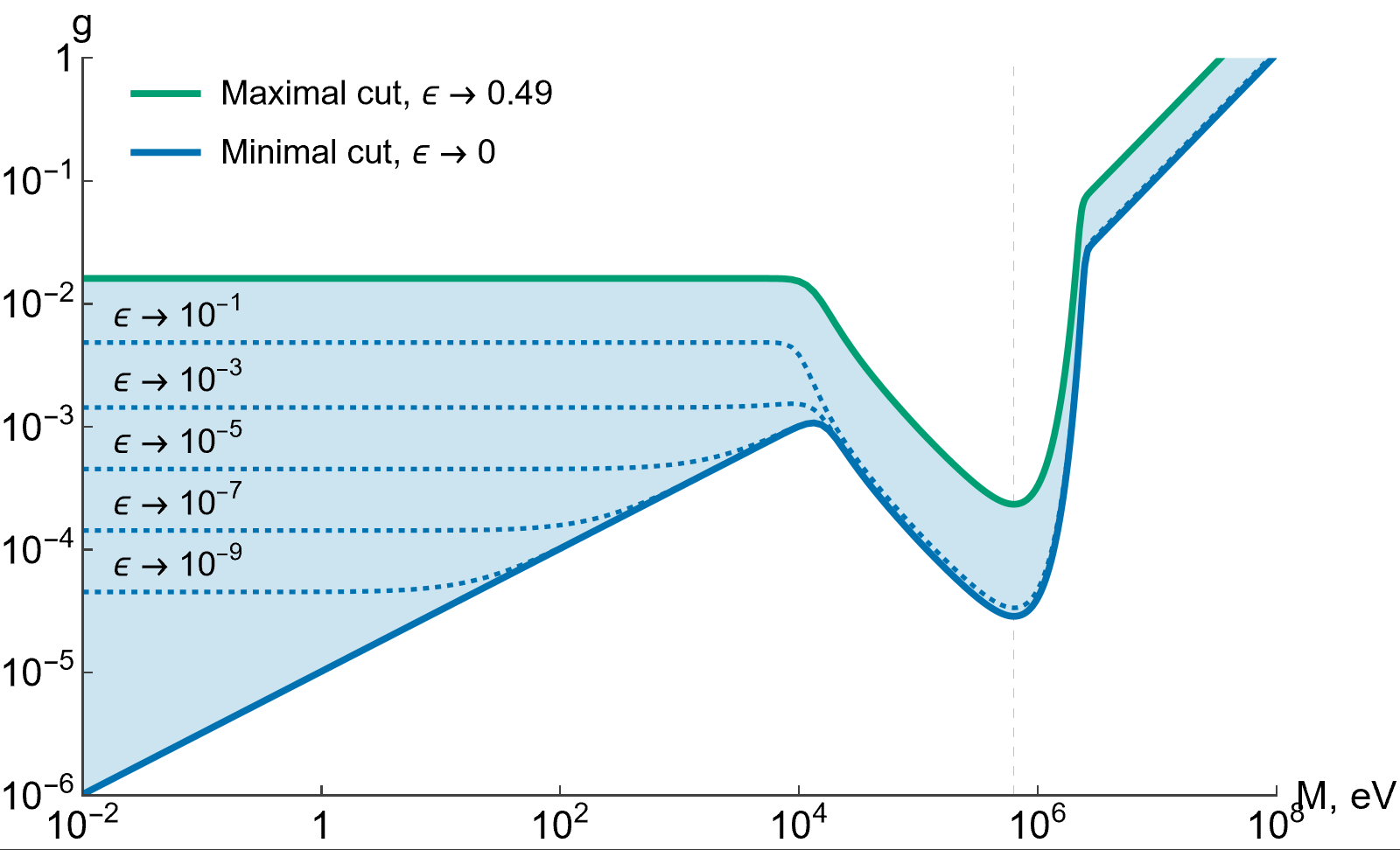}
\end{minipage}
\caption{Impact %
 of the cutoff parameter $\epsilon$ on the single-event survival benchmark for $(V{-}A,D)$ and a degenerate C$\nu$B in the NR (left, $m=2\times10^{-2}~\mathrm{eV}$) and UR (right, $m=10^{-5}~\mathrm{eV}$) regimes, using TXS~0506+056 as the representative source, without~redshift evolution. Only $2\to2$ scattering is included, and~pair production $\nu\bar\nu\to\phi\phi$, which is not subject to the cutoff, is omitted. Dotted curves show intermediate values of $\epsilon$. The~vertical dashed line marks the resonance~position. The shaded regions indicate $\tau>1$. \label{fig:cutoff} }
\end{figure}

Figure~\ref{fig:cutoff} illustrates the cutoff dependence for the $(V{-}A,D)$ benchmark, retaining only the $2\to2$ scattering contribution. Mediator-pair production, $\nu\bar\nu\to\phi\phi$, is not subject to this cutoff and is omitted from the figure to isolate its effect. For~$M^2\lesssim\epsilon s$, the~survival benchmarks depend strongly on $\epsilon$. As~$\epsilon\to0$, the~characteristic spin-one forward enhancement develops, yielding $g\propto M^{1/2}$, whereas for $\epsilon\sim0.1$ this enhancement is removed. For~$\epsilon\lesssim0.1$, the~cutoff mainly affects the light-mediator region, while the curves converge to the $\epsilon\to0$ result at larger mediator masses. The~limiting choice $\epsilon=0.49$ represents an extreme case, requiring the leading outgoing neutrino to lose nearly half its energy and excluding almost the entire $2\to2$ phase space, strongly suppressing the integrated rate, including its $s$-channel contribution. Accordingly, values $0.1\lesssim\epsilon<0.5$ also modify the resonant and heavy-mediator~regions.

The strong cutoff dependence in the light-mediator region demonstrates that survival boundaries derived from the total scattering optical depth need not directly correspond to observable flux attenuation. The~$\epsilon$-dependent curves should therefore be interpreted as sensitivity benchmarks for an imposed energy-transfer criterion, rather than as detector-specific exclusion~limits.
\unskip

\subsection{Relation to Existing~Constraints}
\label{sec:discussion_constraints}

For mediators thermalized with neutrinos during the relevant early-Universe epoch, BBN and CMB observations impose lower mass bounds of order a few MeV under the assumptions discussed in Section~\ref{sec:constraints} \cite{Blinov_2019,Escudero:2018mvt,Sabti:2019mhn}. For~lighter mediators, the~survival benchmarks might therefore be relevant for non-standard thermal histories. For~heavier mediators, the~cosmological constraints on $(V{-}A,D)$ and $(S,M)$ act mainly on $g^2/M^2$ through neutrino free-streaming and still allow relatively large couplings~\cite{Cyr_Racine_2014,Kreisch_2020}. The~resonant features of the highest-energy events, in~particular KM3-230213A for source distances above a Gpc, may therefore extend the propagation sensitivity into a region where cosmological constraints are comparatively~weak.

The propagation calculation assumes the standard relic populations. This is self-consistent for $(S,M)$, since Majorana neutrinos have no additional right-handed states. Unlike a pure-vector interaction with Dirac neutrinos, for~which the right-handed states are thermalized largely independently of the mediator mass~\cite{Luo:2020sho}, the~$(V{-}A,D)$ interaction does not couple to right-handed fields. They can be produced only through chirality flips suppressed by the neutrino mass, dominantly in mediator decays $\phi\to\nu\nu_R$, with~a production rate $\Gamma_R$ satisfying $\Gamma_R/H\sim g^2m_\nu^2M_{\rm Pl}/M^3$. Following Ref.~\cite{Blinov_2019}, they thermalize before BBN only for $g\gtrsim5\times10^{-3}\,(\max(M,T_{\rm dec})/\mathrm{MeV})^{3/2}(0.1~\mathrm{eV}/m_\nu)$, where $T_{\rm dec}\simeq2~\mathrm{MeV}$ is the neutrino decoupling temperature. For~$(V{-}A,D)$, the~decay into right-handed states is suppressed by the same factor $(m_\nu/M)^2$. For~mediators heavier than a few MeV, the~couplings near the resonances remain well below this threshold, and~the standard relic composition is retained, whereas the larger off-resonant couplings can approach it. In~this mass range, the~relevant comparison for $(V{-}A,D)$ is therefore with laboratory bounds. These bounds are particularly strong because a mediator coupled to neutrinos but not to charged leptons does not couple to a conserved current, so its emission rate in meson and $W$ decays grows with the energy released in the decay~\cite{Dror:2017nsg}. For~a coupling to neutrinos only, the~invisible $Z$ width requires $g\lesssim3\times10^{-2}$ for $M<m_Z$, nearly independently of $M$~\cite{Laha:2013xua}. They are strongest for the flavor-universal coupling assumed here (Section~\ref{sec:constraints}).

For $(S,D)$, the~right-handed states are generically thermalized (Section~\ref{sec:constraints}), and~this case serves only as a comparison (Section~\ref{sec:HPvsSA}). For~Dirac neutrinos, this can be avoided by a scalar that carries lepton number and couples only to left-chiral neutrinos, $\phi\,\overline{\nu^c_L}\nu_L$ \cite{Berryman_2018}. Its $s$-channel resonance involves two neutrinos or two antineutrinos rather than a neutrino--antineutrino pair and is not helicity suppressed. This scenario therefore requires a separate~analysis.

The survival benchmarks thus provide a complementary, propagation-based~probe.

\section{Conclusions}

We investigated the sensitivity of astrophysical neutrino propagation to neutrino secret interactions through scattering on the cosmic neutrino background. The~two main scenarios are Dirac neutrinos interacting through a spin-one mediator with a left-chiral $V{-}A$ coupling and Majorana neutrinos interacting through a scalar mediator. For~comparison, we also evaluated scalar exchange with Dirac neutrinos, a~common benchmark in propagation studies, on~the same helicity-polarized relic background. We performed a detailed analysis of $\nu$SI, obtaining single-event survival benchmarks in the UR and NR limiting regimes for nearby sources. Using representative energies and source distances for SN~1987A, NGC~1068, TXS~0506+056 and PKS~0735+178, together with distance hypotheses for KM3-230213A, we included cosmological evolution along the propagation path where~required.

We first considered an illustrative degenerate C$\nu$B mass spectrum to isolate the kinematic differences between the NR and UR regimes and demonstrate the characteristic behaviour of the interaction models. We then combined the three mass-eigenstate contributions for representative non-degenerate neutrino mass spectra, determining the mixed survival probability~numerically.

The resulting survival benchmarks cover the light-mediator, resonant and heavy-mediator regions. Their comparison shows that the interaction structure, the~neutrino nature and the populated relic helicities can produce qualitatively different propagation signatures. Retaining the full mediator-mass dependence reveals a pronounced enhancement of the spin-one interaction rate in the light-mediator region compared with scalar exchange. This enhancement is dominated by near-forward scattering and can be strongly reduced by an energy-transfer cutoff. For~scalar exchange, the~Dirac and Majorana scenarios also differ qualitatively in the UR regime, where the Dirac pole is helicity suppressed and its strength is controlled by the relic-neutrino~mass.

Several parameters relevant to $\nu$SI remain undetermined and may be constrained further by future experiments. In~particular, the~relic-neutrino masses determine whether each background eigenstate is NR or UR and set the positions of its resonant contributions. For~non-degenerate spectra, we provide survival benchmarks for the $(V{-}A,D)$ and $(S,M)$ scenarios using illustrative neutrino mass spectra. The~exact mixed survival probability substantially changes the interpretation of separated resonances. Strong attenuation of a single component leaves the other components available to survive, limiting the effect of an isolated resonance. A~stronger resonant enhancement occurs where nearby resonant contributions overlap and significantly attenuate more than one component at the same mediator~mass.

The event KM3-230213A deserves particular attention because of its extreme energy. Its inclusion shifts the resonant sensitivity to heavier mediators and probes parameter regions beyond those accessible with the lower-energy observations considered here. Within~the present approach, source distances above $1~\mathrm{Gpc}$ may extend the propagation sensitivity into regions where cosmological constraints are comparatively weak. Since the source of KM3-230213A is unknown, the~corresponding curves represent conditional sensitivities that depend on the assumed distance and representative neutrino~energy.

The interaction rates are given in analytic form: helicity-polarized cross sections for each scenario, Fermi--Dirac thermal rates built from Maxwell--Boltzmann expressions through a rapidly converging series and supplemented by closed-form light- and heavy-mediator limits, and~resonant kernels for an arbitrary relic mass. Beyond~their computational convenience, these expressions make the origin of the main propagation features explicit and can be used in transport calculations and for a rapid reassessment of future high-energy~events.

\enlargethispage{2\baselineskip}

Overall, this work provides a systematic comparison of $\nu$SI survival benchmarks across the interaction models, relic-mass spectra and source observations considered. The~survival curves describe the unscattered component of an assumed incident flux and carry no statistical confidence level. Their observational interpretation requires information about the intrinsic source emission, the~detector response, and~energy and angular redistribution during propagation. With~these qualifications and the adopted relic populations, the~results identify characteristic propagation signatures and parameter regions that high-energy neutrino observations can test alongside laboratory and cosmological~probes.

\begin{acknowledgments}
I am grateful to Andreas Trautner for proposing the original idea underlying this work and for helpful comments throughout its development.
An earlier version of this manuscript mistakenly listed him as a coauthor without his consent. I accept responsibility for this oversight and~apologize.
I also thank Adam Smetana for extensive discussions that significantly improved this~manuscript.
This research was funded by the Czech Science Foundation (GA\v{C}R), project No.~24-12702S.
\end{acknowledgments}

\appendix
% Appendix numbering as in the journal version: sections A, B, ...;
% equations, figures and tables numbered A1, A2, ... throughout.
\makeatletter
\renewcommand{\thesubsection}{\thesection.\arabic{subsection}}
\renewcommand{\thesubsubsection}{\thesubsection.\arabic{subsubsection}}
\setcounter{equation}{0}\setcounter{figure}{0}\setcounter{table}{0}
\@removefromreset{equation}{section}
\renewcommand{\theequation}{A\arabic{equation}}
\renewcommand{\thefigure}{A\arabic{figure}}
\renewcommand{\thetable}{A\arabic{table}}
\makeatother

\section{Conventions and Thermal-Rate~Functions}
\label{app:conventions}

These appendices provide the analytic expressions for the cross sections and interaction rates used in the survival benchmarks. For~a Fermi--Dirac momentum distribution, the~exact thermal integrals generally have no closed form in terms of elementary or known special functions. The~UR rates are therefore built from Maxwell--Boltzmann expressions, which can be evaluated analytically, and~the Fermi--Dirac result is recovered as a rapidly converging series (Appendix%
~\ref{app:FD}). Resonances are treated separately (Appendix~\ref{app:pole}).

Appendix~\ref{app:conventions} defines the conventions, the~rate normalization and the common thermal functions, and~gives the UR resonance with a finite relic mass (Appendix~\ref{app:URmass}). \mbox{Appendices~\ref{app:vectorD}, \ref{app:scalarD} and \ref{app:scalarM}} give the cross sections and rates for spin-one exchange with Dirac neutrinos, scalar exchange with Dirac neutrinos and scalar exchange with Majorana neutrinos, respectively. They follow the same order: cross sections, interaction rates, and~expressions used in the survival benchmarks. Their rate tables refer to UR targets unless stated otherwise. Appendix~\ref{app:NR} derives the pole rates for an arbitrary relic mass, including NR and semi-relativistic targets, {and Appendix~\ref{app:refs} collects the comparison cases}.

\subsection{Conventions and Rate~Normalization}

\subsubsection{Kinematics and~Propagators}

We use the dimensionless variables %
\begin{equation}
x=\frac{s}{4ET},
\qquad
r=\frac{M^2}{4ET},
\qquad
\delta=\frac{M\Gamma_\phi}{4ET}.
\end{equation}

Neutrino masses are neglected in unsuppressed cross sections, but~retained
where they give the leading contribution to a helicity-suppressed process.
For massless neutrino final states,
\begin{equation}
s+t+u=0,
\qquad
-s\leq t\leq0.
\label{eq:kin_massless}
\end{equation}
For mediator-pair production $\bar\nu_a\nu_a\to XX$, with~$X=V$ or $\phi$,
the mediator mass is retained:
{\begin{align}
s+t+u&=2M^2,
\qquad s\geq4M^2,
\nonumber\\
t&=M^2-\frac{s}{2}(1-\beta\cos\vartheta),
\qquad
\beta=\sqrt{1-\frac{4M^2}{s}},
\nonumber\\
t_{\min,\max}&=M^2-\frac{s}{2}(1\pm\beta).
\label{eq:VV_kinematics}
\end{align}
}
Here $\vartheta$ is the angle between the incident antineutrino and one
outgoing mediator in the centre-of-mass~frame.

Square brackets in the cross-section tables denote the finite-width
$s$-channel \mbox{propagator},
\begin{equation}
\frac{1}{[s-M^2]}
\equiv
\frac{1}{s-M^2+iM\Gamma_\phi}.
\end{equation}
Thus, $1/[s-M^2]^2$ in a squared amplitude denotes
$|s-M^2+iM\Gamma_\phi|^{-2}$, while $1/[s-M^2]$ in an interference term
denotes the real part of the propagator. The~spacelike $t$- and
$u$-channel propagators need no width.
Near the pole, the~mediator propagator is treated in the narrow-width approximation (NWA),
\begin{equation}
\frac{1}{(s-M^2)^2+M^2\Gamma_\phi^2}
\;\rightarrow\;
\frac{\pi}{M\Gamma_\phi}\,\delta\!\left(s-M^2\right).
\label{eq:NWA}
\end{equation}

\subsubsection{Helicity Conventions and Channel~Counting}

Helicity-polarized cross sections use the populated initial helicities
without an initial-spin average; final helicities are summed.
Spin-averaged quantities are denoted by an overline:
$\overline\sigma$ and $\overline{\mathcal R}$.
Identical-particle symmetry factors are included in the cross~sections.

Dirac rate-table entries refer to one populated target state and, for~the pure $s$-channel contribution, one open final species. Majorana entries include both populated target helicities where stated. For~a fixed incident state $\bar\nu_a$, the~$t$-channel is summed over all populated targets $\nu_i$ and $\bar\nu_i$. The~identical target $\bar\nu_a$ additionally contributes the $u$-channel and the $t$--$u$ interference. The~pure $u$-channel contribution is absorbed into $\mathcal R_{t++}$ through the identical-particle symmetry factor, so that only the interference term enters separately as $\mathcal R_{tu}$. The~$s$-channel, the~$s$--$t$ interference (for $j=a$ only) and mediator-pair production require the target $\nu_a$, and~the $s$-channel is summed over the open final species $\bar\nu_j\nu_j$. For~a non-degenerate background, each target contribution is evaluated with its own mass before~summation.

\subsubsection{Rate~Normalization}

We separate the interaction rate into an off-resonant part and the
$s$-channel pole,
$\mathcal R=\mathcal R_{\mathrm{off}}+\mathcal R_{\mathrm{res}}$.
The latter describes on-shell mediator production at $s=M^2$.
The off-resonant rate is a sum over channel rates,
\begin{equation}
\mathcal R_{\mathrm{off}}
=\sum_X c_X\,\mathcal R_X,
\label{eq:Gamma_nonres}
\end{equation}
where $c_X$ counts target or final species. In~the UR rate tables, each
rate is given in units of
\begin{equation}
\mathcal N_1=\frac{g^4T^2}{16\pi^3E}
\quad\text{(spin one)},
\qquad
\mathcal N_0=\frac{g^4T^2}{64\pi^3E}
\quad\text{(spin zero)},
\end{equation}
so that $\mathcal R_X/\mathcal N$ depends on $E$ and $T$ only through the
dimensionless arguments defined above.
Unless stated otherwise, each $\mathcal R_X$ refers to one target
population, for~example $\bar\nu_{i,+}$ with density $n_i$. The~rate functions are listed in Table~\ref{tab:rate_labels}.
The pole rate scales as $\mathcal R_{\mathrm{res}}\propto g^2$, and~the
off-resonant rate as $g^4$ in the zero-width~limit.

\begin{table}[htbp]
\centering
\setlength{\tabcolsep}{10pt}
\caption{Rate functions and the contributions they describe. Each refers to one target population unless stated~otherwise.}
\label{tab:rate_labels}
\begin{tabular}{ll}
\toprule
\textbf{Rate} & \textbf{Contribution}\\
\midrule
$\mathcal R_{t++}$ &
$t$-channel, target $\bar\nu_{i,+}$\\
$\mathcal R_{t+-}$ &
$t$-channel, target $\nu_{i,-}$\\
$\mathcal R_{tu}$ &
$t$--$u$ interference, identical target $\bar\nu_a$\\
$\mathcal R_s$ &
Off-resonant $s$-channel, per open final species\\
$\mathcal R_{st}$ &
$s$--$t$ interference, final species $j=a$\\
$\mathcal R_{\mathrm{res}}$ &
Resonant $s$-channel pole\\
$\mathcal R_{VV}$, $\mathcal R_{\phi\phi}$ &
Mediator-pair production\\
\bottomrule
\end{tabular}
\end{table}

For Dirac neutrinos the complete identical-target functions are
\begin{equation}
\mathcal R_{t+u}\equiv\mathcal R_{t++}+\mathcal R_{tu},
\qquad
\mathcal R_{s+t}\equiv
\mathcal R_{t+-}+\mathcal R_s+\mathcal R_{st}.
\label{eq:Tcomposite}
\end{equation}
The factor $2\,\mathrm{Re}(\dots)$ is included in the interference
functions, which may be~negative.

\subsection{Thermal Averaging and the FD~Expansion}
\label{app:FD}

The thermal convolution in Equation~\eqref{Gamma} is linear in the target
momentum distribution. Expanding the Fermi--Dirac (FD) distribution,
\begin{equation}
\frac{1}{e^{p/T}+1}
=\sum_{n=1}^{\infty}(-1)^{n+1}e^{-np/T},
\end{equation}
expresses it as a sum of Maxwell--Boltzmann (MB) weights at temperatures
$T/n$. Since the MB integrals can be evaluated analytically, this expansion yields closed-form rates that would otherwise not be available for the FD distribution. Consequently,
\begin{equation}
\mathcal R^{\mathrm{FD}}(E,T)
=\sum_{n=1}^{\infty}(-1)^{n+1}
\mathcal R^{\mathrm{MB}}(E,T/n),
\label{eq:FD_MB}
\end{equation}
with the incident energy, masses, mediator parameters and coupling held
fixed.

For the massless UR functions this gives
\begin{equation}
\left(\frac{\mathcal R}{\mathcal N}\right)^{\mathrm{FD}}\!(r,\delta)
=\mathcal S\left\{\left(\frac{\mathcal R}{\mathcal N}\right)^{\mathrm{MB}}\right\}\!(r,\delta),
\qquad
\mathcal S\{f\}(r,\delta)\equiv\sum_{n=1}^{\infty}
\frac{(-1)^{n+1}}{n^2}\,
f(nr,n\delta).
\label{eq:FDsum}
\end{equation}
The factor $n^{-2}$ follows from $\mathcal N\propto T^2$, while $r$ and
$\delta$ scale as $1/T$. For~finite-mass functions, any additional
mass-dependent arguments must also be evaluated at $T/n$.

\subsection{Common Analytic~Functions}

We use the exponential integrals
\begin{align}
E_1(r)&\equiv\int_r^\infty dt\,\frac{e^{-t}}{t},
\nonumber\\
\operatorname{Ei}(r)&\equiv
\operatorname{PV}\int_{-\infty}^{r}dt\,\frac{e^t}{t},
\qquad r>0.
\end{align}
For complex arguments, $E_1$ denotes its analytic continuation on the
principal~branch.

The MB functions entering the massless UR $t$-channel rates are
\begin{align}
F(r)&\equiv
\int_0^\infty dx\,\frac{e^{-x}}{x+r}
=e^rE_1(r),
\label{eq:F}\\
Q(r)&\equiv
\int_0^\infty dx\,\frac{e^{-x}}{x}
\ln\!\left(1+\frac{x}{r}\right)
=
G^{3,1}_{2,3}\!\left(
r\,\middle|\,
\begin{matrix}
0,1\\
0,0,0
\end{matrix}
\right),
\label{eq:Q}
\end{align}
where $G$ is the Meijer $G$-function. We also use the identity
\begin{equation}
\frac12F(r)^2
=\int_0^\infty dx\,
\frac{e^{-x}\ln(1+x/r)}{x+2r}.
\end{equation}
The $t$-channel and $t$--$u$ interference functions are expressed through
$F$ and $Q$ in the model-specific tables. Their FD forms follow from
Equation~\eqref{eq:FDsum}.

The light- and heavy-mediator limits use
\begin{equation}
\xi_k\equiv
\sum_{n=1}^{\infty}\frac{(-1)^{n+1}}{n^k},
\qquad
\xi_2=\frac{\pi^2}{12},
\qquad
\xi_3=\frac{3\zeta(3)}4,
\qquad
\xi_4=\frac{7\pi^4}{720},
\label{eq:xi_k}
\end{equation}
together with
\begin{equation}
L_r\equiv\ln r+\gamma,
\qquad
c_{\ln}\equiv
\sum_{n=1}^{\infty}\frac{(-1)^{n+1}\ln n}{n^2}
\simeq-0.1013,
\end{equation}
where $\gamma$ is the Euler--Mascheroni~constant.

\subsection{Massless UR Resonance and Pole~Subtraction}
\label{app:pole}

The $s$-channel cross section is proportional to
$g^4s/[(s-M^2)^2+M^2\Gamma_\phi^2]$. With~the species weight $c_s$ of
Appendices~\ref{app:vectorD}--\ref{app:scalarM} factored out, its MB thermal
average gives the finite-width $s$-channel rate
\begin{equation}
\frac{\mathcal R_s+\mathcal R_{\mathrm{res}}}{c_s\mathcal N}
=
\int_0^\infty dx\,
\frac{x^2e^{-x}}{(x-r)^2+\delta^2}
=
\frac{1}{\delta}\operatorname{Im}
\!\left[z+z^2e^{-z}E_1(-z)\right],
\qquad z=r+i\delta,
\label{eq:B}
\end{equation}
with $r>0$ and $\delta>0$. For~$\delta\ll r$, the~Breit--Wigner profile is sharply peaked at $x=r$. The~narrow-width approximation (NWA) gives
\begin{equation}
\frac{\mathcal R_{\mathrm{res}}^{\mathrm{MB}}}{c_s\mathcal N}
=\frac{\pi r^2}{\delta}e^{-r},
\qquad
\frac{\mathcal R_{\mathrm{res}}^{\mathrm{FD}}}{c_s\mathcal N}
=\frac{\pi r^2}{\delta}\ln(1+e^{-r}).
\label{eq:BNWA}
\end{equation}
The pole is enhanced by $1/\delta\propto g^{-2}$.

The off-resonant remainder is obtained by subtracting this pole or,
equivalently, by~evaluating the bare-propagator integral in the finite-part sense:
\begin{align}
\frac{\mathcal R_s^{\mathrm{MB}}}{c_s\mathcal N}(r)
&\equiv
\lim_{\delta\to0^+}
\left[
\frac{\mathcal R_s+\mathcal R_{\mathrm{res}}}{c_s\mathcal N}
-\frac{\pi r^2e^{-r}}{\delta}
\right]
\nonumber\\
&= \int_0^\infty dx\,
\frac{x^2e^{-x}}{(x-r)^2}
=
1-r-r(2-r)e^{-r}\operatorname{Ei}(r).
\label{eq:Boff}
\end{align}
Thus the finite-width rate separates into the NWA pole and the off-resonant remainder up to corrections of order $\delta$.\label{eq:Bsplit}
The model-dependent weights $c_s$ are specified in
Appendices~\ref{app:vectorD}--\ref{app:scalarM}. In~terms of
Equation~\eqref{eq:res_massless}, the~pure-vector helicity-polarized reference
has $\kappa=1$, and~the spin-averaged $(S,D)$ case has
$\kappa=\tfrac14$.

\subsection{UR Resonance with Finite Relic~Mass}
\label{app:URmass}
\label{app:UR}

The pole is evaluated in the NWA with the relic momentum distribution
retained; the derivation and the exact kernels are given in
Appendix~\ref{app:NR}. The~pole $s=M^2$ is accessible only for relic
momenta $p\geq p_{\min}$, or~equivalently for $w=E_i+p\geq w_{\min}$, with~\begin{equation}
\frac{p_{\min}}{T}=\frac{|M^4-4E^2m^2|}{4EM^2T},
\qquad
\frac{w_{\min}}{2T}=\max\!\left(\frac{M^2}{4ET},\,\frac{Em^2}{M^2T}\right).
\label{eq:boundary_main}
\end{equation}
For $m\ll T$, the~MB pole rates are
\begin{align}
\mathcal R^{V-A}_{\rm res}&=\frac{g^2}{4\pi E}\left[\frac{M^2T}{2E}\,
e^{-w_{\min}/2T}+\frac{m^2}{2}\,\mathrm{Ei}\!\left(-\frac{w_{\min}}{2T}\right)\right],
\label{eq:res_massive}\\
\mathcal R^{S,M}_{\rm res}&=\frac{g^2M^2T}{16\pi E^2}\,e^{-p_{\min}/T},\\
\mathcal R^{S,D}_{\rm res}&=-\frac{g^2m^2}{16\pi E}\left[
\left(1+\frac{M^2}{4ET}\right)\mathrm{Ei}\!\left(-\frac{w_{\min}}{2T}\right)
+\frac{M^2}{2Ew_{\min}}\,e^{-w_{\min}/2T}\right].
\label{eq:res_massive_SD}
\end{align}
The FD rates follow from
the series of Equation~\eqref{eq:FD_MB}. For~$m\to0$, the~$(V{-}A,D)$ and
$(S,M)$ rates reduce to Equation~\eqref{eq:res_massless}, while the
helicity-suppressed $(S,D)$ pole vanishes as $m^2$. For~$M^2\ll2Em$, the~boundary grows as $p_{\min}\simeq Em^2/M^2$ and cuts off the low-$M$
side of the~resonance.

These rates are the MB limit of the UR kernels given in Appendix~\ref{app:URkernels}. The~off-resonant UR rates are evaluated in the massless limit (Appendix~\ref{app:pole}), where neutrino-mass corrections are of order $m^2/s$%
.

\subsection{Interference, $s$--$t$}

The interference contains the real part of the $s$-channel propagator,
\begin{equation}
\mathrm{Re}\,\frac{1}{s-M^2+iM\Gamma_\phi}
=
\frac{s-M^2}{(s-M^2)^2+M^2\Gamma_\phi^2}.
\end{equation}
It vanishes at $s=M^2$, changes sign across the pole, and~tends to a
principal value as $\Gamma_\phi\to0$. It has no inverse-width enhancement
and contributes to the off-resonant coefficient $A$.

For light mediators, $\mathcal R_{st}$ is subleading to the $t$-channel
rate, and~near resonance it is subleading to the pole. For~heavy mediators,
it scales as $1/r^2$ like the other non-resonant neutrino-scattering
functions and is retained. For~$(S,D)$ with populated
initial helicities it is additionally helicity~suppressed.

\subsection{Mediator-Pair~Production}

The threshold $s\geq4M^2$ is retained in the pair-production cross
sections. For~heavy mediators it can be reached only by the
high-momentum tail of the relic distribution, so the thermal rate is
exponentially suppressed. The~light- and heavy-mediator expressions used
in the survival benchmarks are collected in
Tables~\ref{tab:V_limits} and \ref{tab:SD_rates}.

\subsection{Computational Prescriptions for the UR~Rates}

Table~\ref{tab:UR_prescriptions} summarizes how the functions enter the
off-resonant coefficient $A$ and the pole coefficient $B$.
The model-specific rate tables in
Appendices~\ref{app:vectorD}--\ref{app:scalarM} give their expressions and
multiplicities. Exact massive-target kernels and their NR behavior are
given in Appendix~\ref{app:NR}.

\begin{table}[htbp]
\centering
\small
\caption{Prescriptions %
 used for the UR benchmark rates.
HP denotes the populated helicities.
Scalar Majorana rates include both populated target helicities.
The FD series is Equation~\eqref{eq:FDsum}. The~massless NWA pole of
Equation~\eqref{eq:BNWA} is the analytic reference, whereas the finite-mass pole coefficients used in the UR benchmarks are
given in Appendix~\ref{app:URmass}. \label{tab:UR_prescriptions}
}
\setlength{\tabcolsep}{8pt}
\begin{tabular}{@{}>{\raggedright\arraybackslash}p{3.4cm}
>{\raggedright\arraybackslash}p{3.6cm}
>{\raggedright\arraybackslash}p{3.4cm}
>{\raggedright\arraybackslash}p{3.7cm}@{}}
\toprule
\textbf{Contribution}
& \textbf{Spin One, HP}
& \textbf{Scalar Dirac, HP}
& \colheadl{\textbf{Scalar Dirac, AV;}\\\textbf{Majorana}}\\
\midrule
$t$, $t$--$u$ ($A$)
& Exact FD series
& Exact FD series
& Exact FD~series\\
$s$, off~resonance ($A$)
& Pole-subtracted FD series
& Neglected, helicity-suppressed
& Pole-subtracted FD~series\\

$s$ pole ($B$)
& NWA: Equation~\eqref{eq:BNWA} ($m=0$); $W^{V{-}A}_{\mathrm{UR}}$ ($m\neq0$)
& NWA: vanishes for $m=0$; $W^{S,D}_{\mathrm{UR}}$ ($m\neq0$)
& NWA: Equation~\eqref{eq:BNWA} ($m=0$); $W^0_{\mathrm{UR}}$ ($m\neq0$)\\

$s$--$t$ ($A$)
& Heavy limit for $r\geq3.5$; negligible for $r<3.5$
& Neglected, helicity-suppressed
& AV: neglected. Majorana: heavy limit for $r\geq3.5$, negligible for $r<3.5$\\

Mediator pairs ($A$)
& Light limit, $r<20$, Heavy limit, $r>20$
& Light limit, $r<1.22$, Heavy limit, $r>1.22$
&  Light limit, $r<1.22$ \newline Heavy limit, $r>1.22$\\
\bottomrule
\end{tabular}
\end{table}
\unskip

\section{Spin-1 Mediator, Dirac~Neutrinos}
\label{app:vectorD}

We consider pure-vector ($V$) and left-chiral ($V{-}A$) vertices,
$g\gamma^\mu$ and $g\gamma^\mu P_L$, respectively, with~$P_L=(1-\gamma^5)/2$. The~incident state is $\bar\nu_{a,+}$, and~the populated targets are $\bar\nu_{i,+}$ and $\nu_{i,-}$. Final helicities are summed. The~$V{-}A$ width is given in Section~\ref{sec:resonant}. For~the pure-vector reference, $\Gamma_\phi^{V,D}\simeq g^2M/(12\pi)$ per neutrino~species.

\subsection{Cross~Sections}

We write both vertices in a common form using $\xi=1$ for $V{-}A$ and
$\xi=0$ for $V$. The~coefficient of the additional right-chiral
$s$-channel contribution is $1-\xi$ in the differential cross section;
after integration, the~pure $s$-channel coefficient is
$c_s=(2-\xi)/3$, i.e.,\ $c_s=1/3$ for $V{-}A$ and $2/3$ for $V$.

For completeness we also give the spin-averaged cross sections, which follow from the helicity-polarized~ones.

\begin{table}[htbp]
\centering
\setlength{\tabcolsep}{10pt}
\caption{Differential %
 cross sections for $V$ and $V{-}A$ mediators and Dirac neutrinos with helicity-polarized active initial states, with~$\xi=0$ ($V$) or $1$ ($V{-}A$).
The pure $s$ row gives the contribution per final species.
Identical-final-particle symmetry factors are included. \label{tab:VA_FH_diff}
}

\begin{tabular}{lcc}
\toprule
\textbf{Process} & \textbf{Channel} &
\boldmath{$\displaystyle
\frac{d\sigma^{V,\,V{-}A}}{dt}\times\frac{4\pi s^2}{g^4}
$}
\\
\midrule
$\bar\nu_{a,+}\bar\nu_{a,+}\to\bar\nu_a\bar\nu_a$
& $t+u$
& $\displaystyle
\frac{s^2}{2}
\left[\frac{1}{t-M^2}+\frac{1}{u-M^2}\right]^2
$
\\ \midrule
$\bar\nu_{a,+}\nu_{a,-}\to\bar\nu_a\nu_a$
& $s+t$
& $\displaystyle
\frac{u^2}{(t-M^2)^2}
+\frac{2u^2}{[s-M^2](t-M^2)}
+\frac{u^2+(1-\xi)t^2}{[s-M^2]^2}
$
\\\midrule
$\bar\nu_{a,+}\nu_{a,-}\to\bar\nu_j\nu_j$, $j\neq a$
& $s$
& $\displaystyle
\frac{u^2+(1-\xi)t^2}{[s-M^2]^2}
$
\\\midrule
$\bar\nu_{a,+}\bar\nu_{i,+}\to\bar\nu_a\bar\nu_i$, $i\neq a$
& $t$
& $\displaystyle
\frac{s^2}{(t-M^2)^2}
$
\\\midrule
$\bar\nu_{a,+}\nu_{i,-}\to\bar\nu_a\nu_i$, $i\neq a$
& $t$
& $\displaystyle
\frac{u^2}{(t-M^2)^2}
$
\\
\midrule
$\bar\nu_{a,+}\nu_{a,-}\to VV$
& $t+u$
& $\displaystyle
\frac12
\left[
\frac{s^2+6M^4-2tu}{tu}
-\frac{M^4(s-2M^2)^2}{t^2u^2}
\right]
$
\\[4pt]
\bottomrule
\end{tabular}
\end{table}
\unskip

\begin{table}[htbp]
\centering
\setlength{\tabcolsep}{10pt}
\caption{Total %
 cross sections for $V$ and $V{-}A$ mediators and Dirac neutrinos with helicity-polarized active initial states, with~$c_s=2/3$ ($V$) or $1/3$ ($V{-}A$).
The pure $s$ contribution is given per final species.
The pair-production cross section vanishes for $s<4M^2$;
its threshold value is understood by continuity.
Identical-final-particle symmetry factors are included. \label{tab:VA_FH_tot}
}

\begin{tabular}{lc}
\toprule
\textbf{Process} &
\boldmath{$\displaystyle
\sigma^{V,\,V{-}A}\times\frac{4\pi s^2}{g^4}
$}
\\
\midrule
$\bar\nu_{a,+}\bar\nu_{a,+}\to\bar\nu_a\bar\nu_a$
& $\displaystyle
\frac{s^3}{M^2(M^2+s)}
-\frac{2s^2}{2M^2+s}\ln\frac{M^2}{M^2+s}
$
\\\midrule
$\bar\nu_{a,+}\nu_{a,-}\to\bar\nu_a\nu_a$
& $\begin{aligned}
\displaystyle
&\frac{s(2M^2+s)}{M^2}
+2(M^2+s)\ln\frac{M^2}{M^2+s}
+\frac{c_s s^3}{[s-M^2]^2}+\\
&+\frac{s(3s+2M^2)+2(s+M^2)^2\ln\frac{M^2}{M^2+s}}{[s-M^2]}
\end{aligned}
$
\\\midrule
$\bar\nu_{a,+}\nu_{a,-}\to\bar\nu_j\nu_j$, $j\neq a$
& $\displaystyle
\frac{c_s s^3}{[s-M^2]^2}
$
\\\midrule
$\bar\nu_{a,+}\bar\nu_{i,+}\to\bar\nu_a\bar\nu_i$, $i\neq a$
& $\displaystyle
\frac{s^3}{M^2(M^2+s)}
$
\\\midrule
$\bar\nu_{a,+}\nu_{i,-}\to\bar\nu_a\nu_i$, $i\neq a$
& $\displaystyle
\frac{s(2M^2+s)}{M^2}
+2(M^2+s)\ln\frac{M^2}{M^2+s}
$
\\
\midrule%
$\bar\nu_{a,+}\nu_{a,-}\to VV$
& $\displaystyle
\frac{2(s^2+4M^4)}{s-2M^2}
\ln\frac{s+\sqrt{s(s-4M^2)}}{s-\sqrt{s(s-4M^2)}}
-2\sqrt{s(s-4M^2)}
$
\\%
\bottomrule
\end{tabular}
\end{table}
\unskip

For pure $V$, the~spin-averaged total cross sections follow from the helicity-polarized ones as
\begin{align}
& \overline\sigma{}_t^{V} =
\frac12
\left(
\sigma_{t,++}^{V}
+\sigma_{t,+-}^{V}
\right),\qquad
\overline\sigma{}_{t+u}^{V} =
\frac12
\left(
\sigma_{t+u}^{V}
+\sigma_{t,+-}^{V}
\right),
\nonumber\\
&\overline\sigma{}_s^{V}
=\frac12\sigma_s^{V},
\qquad
\overline\sigma{}_{st}^{V}
=\frac12\sigma_{st}^{V},\qquad\overline\sigma{}_{VV}^{V}
=\frac12\sigma_{VV}^{V}.
\label{eq:V_av_cross_sections}
\end{align}

The subscripts $++$ and $+-$ denote the initial helicities;
$\sigma_{st}$ denotes the interference contribution.
For massless initial states, every $V{-}A$ spin-averaged cross section is one quarter of its helicity-polarized~counterpart.

\subsection{Interaction~Rates}

For three degenerate target mass eigenstates and three open final species, the~off-resonant rate is
\be
\mathcal R_{\mathrm{off}}=
3\mathcal R_{t++}+3\mathcal R_{t+-}+\mathcal R_{tu}
+3\mathcal R_s+\mathcal R_{st}+\mathcal R_{VV},
\label{eq:V_total}
\ee
and the pole rate $\mathcal R_{\mathrm{res}}$ is summed over the three open final species.
The multiplicities follow the channel counting of Appendix~\ref{app:conventions}.

For the pure-vector spin-averaged (as opposed to helicity-polarized) reference,
$\overline{\mathcal R}_{t++}=\overline{\mathcal R}_{t+-}
=\overline{\mathcal R}_t$, giving
\begin{equation}
\overline{\mathcal R}_{\mathrm{off}}
=
6\overline{\mathcal R}_t
+\overline{\mathcal R}_{tu}
+3\overline{\mathcal R}_s
+\overline{\mathcal R}_{st}
+\overline{\mathcal R}_{VV}.
\label{eq:V_total_SA}
\end{equation}

\begin{table}[htbp]
\centering
\setlength{\tabcolsep}{10pt}
\caption{Rate %
 functions used in the spin-one survival benchmarks for the $V{-}A$
mediator. The~expressions are MB functions, and~$\mathcal S$ denotes the FD
series of Equation~\eqref{eq:FDsum}. For~the $s$--$t$ interference and mediator-pair
production, the~limits used are shown.
$F$, $Q$, $L_r$, $\xi_k$ and $c_{\ln}$ are defined in
\mbox{Appendix~\ref{app:conventions}}. \label{tab:V_limits}}
\begin{tabular}{ll}
\toprule
\textbf{Rate} & \textbf{Expression Used}\\
\midrule
$\mathcal R_{t++}/\mathcal N_1$ & $\displaystyle\mathcal S\left\{\frac1r-1+rF(r)\right\}$\\ \midrule
$\mathcal R_{t+-}/\mathcal N_1$ & $\displaystyle\mathcal S\left\{\frac1r+2-2F(r)-2rQ(r)\right\}$\\ \midrule
$\mathcal R_{tu}/\mathcal N_1$ & $\displaystyle\mathcal S\left\{2\left(F(r)-rF(r)^2\right)\right\}$\\ \midrule
$\mathcal R_{s}/\mathcal N_1$ &
$\displaystyle\frac13\,\mathcal S\left\{1-r-r(2-r)e^{-r}\operatorname{Ei}(r)\right\}$\\ \midrule
$\mathcal R_{st}/\mathcal N_1$ & $\displaystyle r\geq3.5:\ \mathcal S\left\{\frac{4}{3r^2}\right\}$; negligible for $r<3.5$\\ \midrule
$\mathcal R_{\mathrm{res}}/\mathcal N_1$ &
$\begin{aligned}
m=0:&\quad\frac{8\pi^2}{g^2}\,r\ln(1+e^{-r}),\\
m\neq0:&\quad\frac{2\pi g^2}{\mathcal N_1}W^{V{-}A}_{\mathrm{UR}}(\mu)
\end{aligned}$\\
\midrule%
$\mathcal R_{VV}/\mathcal N_1$ &
$\begin{aligned}
r<20:&\quad \mathcal S\left\{2\left[E_1(4r)+e^{-4r}(\ln4-1)\right]\right\}
\simeq -2\left[\xi_2(1+L_r)+c_{\ln}\right],\\
r>20:&\quad \mathcal S\left\{2\sqrt\pi\,\frac{e^{-4r}}{\sqrt r}\right\}
\end{aligned}$\\%
\bottomrule
\end{tabular}
\end{table}

 For helicity-polarized initial states the $V$ and $V{-}A$ mediators give the same functions except in the $s$-channel: $\mathcal R_s$ is doubled ($c_s=\tfrac13\to\tfrac23$,
i.e.,\ $\mathcal R^{V}_s=2\mathcal R^{V{-}A}_s$) and in $\mathcal R_{st}$ the width changes, $\delta_{V{-}A}\to\delta_V=M\Gamma_\phi^V/(4ET)$.
The helicity-polarized resonant rates coincide, $\mathcal R^{V}_{\mathrm{res}} = \mathcal R^{V{-}A}_{\mathrm{res}}$, because~the smaller $V{-}A$ numerator is compensated by its smaller total width.
For massless UR targets, the~pure-vector spin-averaged functions
follow from the helicity-polarized ones as
\begin{alignat}{2}
&\overline{\mathcal R}_t=\tfrac12\left(\mathcal R_{t++}+\mathcal R_{t+-}\right),\qquad
&&\overline{\mathcal R}_{tu}=\tfrac12\mathcal R_{tu},
\nonumber\\
&\overline{\mathcal R}_s=\tfrac12\mathcal R^{V}_s,
&&\overline{\mathcal R}_{st}=\tfrac12\mathcal R_{st}(r,\delta_V),
\nonumber\\
&\overline{\mathcal R}_{VV}=\tfrac12\mathcal R_{VV},
&&\overline{\mathcal R}_{\mathrm{res}}=\tfrac12\mathcal R_{\mathrm{res}}.
\label{eq:V_SA_relations}
\end{alignat}

The pure-vector spin-averaged interference is one half of the helicity-polarized \mbox{contribution}.

\section{Spin-0 Mediator, Dirac~Neutrinos}
\label{app:scalarD}

We consider a real scalar mediator $\phi$ with Yukawa vertex
$g\phi\bar\nu\nu$. The~incident state is $\bar\nu_{a,+}$, and~the
populated targets are $\bar\nu_{i,+}$ and $\nu_{i,-}$.
Final helicities are~summed.

\subsection{Cross~Sections}

For the active initial pair $\bar\nu_{a,+}\nu_{a,-}$, the~scalar
annihilation amplitude contains the spinor bilinear
$\bar v_+(k)u_-(p_i)$, whose squared magnitude is
\begin{equation}
H_{+-}=|\bar v_+(k)u_-(p_i)|^2
=(1+c_\theta)
\left[
EE_i-m_a^2-p_i\sqrt{E^2-m_a^2}
\right].
\label{eq:eta_def}
\end{equation}
Here $E$ and $E_i$ are the incident and target energies,
$p_i=\sqrt{E_i^2-m_a^2}$, and~$c_\theta$ is the cosine of the
angle between their momenta.
We define the helicity-kinematic factor \be\etaa\equiv2H_{+-}/s.\ee
Since $H_{+-}=0$ for $m_a=0$, the~helicity-polarized
$s$-channel vanishes in the massless limit.
To leading order in $E_i/E$,
\begin{equation}
\etaa=
\frac{(1+c_\theta)m_a^2}
{m_a^2+p_i(E_i+p_i)(1-c_\theta)}
\leq 1+c_\theta\leq2.
\label{eq:eta_E}
\end{equation}
For UR targets it reduces to
\begin{equation}
\etaa\simeq
\frac{m_a^2}{2E_i^2}\frac{1+c_\theta}{1-c_\theta},
\qquad
1-c_\theta\gg\frac{m_a^2}{2p_i^2},
\label{eq:eta}
\end{equation}
showing the helicity suppression outside the narrow collinear cone.
For NR targets, $\etaa\to1+c_\theta$, and~its angular average
$\langle\etaa\rangle=1$ gives the same $s$-channel rate for
helicity-polarized and spin-averaged initial~states.

Keeping the neutrino masses in the phase space, the~total
$s$-channel cross section is
\begin{equation}
\sigma^{S,D}_{s,+-}
=
\frac{g^4H_{+-}}{8\pi s[s-M^2]^2}
\frac{(s-4m_j^2)^{3/2}}{(s-4m_a^2)^{1/2}}
\simeq
\frac{g^4m_a^2E(1+c_\theta)}{8\pi(E_i+p_i)[s-M^2]^2}.
\label{eq:sigma_s_FH}
\end{equation}

Tables~\ref{tab:SD_FH_diff} and \ref{tab:SD_FH_tot} collect the
cross sections using massless external-state kinematics, with~the
helicity suppression retained through $\etaa$.
Unlike the spin-one case,
$\mathcal R_{t++}=\mathcal R_{t+-}\equiv\mathcal R_t^S$.

The initial-spin prescription is selected by
\begin{equation}
\eta=
\begin{cases}
\etaa, & \text{helicity-polarized},\\
1, & \text{spin-averaged},
\end{cases}
\qquad
\chi=
\begin{cases}
1, & \text{helicity-polarized},\\
0, & \text{spin-averaged}.
\end{cases}
\end{equation}
Below we present the differential and total cross sections for both
initial-spin prescriptions, selected by $\eta$ and $\chi$.

\begin{table}[htbp]
\centering
\setlength{\tabcolsep}{10pt}
\caption{Differential %
 cross sections for a scalar mediator and Dirac neutrinos.
The initial-spin prescription is selected by $\eta$ and $\chi$.
For helicity-polarized initial states, the~incident state is
$\bar\nu_{a,+}$ and the populated targets are
$\bar\nu_{i,+}$ and $\nu_{i,-}$.
Distinct species satisfy $i\neq a$.
The factor $\etaa$ is given in Equation~\eqref{eq:eta_E}, to~leading order in $E_i/E$.
The pair-production row uses Equation~\eqref{eq:VV_kinematics}. \label{tab:SD_FH_diff}
}
\begin{tabular}{lcc}
\toprule
\textbf{Process}
& \textbf{Channel}
& \boldmath{$\displaystyle\frac{16\pi s^2}{g^4}
\frac{d\sigma^{S,D}}{dt}$}\\
\midrule

$\anu_a\anu_a\to\anu_a\anu_a$
& $t+u$
& $\displaystyle
\frac12
\left[
\frac{t^2}{(t-M^2)^2}
+\frac{u^2}{(u-M^2)^2}
+(1+\chi)\frac{tu}{(t-M^2)(u-M^2)}
\right]$\\
\midrule

$\anu_a\nu_a\to\anu_a\nu_a$
& $s+t$
& $\displaystyle
\frac{t^2}{(t-M^2)^2}
+\eta
\left[
\frac{st}{[s-M^2](t-M^2)}
+\frac{s^2}{[s-M^2]^2}
\right]$\\
\midrule

$\anu_a\nu_a\to\anu_i\nu_i$
& $s$
& $\displaystyle
\eta\frac{s^2}{[s-M^2]^2}$\\
\midrule

\shortstack[l]{
$\anu_a\anu_i\to\anu_a\anu_i$\\
$\anu_a\nu_i\to\anu_a\nu_i$
}
& $t$
& $\displaystyle\frac{t^2}{(t-M^2)^2}$\\

\midrule
$\anu_a\nu_a\to\phi\phi$
& $t+u$
& $\displaystyle
(1+\chi)\frac{(t-u)^2(tu-M^4)}{4t^2u^2}$\\

\bottomrule
\end{tabular}
\end{table}
\unskip

\begin{table}[htbp]
\centering
\setlength{\tabcolsep}{10pt}
\caption{Total %
 cross sections corresponding to
Table~\ref{tab:SD_FH_diff}.
The pair-production cross section includes $1/2!$ and
vanishes for $s<4M^2$. \label{tab:SD_FH_tot}
}
\begin{tabular}{lc}
\toprule
\textbf{Process}
& \boldmath{$\displaystyle\frac{16\pi s^2}{g^4}\sigma^{S,D}$}\\
\midrule

$\anu_a\anu_a\to\anu_a\anu_a$
& $\displaystyle
\left[(5+\chi)M^2+(3+\chi)s\right]
\left[
\frac{s}{2(M^2+s)}
+\frac{M^2}{2M^2+s}\ln\frac{M^2}{M^2+s}
\right]$\\
\midrule

$\anu_a\nu_a\to\anu_a\nu_a$
& $\begin{aligned}
&\frac{s(s+2M^2)}{M^2+s}
+2M^2\ln\frac{M^2}{M^2+s}\\
&\quad+\eta
\left\{
\frac{s^3}{[s-M^2]^2}
+\frac{s}{[s-M^2]}
\left[
s+M^2\ln\frac{M^2}{M^2+s}
\right]
\right\}
\end{aligned}$\\
\midrule

$\anu_a\nu_a\to\anu_i\nu_i$
& $\displaystyle
\eta\frac{s^3}{[s-M^2]^2}$\\
\midrule

\shortstack[l]{
$\anu_a\anu_i\to\anu_a\anu_i$\\
$\anu_a\nu_i\to\anu_a\nu_i$
}
& $\displaystyle
\frac{s(s+2M^2)}{M^2+s}
+2M^2\ln\frac{M^2}{M^2+s}$\\
\midrule
$\anu_a\nu_a\to\phi\phi$
& $\begin{aligned}
(1+\chi)\Biggl[
\frac{s^2-4sM^2+6M^4}{s-2M^2}
\ln\frac{s+\sqrt{s(s-4M^2)}}{s-\sqrt{s(s-4M^2)}}-\frac32\sqrt{s(s-4M^2)}
\Biggr]
\end{aligned}$\\
\bottomrule
\end{tabular}
\end{table}
\unskip

\subsection{Interaction~Rates}

For three degenerate target mass eigenstates and three open final
species, the~helicity-polarized off-resonant rate is
\begin{equation}
\mathcal R^{S,D}_{\mathrm{off}}
=
6\mathcal R_t^S+\mathcal R_{tu}^S
+3\mathcal R_s+\mathcal R_{st}^S
+\mathcal R_{\phi\phi},
\label{eq:SD_total_FH}
\end{equation}
and the pole rate is $\mathcal R^{S}_{\mathrm{res}}$.
The spin-averaged rates have the same form, with~the overlined functions of Equation~\eqref{eq:SD_SA_relations} and Table~\ref{tab:SD_rates}.

\begin{table}[htbp]
\centering
\setlength{\tabcolsep}{10pt}
\small %
\caption{Rate %
 functions used in the spin-zero survival benchmarks. The~expressions are
MB functions, and~$\mathcal S$ denotes the FD series of Equation~\eqref{eq:FDsum}.
For mediator-pair production, the~limits used are shown.
The kernels $W^{S,D}_{\mathrm{UR}}$ and $W^0_{\mathrm{UR}}$ are defined in
Appendix~\ref{app:URkernels}; $F$, $Q$, $L_r$, $\xi_k$ and $c_{\ln}$ in
Appendix~\ref{app:conventions}. \label{tab:SD_rates}
}
\begin{tabular}{ll}
\toprule
\textbf{Rate} & \textbf{Expression Used}\\
\midrule
$\mathcal R_t^S/\mathcal N_0$
& $\displaystyle\mathcal S\left\{1+rF(r)-2rQ(r)\right\}$\\
\midrule
$\mathcal R_{tu}^S/\mathcal N_0$
& $\displaystyle\mathcal S\left\{1-rQ(r)-\frac r2F(r)^2\right\}$\\
\midrule
$\mathcal R_s$, $\mathcal R_{st}^S$ (HP)
& neglected for $m_a\ll T$ (helicity suppressed)\\
\midrule
$\overline{\mathcal R}_s/\mathcal N_0$
& $\mathcal S\left\{1-r-r(2-r)e^{-r}\operatorname{Ei}(r)\right\}$\\
\midrule
$\overline{\mathcal R}{}_{st}^S$
& neglected\\
\midrule
$\mathcal R_{\mathrm{res}}^S/\mathcal N_0$ (HP)
& $\displaystyle\frac{\pi g^2}{\mathcal N_0}W^{S,D}_{\mathrm{UR}}(\mu)$;
vanishes for $m_a=0$\\
\midrule
$\overline{\mathcal R}{}_{\mathrm{res}}^S/\mathcal N_0$
& $\begin{aligned}
m=0:&\quad\frac{8\pi^2}{g^2}\,r\ln(1+e^{-r}),\\
m\neq0:&\quad\frac{\pi g^2}{2\mathcal N_0}W^0_{\mathrm{UR}}(\mu)
\end{aligned}$\\
\midrule
$\mathcal R_{\phi\phi}/\mathcal N_0$
& $\begin{aligned}
r<1.22:&\quad\mathcal S\left\{2\left[E_1(4r)+e^{-4r}\left(\ln4-\frac32\right)\right]\right\}
\simeq-2\left[\xi_2\left(\frac32+L_r\right)+c_{\ln}\right],\\
r>1.22:&\quad\mathcal S\left\{\frac{\sqrt\pi}{4}\frac{e^{-4r}}{r^{5/2}}\right\}
\end{aligned}$\\
\bottomrule
\end{tabular}
\end{table}

The helicity factor $\etaa$ must remain inside the thermal convolution
because it depends on the target momentum and collision angle.
In the NWA, angular integration expresses the helicity-polarized $s$-channel pole through the kernel $W^{S,D}$ of
Equation~\eqref{eq:K_def}. The~off-resonant $s$-channel and the $s$--$t$ interference carry the same helicity factor. For~$m_a\ll T$ they are
helicity suppressed and small compared with the other contributions,
and they are neglected in the UR benchmarks. The~resonant $s$-channel
is retained, because~its inverse-width enhancement can compensate for
the helicity suppression. Table~\ref{tab:SD_rates} gives the rate
functions used for the scalar~mediator.

For spin-averaged initial states, the~corresponding functions are
\begin{equation}
\overline{\mathcal R}_{t}^S=\mathcal R_{t}^S,
\qquad
\overline{\mathcal R}_{tu}^S=\frac12\mathcal R_{tu}^S,
\qquad
\overline{\mathcal R}_{\phi\phi}=\frac12\mathcal R_{\phi\phi},
\label{eq:SD_SA_relations}
\end{equation}
while $\overline{\mathcal R}_s$, $\overline{\mathcal R}{}_{st}^S$ and $\overline{\mathcal R}{}_{\mathrm{res}}^S$ are given in Table~\ref{tab:SD_rates}.

\section{Spin-0 Mediator, Majorana~Neutrinos}
\label{app:scalarM}

For Majorana neutrinos the field is self-conjugate.
The scalar vertex is $\tfrac12 g\phi\bar\nu^c\nu$, with~the
symmetry factor absorbed in the coupling.
The mediator width is
$\Gamma_\phi^M=\Gamma_\phi^D/2$.

\subsection{Cross~Sections}
\label{sec:SM_fh}

Each mass eigenstate $\nu_i$ has two equally populated target
helicities in the C$\nu$B.
For a parity-even scalar interaction,
$\sigma_{++}=\sigma_{--}$ and $\sigma_{+-}=\sigma_{-+}$.
For either incident helicity $h=\pm$, averaging over the target
helicities therefore gives
\begin{equation}
\frac{\sigma_{h+}+\sigma_{h-}}{2}
=\overline{\sigma}.
\end{equation}
The fixed-helicity beam rate is consequently equal to the
spin-averaged rate evaluated with the total target density $2n_i$.
This target-helicity sum is included in the rates of
\mbox{Equation~\eqref{eq:SM_total}}.

In the UR limit, the~$s$-channel selects equal initial helicities,
$\sigma_{s,++}^{S,M}=\sigma_{s,--}^{S,M}
\simeq2\overline{\sigma}_s^{S,M}$, while
$\sigma_{s,+-}^{S,M}\to0$.
The target helicity required for the resonance is populated,
and the Majorana pole is not suppressed by $m_a^2/E_i^2$.
Mediator-pair production selects opposite initial helicities,
which are also~populated.

Below we present the spin-averaged differential and total cross
sections. Final helicities are summed, and~identical-particle
symmetry factors are~included.

\begin{table}[htbp]
\centering
\setlength{\tabcolsep}{10pt}
\small
\caption{\label{tab:ScalarMajorana}
Differential cross sections for a scalar mediator and Majorana
neutrinos in the massless limit.
The spin-averaged pair-production cross section equals the
$(S,D)$ result in Table~\ref{tab:SD_FH_diff} and uses
Equation~\eqref{eq:VV_kinematics}.
}
\begin{tabular}{lcc}
\toprule
\textbf{Process}
& \textbf{Channel}
& \boldmath{$\displaystyle
\frac{16\pi s^2}{g^4}
\frac{d\overline{\sigma}^{S,M}}{dt}$}\\
\midrule

$\nu_i\nu_i\to\nu_j\nu_j$, $j\neq i$
& $s$
& $\displaystyle\frac12\frac{s^2}{[s-M^2]^2}$\\
\midrule

$\nu_i\nu_j\to\nu_i\nu_j$, $i\neq j$
& $t$
& $\displaystyle\frac{t^2}{(t-M^2)^2}$\\
\midrule

$\nu_i\nu_i\to\nu_i\nu_i$
& $s+t+u$
& $\begin{aligned}
&\frac12
\left[
\frac{u^2}{(u-M^2)^2}
+\frac{t^2}{(t-M^2)^2}
+\frac{s^2}{[s-M^2]^2}
\right.\\
&\left.+\frac{st}{(t-M^2)[s-M^2]}
+\frac{tu}{(t-M^2)(u-M^2)}
+\frac{su}{[s-M^2](u-M^2)}\right]
\end{aligned}$\\
\midrule
$\nu_a\nu_a\to\phi\phi$
& $t+u$
& $\displaystyle\frac{(t-u)^2(tu-M^4)}{4t^2u^2}$\\

\bottomrule
\end{tabular}
\end{table}
\vspace{-6pt}

\begin{table}[htbp]
\centering
\setlength{\tabcolsep}{10pt}
\caption{\label{tab:SM_av_tot}
Total cross sections corresponding to
Table~\ref{tab:ScalarMajorana}.
The pair-production cross section vanishes for $s<4M^2$. In~the heavy-mediator limit, $M^2\gg s$, the~same-state cross section approaches the contact form $\overline{\sigma}\to\frac{5}{12}\,g^4s/(16\pi M^4)$.
}
\begin{tabular}{lc}
\toprule
\textbf{Process}
& \boldmath{$\displaystyle
\frac{16\pi s^2}{g^4}\overline{\sigma}^{S,M}$}\\
\midrule

$\nu_i\nu_i\to\nu_j\nu_j$, $j\neq i$
& $\displaystyle\frac{s^3}{2[s-M^2]^2}$\\
\midrule

$\nu_i\nu_j\to\nu_i\nu_j$, $i\neq j$
& $\displaystyle
\frac{s(s+2M^2)}{s+M^2}
+2M^2\ln\frac{M^2}{M^2+s}$\\
\midrule
$\nu_i\nu_i\to\nu_i\nu_i$
& $\begin{aligned}
&\frac{s(3s+5M^2)}{2(s+M^2)}
+\frac{s^2}{[s-M^2]}
+\frac{s^3}{2[s-M^2]^2}+\\
&\quad+M^2
\left[
\frac{3s+5M^2}{s+2M^2}
+\frac{s}{[s-M^2]}
\right]
\ln\frac{M^2}{M^2+s}
\end{aligned}$\\

\midrule
$\nu_a\nu_a\to\phi\phi$
& $\begin{aligned}
\frac{s^2-4sM^2+6M^4}{s-2M^2}
\ln\frac{s+\sqrt{s(s-4M^2)}}{s-\sqrt{s(s-4M^2)}} -\frac32\sqrt{s(s-4M^2)}
\end{aligned}$\\

\bottomrule
\end{tabular}
\end{table}
\unskip

\subsection{Interaction~Rates}

For three degenerate target mass eigenstates and three open final
species, the~scalar Majorana off-resonant rate is
\begin{equation}
\mathcal R_{\mathrm{off}}^{S,M}
=
6\mathcal R_t^S+\mathcal R_{tu}^S+3\mathcal R_s
+\mathcal R_{st}^{S,M}+\mathcal R_{\phi\phi},
\label{eq:SM_total}
\end{equation}
and the pole rate is $\mathcal R_{\mathrm{res}}^{S,M}$.
The functions $\mathcal R_t^S$, $\mathcal R_{tu}^S$ and
$\mathcal R_{\phi\phi}$ are the $(S,D)$ functions in
Table~\ref{tab:SD_rates}.
These $(S,D)$ functions refer to one target population, so the coefficients in Equation~\eqref{eq:SM_total} count both populated target helicities and the open final species, whereas $\mathcal R_{st}^{S,M}$ and $\mathcal R_{\mathrm{res}}^{S,M}$ in Table~\ref{tab:SM_rates} already include both target~helicities.

Table~\ref{tab:SM_rates} gives the remaining functions used in the
survival benchmarks. The~resonant rate is $\mathcal R_{\mathrm{res}}^{S,M}
=g^2\pi W^0_{\mathrm{UR}}$, with~the finite-mass UR kernel
defined in Appendix~\ref{app:URkernels}.

\begin{table}[htbp]
\centering
\centering
\setlength{\tabcolsep}{10pt}
\small
\renewcommand{\arraystretch}{1.5}
\caption{
\label{tab:SM_rates}
Majorana-specific rate functions used in Equation~\eqref{eq:SM_total}. The~remaining functions are given in Table~\ref{tab:SD_rates}. The~expressions
are MB functions, and~$\mathcal S$ denotes the FD series of
Equation~\eqref{eq:FDsum}.
}
\begin{tabular}{ll}
\toprule
\textbf{Rate} & \textbf{Expression Used}\\
\midrule
$\mathcal R_s/\mathcal N_0$
& $\mathcal S\left\{1-r-r(2-r)e^{-r}\operatorname{Ei}(r)\right\}$\\
\midrule
$\mathcal R_{st}^{S,M}/\mathcal N_0$
& $\displaystyle r\geq3.5:\ \mathcal S\left\{-\frac{2}{r^2}\right\}$; negligible for $r<3.5$\\
\midrule
$\mathcal R_{\mathrm{res}}^{S,M}/\mathcal N_0$
& $\begin{aligned}
m_a=0:&\quad\frac{16\pi^2}{g^2}r\ln(1+e^{-r}),\\
m_a\neq0:&\quad\frac{\pi g^2}{\mathcal N_0}W_{\mathrm{UR}}^0(\mu)
\end{aligned}$\\
\bottomrule
\end{tabular}
\end{table}
\unskip

\section{Massive-Target Kernels and the NR~Resonance}
\label{app:NR}
\label{app:massive}

This appendix derives the pole rates for an arbitrary relic mass. They are needed for the NR benchmarks, for~the finite-mass UR pole of Appendix~\ref{app:URmass}, and~for relic states that become semi-relativistic along the propagation path of distant~sources.

\subsection{Pole in the Narrow-Width~Approximation}
\label{app:NWApole}
For $E\gg m,p$, the~invariant mass and the M\o ller velocity are
$s\simeq2E(E_i-p\cos\theta)$ and $v_{M\text{\o}ller}\simeq s/(2EE_i)$,
with $E_i=\sqrt{p^2+m^2}$. Replacing the angular integration in
Equation~\eqref{Gamma} by an integration over $s$,
$ds=-2Ep\,d\cos\theta$, gives
\begin{equation}
\mathcal R
=\frac{1}{16\pi^2E^2}\int_0^\infty dp\,\frac{p}{E_i}\,f(p)
\int_{s_-}^{s_+}ds\;s\,\sigma(s),
\qquad
s_\pm=2E\left(E_i\pm p\right).
\label{eq:R_s_p}
\end{equation}
The pole is evaluated in the NWA of Equation~\eqref{eq:NWA}. After~summation over the open final species, the~total width cancels
against the final-state partial widths. The~$\delta$-function removes the $s$-integration, fixes the
collision angle, $\cos\theta^*=(E_i-\mu)/p$ with $\mu=M^2/(2E)$, and~restricts the relic momenta by $E_i-p\le\mu\le E_i+p$, i.e.,
\begin{equation}
p\geq p_{\min}=\frac{|\mu^2-m^2|}{2\mu}
\quad\Longleftrightarrow\quad
w\equiv E_i+p\geq w_{\min}=\max\!\left(\mu,\frac{m^2}{\mu}\right),
\label{eq:res_boundary}
\end{equation}
where $E_i-p=m^2/w$, $p=(w^2-m^2)/(2w)$ and $w(p_{\min})=w_{\min}$.
The two forms describe the same kinematic boundary. We use $p$ when
the numerator of the cross section does not depend on the helicities,
and $w$ when it depends on $E_i\pm p$.

\subsection{Exact~Kernels}
With this boundary, the~remaining momentum integration defines the kernels through which the pole rates are expressed,
\begin{align}
W^{0}(\mu)&=
\frac{\mu}{8\pi^2E}
\int_{p_{\min}}^\infty dp\,\frac{p}{E_i}f(p),
\label{eq:WU_exact}\\
W^{V-A}(\mu)&=
\frac{1}{16\pi^2E}
\int_{w_{\min}}^\infty dw\,
\left(\mu-\frac{m^2}{w}\right)
f\!\left(\frac{w^2-m^2}{2w}\right),
\label{eq:WVA_exact}\\
W^{S,D}(\mu)&=
\frac{m^2}{16\pi^2E}
\int_{w_{\min}}^\infty dw\,
\frac{w-\mu}{w^2}
f\!\left(\frac{w^2-m^2}{2w}\right),
\label{eq:K_def}
\end{align}
where $f(p)=(e^{p/T}+1)^{-1}$. The~kernel $W^0$ has a
helicity-independent numerator, while $W^{V-A}$ and $W^{S,D}$ contain the
helicity-dependent numerators at $\cos\theta^*$; the factor $m^2$ in
$W^{S,D}$ reflects the helicity flip required for the populated initial
states. The~pole rates are
\begin{equation}
\begin{gathered}
\mathcal R^{V-A}_{\mathrm{res}}=2\pi g^2W^{V-A},
\qquad
\mathcal R^{S,D}_{\mathrm{res}}=\pi g^2W^{S,D},
\\
\mathcal R^{S,M}_{\mathrm{res}}=\pi g^2W^0,
\qquad
\overline{\mathcal R}{}^{S,D}_{\mathrm{res}}=\frac{\pi}{2}g^2W^0,
\qquad
\overline{\mathcal R}{}^{V}_{\mathrm{res}}=\pi g^2W^0.
\end{gathered}
\label{eq:massive_B}
\end{equation}
For $m\ll T$ and $m\gg T$, the~kernels simplify as~follows.

\subsection{UR~Targets}
\label{app:URkernels}
For $m\ll T$, we retain the mass-dependent boundary and the helicity-dependent numerators, while using $p\simeq w/2$ and $p/E_i\simeq1$ in the thermal weight. The~resulting correction to the thermal weight is of order $m/T$. With~\begin{equation}
F_{-k}(u)=
\int_u^\infty\frac{dv}{v^k(e^v+1)},\quad k=1,2,
\label{eq:UR_boundary}
\end{equation}
the UR kernels are
\begin{align}
W^{V-A}_{\mathrm{UR}}
&=
\frac{T}{8\pi^2E}
\left[
\mu\ln\!\left(1+e^{-w_{\min}/2T}\right)
-\frac{m^2}{2T}F_{-1}\!\left(\frac{w_{\min}}{2T}\right)
\right],
\label{eq:Bres_VA}\\
W^{S,D}_{\mathrm{UR}}
&=
\frac{m^2}{16\pi^2E}
\left[F_{-1}\!\left(\frac{w_{\min}}{2T}\right)-rF_{-2}\!\left(\frac{w_{\min}}{2T}\right)\right],
\label{eq:K_UR}\\
W^{0}_{\mathrm{UR}}
&=\frac{\mu T}{8\pi^2E}\ln\!\left(1+e^{-p_{\min}/T}\right),
\label{eq:W0_UR}
\end{align}
and the pole rates follow from Equation~\eqref{eq:massive_B}. These FD expressions are equivalent to applying the series of Equation~\eqref{eq:FD_MB} to the MB rates. In~the MB limit, $\ln(1+e^{-u})\to e^{-u}$, $F_{-1}(u)\to-\mathrm{Ei}(-u)$ and $F_{-2}(u)\to e^{-u}/u+\mathrm{Ei}(-u)$, which gives Equations~\eqref{eq:res_massive}--\eqref{eq:res_massive_SD}.

\subsection{NR~Targets}
In the opposite limit, $m\gg T$, the~kernels simplify to closed forms. The~relic energy is then $E_i\simeq m$, so that $w\simeq m+p$ and $m^2/w\simeq m-p$, and~the resonance lies at $\mu\simeq m$, i.e.,~$M^2\simeq2Em$. The~two sides of the resonance are distinguished by $\varsigma=\mathrm{sign}(\mu-m)=\mathrm{sign}(M^2-2Em)$, with~$\mu-m\simeq\varsigma\,p_{\min}$. Defining
\begin{equation}
\begin{gathered}
\frac{p_{\min}}{T}\simeq\frac{|M^2-2Em|}{2ET},
\\
\ell_0=\ln\!\left(1+e^{-p_{\min}/T}\right)
\;\xrightarrow{\ \mathrm{MB}\ }\;
e^{-p_{\min}/T},
\\
\ell_1=\frac{p_{\min}}{T}\ln\!\left(1+e^{-p_{\min}/T}\right)-\mathrm{Li}_2\!\left(-e^{-p_{\min}/T}\right)
\;\xrightarrow{\ \mathrm{MB}\ }\;
\left(1+\frac{p_{\min}}{T}\right)e^{-p_{\min}/T},
\end{gathered}
\end{equation}
the kernels become, to~leading order in $p/m$,
\begin{equation}
\begin{gathered}
W^0\simeq\frac{T^2}{8\pi^2E}\,\ell_1,
\\
W^{V-A}\simeq\frac{T^2}{16\pi^2E}\left[\ell_1+\varsigma\frac{p_{\min}}{T}\ell_0\right],
\\
W^{S,D}\simeq\frac{T^2}{16\pi^2E}\left[\ell_1-\varsigma\frac{p_{\min}}{T}\ell_0\right].
\end{gathered}
\label{eq:NR_kernels}
\end{equation}
The NR resonance is centred at $M^2=2Em$ with a relative thermal width
$\delta M/M\sim T/m$. The~mass factor in $W^{S,D}$ cancels against
$w^2\simeq m^2$, so the $(S,D)$ pole is not helicity suppressed for
NR targets. In~the numerical evaluation, two further orders in
$(p/m)^2$ are retained. Figure~\ref{fig:HPSA} (right) illustrates how the thermal width of the resonance evolves for relic masses spanning the UR, semi-relativistic and NR~regimes.

\subsection{Validity of the~NWA}
The NWA requires the resonance to be narrow compared with its thermal smearing. The~thermal motion of the relics spreads $s$ by $\pm2Ep$, i.e.,
$\delta M/M\sim T/m$, which exceeds $\Gamma_\phi/M$ by many orders of
magnitude for the couplings considered. The~NWA is therefore accurate up
to corrections of order $M\Gamma_\phi/\delta s$. For~$(V{-}A,D)$ and $(S,M)$ with NR targets, $m=2\times10^{-2}$~eV, the~sources of Table~\ref{tab:2}, and~$M/\sqrt{2Em}=0.999$, 1 and 1.001, the~sum of the NWA pole and the finite part agrees with the finite-width result to better than $10^{-6}$ for $\Gamma_\phi/M=10^{-5}$--$10^{-7}$.

\subsection{Off-Resonant NR~Rates}
Away from the pole, the~NR rates are simpler still. For~$m\gg T$, the~non-resonant rates factorise,
$\mathcal R_i\simeq n_i^{\rm eff}\sigma(s_0)$ with $s_0=2Em_i$, where
$n_i^{\rm eff}=n_i$ for the helicity-independent cases and
$n_i\langle(1+v)/2\rangle$ for $V{-}A$. The~off-resonant $s$-channel and
the $s$--$t$ interference are evaluated at $s_0$ in the zero-width
limit. Near~the pole, where this factorized form diverges, the~double pole is regularized by the thermal spread of $s$,
\begin{equation}
(s_0-M^2)^2\;\rightarrow\;(s_0-M^2)^2+(\Delta s)^2,
\qquad
\Delta s=6ET\simeq2E\langle p\rangle_{\rm FD}.
\end{equation}
This prescription reproduces the full thermal average at the pole-dominated minimum and far from the resonance. Differences of a few per cent in $g$ appear only in narrow transition regions next to the minimum, $0.8\lesssim M/\sqrt{2Em}\lesssim1.5$.

\subsection{Distant~Sources}
For distant sources, the~rates are evaluated numerically along the propagation path, with~$T(z)=T_0(1+z)$. The~ratio $m/T(z)$ determines which description of the relic kinematics applies: the NR expansion for $m/T\geq20$, the~exact one-dimensional kernels of Equations~\eqref{eq:WU_exact}--\eqref{eq:K_def} for $3\leq m/T<20$, and~the UR expressions for $m/T<3$. Using the exact kernels everywhere changes $g$ by a relative amount below $10^{-4}$ for the NR degenerate benchmark and below $0.6\%$ for the 3 NR spectrum of TXS~0506+056. For~the UR benchmark, the~UR kernels deviate from the exact ones at order $m/T$, which changes $g$ by less than $0.1\%$ for $(V{-}A,D)$ and by up to $2\%$ for $(S,M)$ near the low-mass edge of the resonance, $M\simeq\sqrt{2Em}$.

\section{Comparison~Cases}
\label{app:refs}
\unskip

\subsection{Spin-Averaged $(S,D)$ Case}

\begin{figure}[htbp]
\centering
\includegraphics[width=0.48\linewidth]{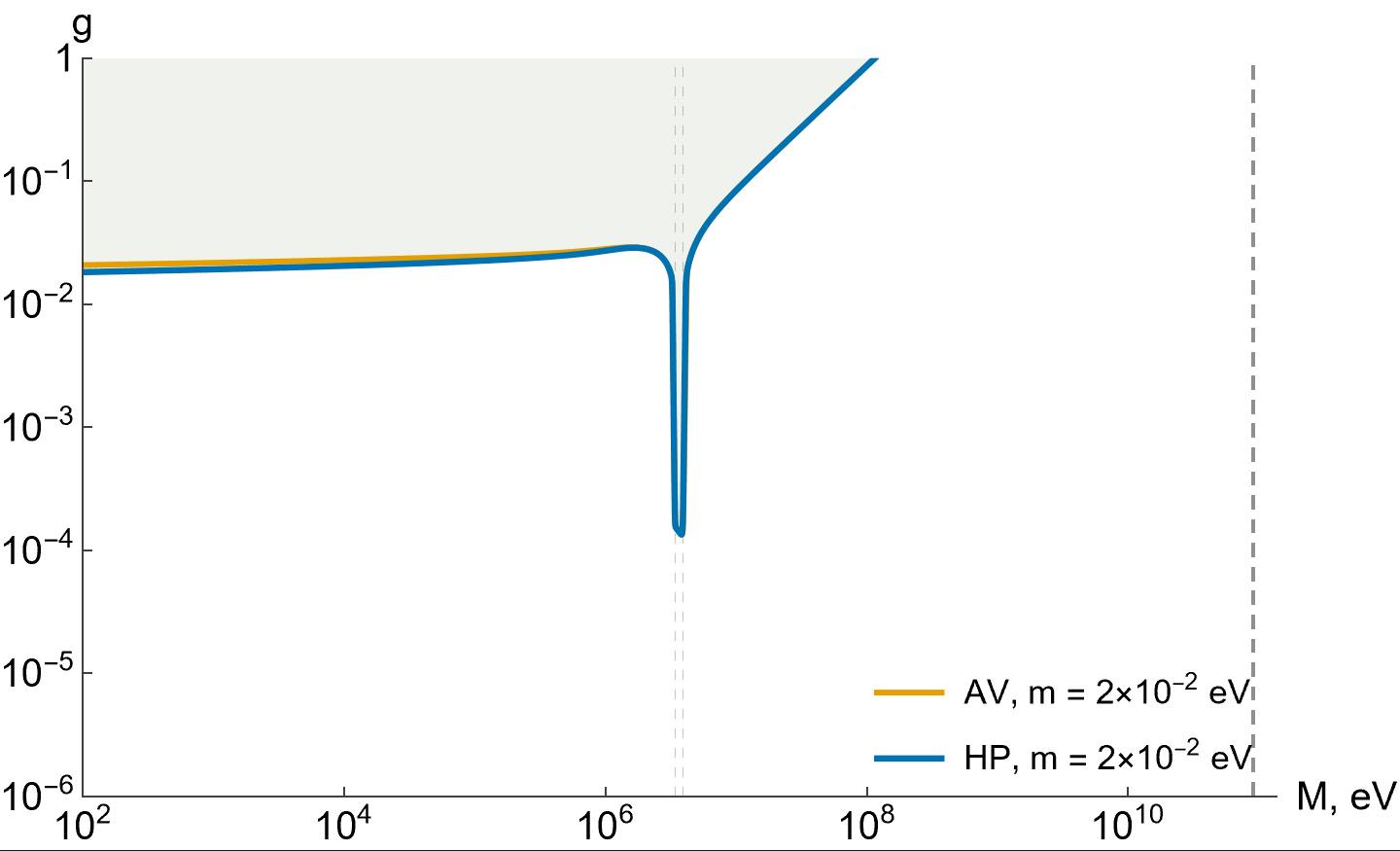}\hfill
\includegraphics[width=0.48\linewidth]{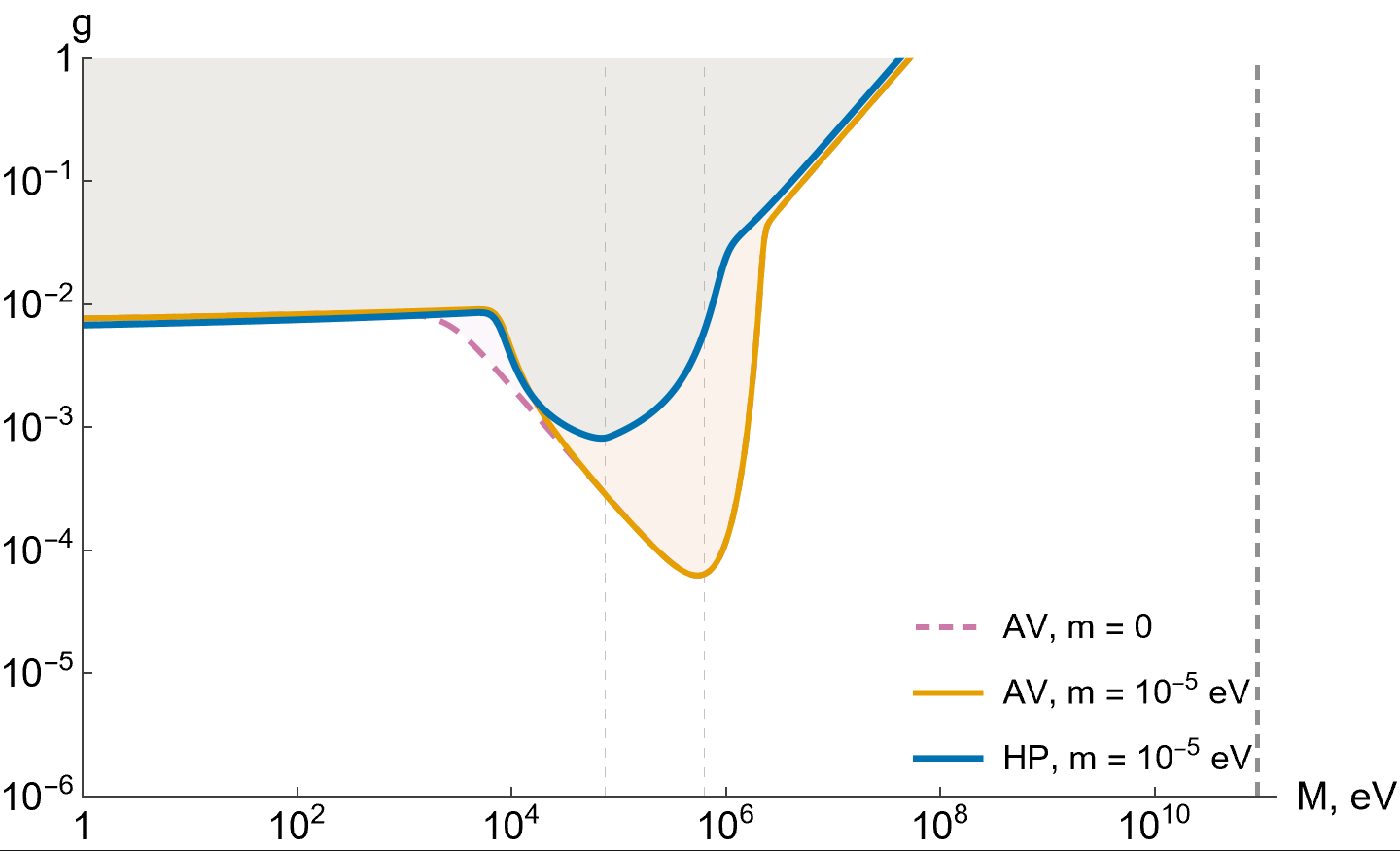}
\caption{Helicity-polarized %
 (HP) and spin-averaged (AV, both helicities equally populated) $(S,D)$ survival benchmarks for TXS~0506+056. Left: NR benchmark with $m=2\times10^{-2}~\mathrm{eV}$ for both curves. Right: UR benchmark, with~HP and AV curves at $m=10^{-5}~\mathrm{eV}$ and a massless AV curve. The~thick vertical dashed line marks the $Z$-boson~mass. The shaded regions indicate $\tau>1$.}
\label{fig:scalar_spin}
\end{figure}

For comparison with the helicity-polarized case, we also show the spin-averaged curves, $\overline{(S,D)}$. For~a parity-even scalar interaction, $\sigma_{++}=\sigma_{--}$ and $\sigma_{+-}=\sigma_{-+}$, so averaging over the target helicities reproduces $\overline\sigma$ for either incident helicity (Appendix~\ref{sec:SM_fh}). The~spin-averaged rate therefore corresponds to a background in which both helicities are equally populated, with~the same total density $n_i$ per species. A~different total density rescales the rate. Since both coefficients $A$ and $B$ in Equation~\eqref{eq:g_general} are proportional to $n_i$, the~survival benchmarks scale as $g\propto n_i^{-1/4}$ away from the resonance and as $g\propto n_i^{-1/2}$ where the pole dominates. Figure~\ref{fig:scalar_spin} compares the two prescriptions. The~spin-averaged pole also depends on the relic mass through the finite-mass kinematic boundary, although~it has no helicity-suppression factor. For~NR targets, $\langle\etaa\rangle=1$, so the UR helicity suppression does not persist in the angularly averaged $s$-channel rate, and~the two prescriptions nearly~coincide.
\unskip

\subsection{Spin-One Reference~Comparisons}
\label{sec:vecref}\label{app:vecref}

The pure-vector coupling with Dirac neutrinos, $(V,D)$, is
\begin{equation}
\mathcal{L}_{V}^{D}
=
{g_{\alpha\beta}\,\anu_{\alpha}\gamma_{\mu}\nu_{\beta}\,V^{\mu}.}
\end{equation}

\begin{figure}[htbp]
\centering
\includegraphics[width=0.48\linewidth]{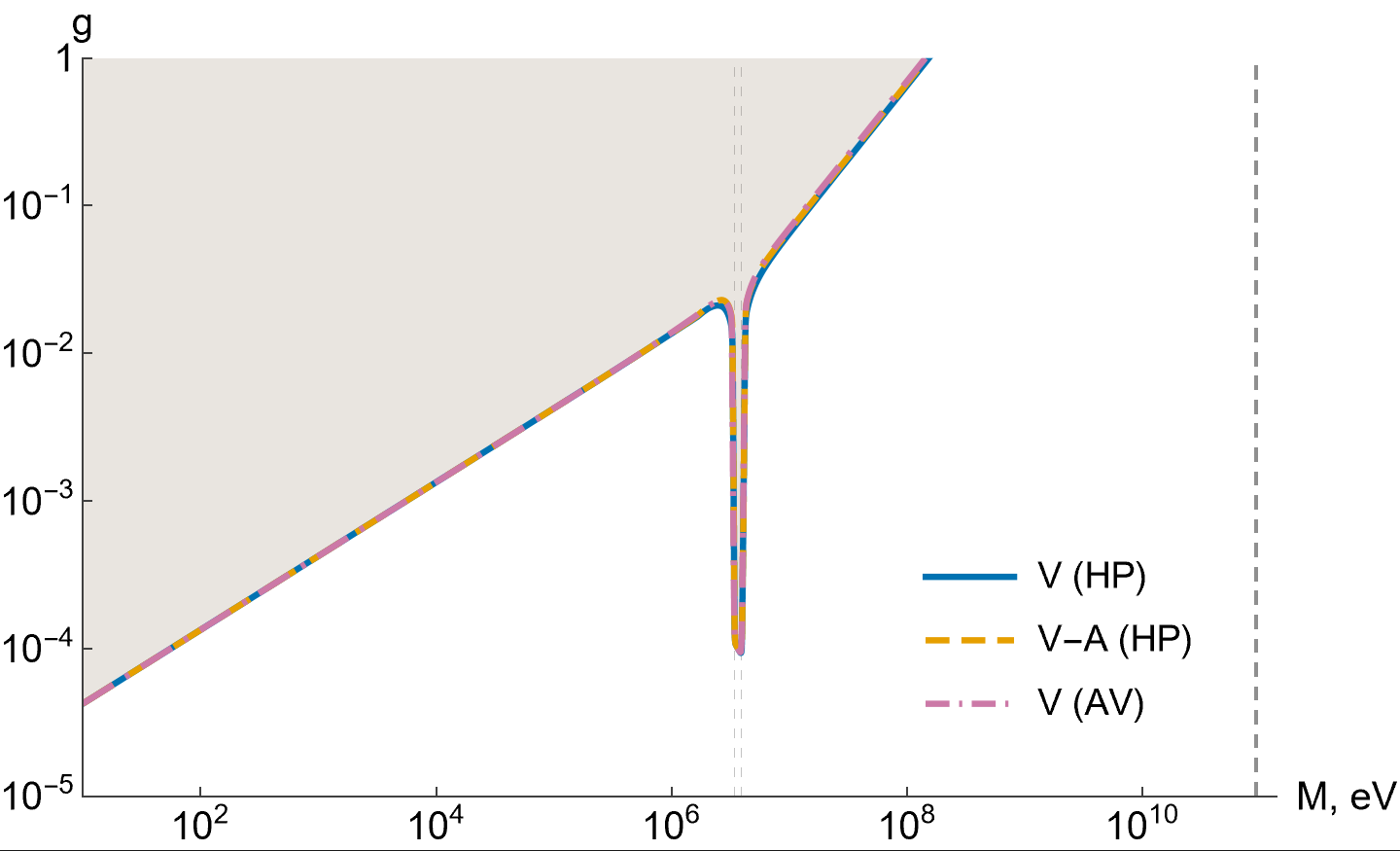}\hfill
\includegraphics[width=0.48\linewidth]{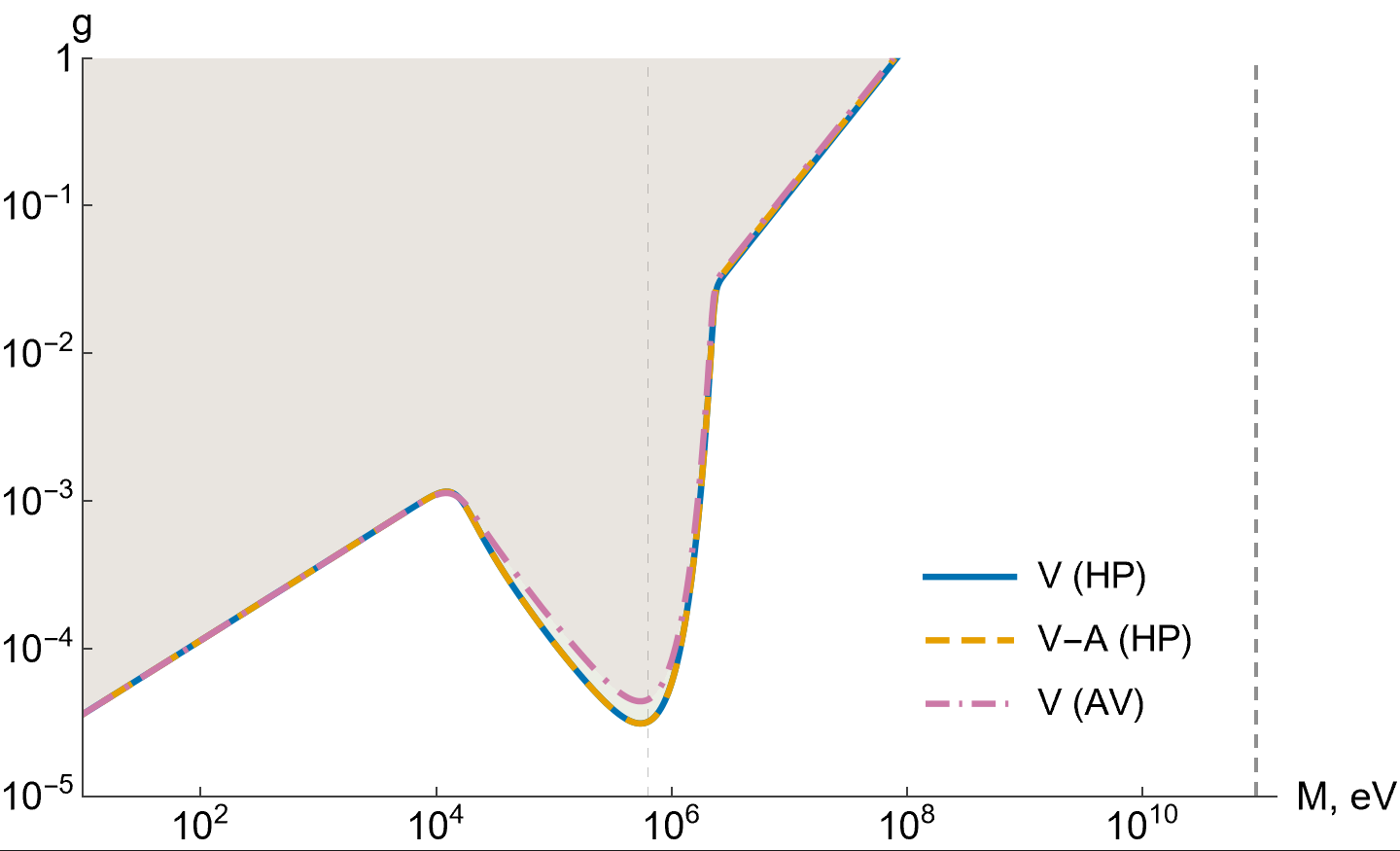}
\caption{Spin-one %
 reference comparisons for TXS~0506+056 and a degenerate C$\nu$B, in~the NR (Left, $m=2\times10^{-2}~\mathrm{eV}$) and UR (Right, $m=10^{-5}~\mathrm{eV}$) regimes. The~solid and dashed curves show pure-vector and $V{-}A$ interactions with populated initial helicities (HP). The~dash-dotted curves show the pure-vector spin-averaged case (AV), denoted by $\overline{V}$ in the text. The~thick vertical dashed line marks the $Z$-boson~mass. The shaded regions indicate $\tau>1$.}
\label{fig:vector_reference}
\end{figure}

In the negligible-neutrino-mass limit, the~populated states $\nu_-$ and $\bar\nu_+$ belong to left-chiral fields, and~chirality is conserved along each fermion line. On~these states the vertices $\gamma^\mu$, $-\gamma^\mu\gamma_5$ and $\gamma^\mu P_L$ coincide. The~pure-vector and pure axial-vector interactions therefore give identical helicity-polarized cross sections, which coincide with the $V{-}A$ ones up to contributions involving right-chiral final states. These enter only through the $s$-channel and the mediator width. Accordingly, the~pure-vector and $V{-}A$ curves in Figure~\ref{fig:vector_reference} nearly coincide, with~slight differences near and above the resonance, while the spin-averaged pure-vector case shows a slightly less pronounced resonance. Since the pure-vector interaction would populate right-handed states in a standard thermal history (Section~\ref{sec:constraints}), these curves serve only as a link to earlier~calculations.

\bibliography{main}

@article{Berryman_2023,
  author = {Berryman, J. M. and Blinov, N. and Brdar, V. and Brinckmann, T. and Bustamante, M. and Cyr-Racine, F. Y. and Das, A. and De Gouvêa, A. and Denton, P. B. and Dev, P. B. and others},
  title = {{Neutrino self-interactions: A white paper}},
  journal = {Phys. Dark Universe},
  volume = {42},
  pages = {101267},
  year = {2023},
  doi = {10.1016/j.dark.2023.101267},
}

@article{Carena_2003,
  author = {Carena, M. and de Gouvêa, A. and Freitas, A. and Schmitt, M.},
  title = {{Invisible {Z}-boson decays at $e^+e^-$ colliders}},
  journal = {Phys. Rev. D},
  volume = {68},
  pages = {113007},
  year = {2003},
  doi = {10.1103/physrevd.68.113007},
}

@article{Flowers:1976kb,
  author = {Flowers, E. G. and Sutherland, P. G.},
  title = {{Neutrino Neutrino Scattering and Supernovae}},
  journal = {Astrophys. J. Lett.},
  volume = {208},
  pages = {L19--L21},
  year = {1976},
  doi = {10.1086/182223},
}

@article{Berbig:2020wve,
  author = {Berbig, M. and Jana, S. and Trautner, A.},
  title = {{The Hubble tension and a renormalizable model of gauged neutrino self-interactions}},
  journal = {Phys. Rev. D},
  volume = {102},
  pages = {115008},
  year = {2020},
  doi = {10.1103/PhysRevD.102.115008},
}

@article{Blinov_2019,
  author = {Blinov, N. and Kelly, K. J. and Krnjaic, G. and McDermott, S. D.},
  title = {{Constraining the Self-Interacting Neutrino Interpretation of the Hubble Tension}},
  journal = {Phys. Rev. Lett.},
  volume = {123},
  pages = {191102},
  year = {2019},
  doi = {10.1103/physrevlett.123.191102},
}

@article{Venzor_2023,
  author = {Venzor, J. and Garcia-Arroyo, G. and De-Santiago, J. and Pérez-Lorenzana, A.},
  title = {{Resonant neutrino self-interactions and the $H_0$ tension}},
  journal = {Phys. Rev. D},
  volume = {108},
  pages = {043536},
  year = {2023},
  doi = {10.1103/physrevd.108.043536},
}

@article{Abreu_2023,
  author = {Abreu, H. and Anders, J. and Antel, C. and Ariga, A. and Ariga, T. and Atkinson, J. and Bernlochner, F. U. and Blesgen, T. and Boeckh, T. and Boyd, J. and others},
  title = {{First Direct Observation of Collider Neutrinos with FASER at the LHC}},
  journal = {Phys. Rev. Lett.},
  volume = {131},
  pages = {031801},
  year = {2023},
  doi = {10.1103/physrevlett.131.031801},
}

@article{Cyr_Racine_2014,
  author = {Cyr-Racine, F. Y. and Sigurdson, K.},
  title = {{Limits on neutrino-neutrino scattering in the early Universe}},
  journal = {Phys. Rev. D},
  volume = {90},
  pages = {123533},
  year = {2014},
  doi = {10.1103/physrevd.90.123533},
}

@article{Esteban_2021,
  author = {Esteban, I. and Pandey, S. and Brdar, V. and Beacom, J. F.},
  title = {{Probing secret interactions of astrophysical neutrinos in the high-statistics era}},
  journal = {Phys. Rev. D},
  volume = {104},
  pages = {123014},
  year = {2021},
  doi = {10.1103/physrevd.104.123014},
}

@article{Das_2022,
  author = {Das, A. and Perez-Gonzalez, Y. F. and Sen, M.},
  title = {{Neutrino secret self-interactions: A booster shot for the cosmic neutrino background}},
  journal = {Phys. Rev. D},
  volume = {106},
  pages = {095042},
  year = {2022},
  doi = {10.1103/physrevd.106.095042},
}

@article{Camarena_2023,
  author = {Camarena, D. and Cyr-Racine, F. Y. and Houghteling, J.},
  title = {{Confronting self-interacting neutrinos with the full shape of the galaxy power spectrum}},
  journal = {Phys. Rev. D},
  volume = {108},
  pages = {103535},
  year = {2023},
  doi = {10.1103/physrevd.108.103535},
}

@article{Kolb:1987qy,
  author = {Kolb, E. W. and Turner, M. S.},
  title = {{Supernova SN 1987a and the Secret Interactions of Neutrinos}},
  journal = {Phys. Rev. D},
  volume = {36},
  pages = {2895},
  year = {1987},
  doi = {10.1103/PhysRevD.36.2895},
}

@article{Bustamante:2020mep,
  author = {Bustamante, M. and Rosenstr\o{}m, C. and Shalgar, S. and Tamborra, I.},
  title = {{Bounds on secret neutrino interactions from high-energy astrophysical neutrinos}},
  journal = {Phys. Rev. D},
  volume = {101},
  pages = {123024},
  year = {2020},
  doi = {10.1103/PhysRevD.101.123024},
}

@article{Doring:2023vmk,
  author = {D\"oring, C. and Vogl, S.},
  title = {{Testing secret interaction with astrophysical neutrino point sources}},
  journal = {J. Cosmol. Astropart. Phys.},
  volume = {2024},
  number = {07},
  pages = {015},
  year = {2024},
  doi = {10.1088/1475-7516/2024/07/015},
}

@article{Chang_2023,
  author = {Chang, P. W. and Esteban, I. and Beacom, J. F. and Thompson, T. A. and Hirata, C. M.},
  title = {{Toward Powerful Probes of Neutrino Self-Interactions in Supernovae}},
  journal = {Phys. Rev. Lett.},
  volume = {131},
  pages = {071002},
  year = {2023},
  doi = {10.1103/physrevlett.131.071002},
}

@article{Wu_2024,
  author = {Wu, Q.-f. and Xu, X. J.},
  title = {{Shedding light on neutrino self-interactions with solar antineutrino searches}},
  journal = {J. Cosmol. Astropart. Phys.},
  volume = {2024},
  number = {02},
  pages = {037},
  year = {2024},
  doi = {10.1088/1475-7516/2024/02/037},
}

@article{Bakhti_2019,
  author = {Bakhti, P. and Farzan, Y. and Rajaee, M.},
  title = {{Secret interactions of neutrinos with light gauge boson at the DUNE near detector}},
  journal = {Phys. Rev. D},
  volume = {99},
  pages = {055019},
  year = {2019},
  doi = {10.1103/physrevd.99.055019},
}

@article{Berryman_2018,
  author = {Berryman, J. M. and de Gouvêa, A. and Kelly, K. J. and Zhang, Y.},
  title = {{Lepton-number-charged scalars and neutrino beamstrahlung}},
  journal = {Phys. Rev. D},
  volume = {97},
  pages = {075030},
  year = {2018},
  doi = {10.1103/physrevd.97.075030},
}

@article{Gelmini:1980re,
  author = {Gelmini, G. B. and Roncadelli, M.},
  title = {{Left-Handed Neutrino Mass Scale and Spontaneously Broken Lepton Number}},
  journal = {Phys. Lett. B},
  volume = {99},
  pages = {411--415},
  year = {1981},
  doi = {10.1016/0370-2693(81)90559-1},
}

@article{Forastieri_2019,
  author = {Forastieri, F. and Lattanzi, M. and Natoli, P.},
  title = {{Cosmological constraints on neutrino self-interactions with a light mediator}},
  journal = {Phys. Rev. D},
  volume = {100},
  pages = {103526},
  year = {2019},
  doi = {10.1103/physrevd.100.103526},
}

@article{Lessa_2007,
  author = {Lessa, A. P. and Peres, O. L. G.},
  title = {{Revising limits on neutrino-Majoron couplings}},
  journal = {Phys. Rev. D},
  volume = {75},
  pages = {094001},
  year = {2007},
  doi = {10.1103/PhysRevD.75.094001},
}

@misc{Barik:2024kwv,
  author = {Barik, A. K. and Rai, S. K. and Srivastava, A.},
  title = {{Discovering an invisible Z' at the muon collider}},
  year = {2024},
  eprint = {2408.14396},
  archiveprefix = {arXiv},
}

@article{Kelly:2018tyg,
  author = {Kelly, K. J. and Machado, P. A.},
  title = {{Multimessenger astronomy and new neutrino physics}},
  journal = {J. Cosmol. Astropart. Phys.},
  volume = {2018},
  number = {10},
  pages = {048},
  year = {2018},
  doi = {10.1088/1475-7516/2018/10/048},
}

@article{Escudero_2020,
  author = {Escudero, M. and Witte, S. J.},
  title = {{A CMB search for the neutrino mass mechanism and its relation to the Hubble tension}},
  journal = {Eur. Phys. J. C},
  volume = {80},
  pages = {294},
  year = {2020},
  doi = {10.1140/epjc/s10052-020-7854-5},
}

@article{Heeck:2018nzc,
  author = {Heeck, J. and Lindner, M. and Rodejohann, W. and Vogl, S.},
  title = {{Non-Standard Neutrino Interactions and Neutral Gauge Bosons}},
  journal = {SciPost Phys.},
  volume = {6},
  pages = {038},
  year = {2019},
  doi = {10.21468/SciPostPhys.6.3.038},
}

@article{Farzan2016,
  author = {Farzan, Y. and Heeck, J.},
  title = {{Neutrinophilic nonstandard interactions}},
  journal = {Phys. Rev. D},
  volume = {94},
  pages = {053010},
  year = {2016},
  doi = {10.1103/PhysRevD.94.053010},
}

@article{OKADA2020135845,
  author = {Okada, N. and Okada, S. and Shafi, Q.},
  title = {{Light Z' and dark matter from U(1)X gauge symmetry}},
  journal = {Phys. Lett. B},
  volume = {810},
  pages = {135845},
  year = {2020},
  doi = {10.1016/j.physletb.2020.135845},
}

@article{Kelly2020,
  author = {Kelly, K. J. and Sen, M. and Tangarife, W. and Zhang, Y.},
  title = {{Origin of sterile neutrino dark matter via secret neutrino interactions with vector bosons}},
  journal = {Phys. Rev. D},
  volume = {101},
  pages = {115031},
  year = {2020},
  doi = {10.1103/PhysRevD.101.115031},
}

@article{Kreisch_2020,
  author = {Kreisch, C. D. and Cyr-Racine, F. Y. and Dor\'e, O.},
  title = {{Neutrino puzzle: Anomalies, interactions, and cosmological tensions}},
  journal = {Phys. Rev. D},
  volume = {101},
  pages = {123505},
  year = {2020},
  doi = {10.1103/PhysRevD.101.123505},
}

@article{Deppisch:2020sqh,
  author = {Deppisch, F. F. and Graf, L. and Rodejohann, W. and Xu, X. J.},
  title = {{Neutrino Self-Interactions and Double Beta Decay}},
  journal = {Phys. Rev. D},
  volume = {102},
  pages = {051701},
  year = {2020},
  doi = {10.1103/PhysRevD.102.051701},
}

@article{Deppisch_2020,
  author = {Deppisch, F. F. and Graf, L. and Šimkovic, F.},
  title = {{Searching for New Physics in Two-Neutrino Double Beta Decay}},
  journal = {Phys. Rev. Lett.},
  volume = {125},
  pages = {171801},
  year = {2020},
  doi = {10.1103/physrevlett.125.171801},
}

@article{Blum_2018,
  author = {Blum, K. and Nir, Y. and Shavit, M.},
  title = {{Neutrinoless double-beta decay with massive scalar emission}},
  journal = {Phys. Lett. B},
  volume = {785},
  pages = {354--361},
  year = {2018},
  doi = {10.1016/j.physletb.2018.08.022},
}

@article{de_Gouv_a_2020,
  author = {de Gouvêa, A. and Dev, P. B. and Dutta, B. and Ghosh, T. and Han, T. and Zhang, Y.},
  title = {{Leptonic scalars at the LHC}},
  journal = {J. High Energy Phys.},
  volume = {2020},
  number = {07},
  pages = {142},
  year = {2020},
  doi = {10.1007/jhep07(2020)142},
}

@article{BelleII,
  author = {Adachi, I. and Adamczyk, K. and Aggarwal, L. and Ahmed, H. and Aihara, H. and Akopov, N. and Aloisio, A. and Anh Ky, N. and Asner, D. M. and Aushev, T. and others},
  title = {{Search for Lepton-Flavor-Violating $\ensuremath{\tau}$ Decays to a Lepton and an Invisible Boson at Belle II}},
  journal = {Phys. Rev. Lett.},
  volume = {130},
  pages = {181803},
  year = {2023},
  doi = {10.1103/PhysRevLett.130.181803},
}

@article{Cortina_Gil_2021,
  author = {Gil, E. C. and Kleimenova, A. and Minucci, E. and Padolski, S. and Petrov, P. and Shaikhiev, A. and Volpe, R. and Numao, T. and Petrov, Y. and Velghe, B. and others},
  title = {{Search for K+ decays to a muon and invisible particles}},
  journal = {Phys. Lett. B},
  volume = {816},
  pages = {136259},
  year = {2021},
  doi = {10.1016/j.physletb.2021.136259},
}

@article{Brdar_2020,
  author = {Brdar, V. and Lindner, M. and Vogl, S. and Xu, X. J.},
  title = {{Revisiting neutrino self-interaction constraints from $Z$ and $\ensuremath{\tau}$ decays}},
  journal = {Phys. Rev. D},
  volume = {101},
  pages = {115001},
  year = {2020},
  doi = {10.1103/PhysRevD.101.115001},
}

@misc{Dev:2024twk,
  author = {Dev, P. S. B. and Kim, D. and Sathyan, D. and Sinha, K. and Zhang, Y.},
  title = {{New Laboratory Constraints on Neutrinophilic Mediators}},
  year = {2024},
  eprint = {2407.12738},
  archiveprefix = {arXiv},
}

@article{Pasquini:2015fjv,
  author = {Pasquini, P. S. and Peres, O. L. G.},
  title = {{Bounds on Neutrino-Scalar Yukawa Coupling}},
  journal = {Phys. Rev. D},
  volume = {93},
  pages = {053007},
  year = {2016},
  doi = {10.1103/PhysRevD.93.053007},
}

@article{Laha:2013xua,
  author = {Laha, R. and Dasgupta, B. and Beacom, J. F.},
  title = {{Constraints on New Neutrino Interactions via Light Abelian Vector Bosons}},
  journal = {Phys. Rev. D},
  volume = {89},
  pages = {093025},
  year = {2014},
  doi = {10.1103/PhysRevD.89.093025},
}

@article{Grohs_2020,
  author = {Grohs, E. and Fuller, G. M. and Sen, M.},
  title = {{Consequences of neutrino self-interactions for weak decoupling and big bang nucleosynthesis}},
  journal = {J. Cosmol. Astropart. Phys.},
  volume = {2020},
  number = {07},
  pages = {001},
  year = {2020},
  doi = {10.1088/1475-7516/2020/07/001},
}

@article{Huang_2018,
  author = {Huang, G.-y. and Ohlsson, T. and Zhou, S.},
  title = {{Observational constraints on secret neutrino interactions from big bang nucleosynthesis}},
  journal = {Phys. Rev. D},
  volume = {97},
  pages = {075009},
  year = {2018},
  doi = {10.1103/PhysRevD.97.075009},
}

@article{Escudero:2018mvt,
  author = {Escudero, M.},
  title = {{{Neutrino decoupling beyond the Standard Model: CMB constraints on the Dark Matter mass with a fast and precise $N_{\rm eff}$ evaluation}}},
  journal = {J. Cosmol. Astropart. Phys.},
  volume = {2019},
  number = {02},
  pages = {007},
  year = {2019},
  doi = {10.1088/1475-7516/2019/02/007},
}

@article{Sabti:2019mhn,
  author = {Sabti, N. and Alvey, J. and Escudero, M. and Fairbairn, M. and Blas, D.},
  title = {{Refined Bounds on MeV-scale Thermal Dark Sectors from BBN and the CMB}},
  journal = {J. Cosmol. Astropart. Phys.},
  volume = {2020},
  number = {01},
  pages = {004},
  year = {2020},
  doi = {10.1088/1475-7516/2020/01/004},
}

@article{Luo:2020sho,
  author = {Luo, X. and Rodejohann, W. and Xu, X. J.},
  title = {{{Dirac neutrinos and $N_{{\rm eff}}$}}},
  journal = {J. Cosmol. Astropart. Phys.},
  volume = {2020},
  number = {06},
  pages = {058},
  year = {2020},
  doi = {10.1088/1475-7516/2020/06/058},
}

@article{Luo:2020fdt,
  author = {Luo, X. and Rodejohann, W. and Xu, X. J.},
  title = {{{Dirac neutrinos and N$_{eff}$. Part II. The freeze-in case}}},
  journal = {J. Cosmol. Astropart. Phys.},
  volume = {2021},
  number = {03},
  pages = {082},
  year = {2021},
  doi = {10.1088/1475-7516/2021/03/082},
}

@article{Abazajian:2019oqj,
  author = {Abazajian, K. N. and Heeck, J.},
  title = {{Observing Dirac neutrinos in the cosmic microwave background}},
  journal = {Phys. Rev. D},
  volume = {100},
  pages = {075027},
  year = {2019},
  doi = {10.1103/PhysRevD.100.075027},
}

@article{Kamada_2015,
  author = {Kamada, A. and Yu, H. B.},
  title = {{Coherent propagation of PeV neutrinos and the dip in the neutrino spectrum at IceCube}},
  journal = {Phys. Rev. D},
  volume = {92},
  pages = {113004},
  year = {2015},
  doi = {10.1103/physrevd.92.113004},
}

@article{Ioka_2014,
  author = {Ioka, K. and Murase, K.},
  title = {{IceCube PeV-EeV neutrinos and secret interactions of neutrinos}},
  journal = {Prog. Theor. Exp. Phys.},
  volume = {2014},
  pages = {061E01},
  year = {2014},
  doi = {10.1093/ptep/ptu090},
}

@article{Ng_2014,
  author = {Ng, K. C. and Beacom, J. F.},
  title = {{Cosmic neutrino cascades from secret neutrino interactions}},
  journal = {Phys. Rev. D},
  volume = {90},
  pages = {065035},
  year = {2014},
  doi = {10.1103/physrevd.90.065035},
}

@article{Ibe_2014,
  author = {Ibe, M. and Kaneta, K.},
  title = {{Cosmic neutrino background absorption line in the neutrino spectrum at IceCube}},
  journal = {Phys. Rev. D},
  volume = {90},
  pages = {053011},
  year = {2014},
  doi = {10.1103/physrevd.90.053011},
}

@article{Ambrosone:2024zrf,
  author = {Ambrosone, A.},
  title = {{Berezinsky hidden sources: An emergent tension in the high-energy neutrino sky?}},
  journal = {J. Cosmol. Astropart. Phys.},
  volume = {2024},
  number = {09},
  pages = {075},
  year = {2024},
  doi = {10.1088/1475-7516/2024/09/075},
}

@article{He:2025bex,
  author = {He, Y. and Liu, J. and Wang, X. P. and Zhong, Y. M.},
  title = {{Implications of the KM3NeT Ultrahigh-energy Event on Neutrino Self-interactions}},
  journal = {Phys. Rev. D},
  volume = {113},
  pages = {043022},
  year = {2026},
  doi = {10.1103/bb85-7kyf},
}

@article{Troitsky:2023nli,
  author = {Troitsky, S.},
  title = {{Origin of high-energy astrophysical neutrinos: New results and prospects}},
  journal = {Usp. Fiz. Nauk},
  volume = {194},
  pages = {371--383},
  year = {2024},
  doi = {10.3367/UFNr.2023.04.039581},
}

@article{PhysRevD.107.042005,
  author = {Allakhverdyan, V. A. and Avrorin, A. D. and Avrorin, A. V. and Aynutdinov, V. M. and Bardačová, Z. and Belolaptikov, I. A. and Borina, I. V. and Budnev, N. M. and Dik, V. Y. and Domogatsky, G. V. and others},
  title = {{Diffuse neutrino flux measurements with the Baikal-GVD neutrino telescope}},
  journal = {Phys. Rev. D},
  volume = {107},
  pages = {042005},
  year = {2023},
  doi = {10.1103/PhysRevD.107.042005},
}

@misc{Wang:2025qap,
  author = {Wang, I. R. and Xu, X. J. and Zhou, B.},
  title = {{Widen the Resonance: Probing a New Regime of Neutrino Self-Interactions with Astrophysical Neutrinos}},
  year = {2025},
  eprint = {2501.07624},
  archiveprefix = {arXiv},
}

@article{Bally2020,
  author = {Bally, A. and Jana, S. and Trautner, A.},
  title = {{Neutrino Self-Interactions and XENON1T Electron Recoil Excess}},
  journal = {Phys. Rev. Lett.},
  volume = {125},
  pages = {161802},
  year = {2020},
  doi = {10.1103/PhysRevLett.125.161802},
}

@article{DiFranzo:2015qea,
  author = {DiFranzo, A. and Hooper, D.},
  title = {{Searching for MeV-Scale Gauge Bosons with IceCube}},
  journal = {Phys. Rev. D},
  volume = {92},
  pages = {095007},
  year = {2015},
  doi = {10.1103/PhysRevD.92.095007},
}

@misc{rozhkov2024,
  author = {Rozhkov, V. and Troitsky, S.},
  title = {{Multimessenger astronomy}},
  year = {2024},
  eprint = {2409.11818},
  archiveprefix = {arXiv},
}

@article{KM3NeT:2025npi,
  author = {{KM3NeT Collaboration}},
  title = {{Observation of an ultra-high-energy cosmic neutrino with KM3NeT}},
  journal = {Nature},
  volume = {638},
  pages = {376--382},
  year = {2025},
  doi = {10.1038/s41586-024-08543-1},
}

@article{Aglietta:1987mc,
  author = {Aglietta, M. and Badino, G. and Bologna, G. and Castagnoli, C. and Castellina, A. and Dadykin, V. L. and Fulgione, W. and Galeotti, P. and Kalchukov, F. F. and Kortchaguin, B. and others},
  title = {{On the Event Observed in the Mont Blanc Underground Neutrino Observatory during the Occurrence of Supernova 1987A}},
  journal = {Europhys. Lett.},
  volume = {3},
  pages = {1315--1320},
  year = {1987},
}

@article{Alekseev:1987qf,
  author = {Alekseev, E. N. and Alekseeva, L. N. and Volchenko, V. I. and Krivosheina, I. V.},
  title = {{Possible Detection of a Neutrino Signal from SN 1987A in the Baksan Experiment}},
  journal = {JETP Lett.},
  volume = {45},
  pages = {589--592},
  year = {1987},
}

@article{Bionta:1987qt,
  author = {Bionta, R. M. and Blewitt, G. and Bratton, C. B. and Casper, D. and Ciocio, A. and Claus, R. and Cortez, B. and Crouch, M. and Dye, S. T. and Errede, S. and others},
  title = {{Observation of a Neutrino Burst in Coincidence with Supernova 1987A in the Large Magellanic Cloud}},
  journal = {Phys. Rev. Lett.},
  volume = {58},
  pages = {1494},
  year = {1987},
  doi = {10.1103/PhysRevLett.58.1494},
}

@article{Hirata:1987hu,
  author = {Hirata, K. and Kajita, T. and Koshiba, M. and Nakahata, M. and Oyama, Y. and Sato, N. and Suzuki, A. and Takita, M. and Totsuka, Y. and Kifune, T. and others},
  title = {{Observation of a Neutrino Burst from the Supernova SN 1987A}},
  journal = {Phys. Rev. Lett.},
  volume = {58},
  pages = {1490},
  year = {1987},
  doi = {10.1103/PhysRevLett.58.1490},
}

@inproceedings{Panagia:2003rt,
  author = {Panagia, N.},
  title = {{A Geometric Determination of the Distance to SN 1987A and the LMC}},
  booktitle = {Supernovae and Gamma-Ray Bursters},
  editor = {Weiler, K.},
  series = {Lecture Notes in Physics},
  volume = {598},
  pages = {113--144},
  year = {2003},
  publisher = {Springer},
}

@article{Scholberg_2012,
  author = {Scholberg, K.},
  title = {{Supernova Neutrino Detection}},
  journal = {Annu. Rev. Nucl. Part. Sci.},
  volume = {62},
  pages = {81--103},
  year = {2012},
  doi = {10.1146/annurev-nucl-102711-095006},
}

@article{Markham:2026ngc,
  author = {Markham, M. and Bentz, M. C. and Ferrarese, L. and Onken, C. A. and Vestergaard, M.},
  title = {{Cepheid-based Distances to NGC 4303 and NGC 1068}},
  journal = {Astrophys. J.},
  volume = {1000},
  pages = {78},
  year = {2026},
  doi = {10.3847/1538-4357/ae47d8},
}

@article{2022IceCube,
  author = {{IceCube Collaboration} and Abbasi, R. and Ackermann, M. and Adams, J. and Aguilar, J. A. and Ahlers, M. and Ahrens, M. and Alameddine, J. M. and Alispach, C. and Alves, Jr., A. A. and others},
  title = {{Evidence for neutrino emission from the nearby active galaxy NGC 1068}},
  journal = {Science},
  volume = {378},
  pages = {538--543},
  year = {2022},
  doi = {10.1126/science.abg3395},
}

@article{Urry:1995mg,
  author = {Urry, C. M. and Padovani, P.},
  title = {{Unified schemes for radio-loud active galactic nuclei}},
  journal = {Publ. Astron. Soc. Pac.},
  volume = {107},
  pages = {803},
  year = {1995},
  doi = {10.1086/133630},
}

@article{IceCube:2018dnn,
  author = {{IceCube, Fermi-LAT, MAGIC, AGILE, ASAS-SN, HAWC, H.E.S.S., INTEGRAL, Kanata, Kiso, Kapteyn, Liverpool Telescope, Subaru, Swift/NuSTAR, VERITAS, and VLA/17B-403 Collaborations}},
  title = {{Multimessenger observations of a flaring blazar coincident with high-energy neutrino IceCube-170922A}},
  journal = {Science},
  volume = {361},
  pages = {eaat1378},
  year = {2018},
  doi = {10.1126/science.aat1378},
}

@article{Paiano:2018qeq,
  author = {Paiano, S. and Falomo, R. and Treves, A. and Scarpa, R.},
  title = {{The redshift of the BL Lac object TXS 0506+056}},
  journal = {Astrophys. J. Lett.},
  volume = {854},
  pages = {L32},
  year = {2018},
  doi = {10.3847/2041-8213/aaad5e},
}

@article{2018,
  author = {{IceCube Collaboration} and Aartsen, M. and Ackermann, M. and Adams, J. and Aguilar, J. A. and Ahlers, M. and Ahrens, M. and Al Samarai, I. and Altmann, D. and Andeen, K. and others},
  title = {{Neutrino emission from the direction of the blazar TXS 0506+056 prior to the IceCube-170922A alert}},
  journal = {Science},
  volume = {361},
  pages = {147--151},
  year = {2018},
  doi = {10.1126/science.aat2890},
}

@article{VERITAS:2023eso,
  author = {Acharyya, A. and Adams, C. B. and Archer, A. and Bangale, P. and Bartkoske, J. T. and Batista, P. and Benbow, W. and Brill, A. and Buckley, J. H. and Christiansen, J. L. and others},
  title = {{Multiwavelength Observations of the Blazar PKS 0735+178 in Spatial and Temporal Coincidence with an Astrophysical Neutrino Candidate IceCube-211208A}},
  journal = {Astrophys. J.},
  volume = {954},
  pages = {70},
  year = {2023},
  doi = {10.3847/1538-4357/ace327},
}

@inproceedings{Dik_2023,
  author = {Dik, V. Y.},
  title = {{Baikal-GVD real-time data processing and follow-up analysis of GCN notices}},
  booktitle = {Proceedings of the 38th International Cosmic Ray Conference (ICRC2023)},
  series = {PoS},
  volume = {ICRC2023},
  pages = {1458},
  year = {2023},
  publisher = {Sissa Medialab},
  doi = {10.22323/1.444.1458},
}

@article{Petkov:2022fnz,
  author = {Petkov, V. B. and Gorbacheva, E. A. and Guliev, R. M. and Dzaparova, I. M. and Kochkarov, M. M. and Kostyuk, M. G. and Kurenya, A. N. and Martakov, E. S. and Novoseltsev, Y. F. and Novoseltseva, R. V. and others},
  title = {{The Search for Neutrino Events from the Blazar PKS 0735+17 at the Baksan Underground Scintillation Telescope}},
  journal = {PoS},
  volume = {MUTO2022},
  pages = {033},
  year = {2022},
  doi = {10.22323/1.425.0033},
}

@misc{Omeliukh:2024kgk,
  author = {Omeliukh, A. and Garrappa, S. and Ramazani, V. F. and Franckowiak, A. and Winter, W. and Lindfors, E. and Nilsson, K. and Jormanainen, J. and Wierda, F. and Filippenko, A. V. and others},
  title = {{Multi-epoch leptohadronic modeling of neutrino source candidate blazar PKS 0735+178}},
  year = {2024},
  eprint = {2409.04165},
  archiveprefix = {arXiv},
}

@article{Sahakyan_2022,
  author = {Sahakyan, N. and Giommi, P. and Padovani, P. and Petropoulou, M. and Bégué, D. and Boccardi, B. and Gasparyan, S.},
  title = {{A multimessenger study of the blazar PKS 0735+178: A new major neutrino source candidate}},
  journal = {Mon. Not. R. Astron. Soc.},
  volume = {519},
  pages = {1396--1408},
  year = {2022},
  doi = {10.1093/mnras/stac3607},
}

@article{Falomo2021,
  author = {Falomo, R. and Treves, A. and Paiano, S.},
  journal = {Astron. Telegr.},
  volume = {15132},
  pages = {1},
  year = {2021},
}

@article{Nilsson2012,
  author = {Nilsson, K. and Pursimo, T. and Villforth, C. and Lindfors, E. and Takalo, L. O. and Sillanpää, A.},
  title = {{Redshift constraints for RGB 0136+391 and PKS 0735+178 from deep optical imaging}},
  journal = {Astron. Astrophys.},
  volume = {547},
  pages = {A1},
  year = {2012},
  doi = {10.1051/0004-6361/201219848},
}

@article{KM3NeT:2026blazars,
  author = {Adriani, O. and Albert, A. and Alhebsi, A. R. and Alshalloudi, S. and Alves Garre, S. and Ambrosone, A. and Ameli, F. and Andre, M. and Aphecetche, L. and Ardid, M. and others},
  title = {{Blazars as a Potential Origin of the KM3-230213A Event}},
  journal = {J. Cosmol. Astropart. Phys.},
  volume = {2026},
  number = {03},
  pages = {033},
  year = {2026},
  doi = {10.1088/1475-7516/2026/03/033},
}

@article{KM3NeT:2026galactic,
  author = {Adriani, O. and Aiello, S. and Albert, A. and Alhebsi, A. R. and Alshamsi, M. and Alves Garre, S. and Ambrosone, A. and Ameli, F. and Andre, M. and Aphecetche, L. and others},
  title = {{On the Potential Galactic Origin of the Ultra-high-energy Event KM3-230213A}},
  journal = {Astrophys. J.},
  volume = {1003},
  pages = {157},
  year = {2026},
  doi = {10.3847/1538-4357/ade3b0},
}

@article{KM3NeT:2025bxl,
  author = {Adriani, O. and Aiello, S. and Albert, A. and Alhebsi, A. R. and Alshamsi, M. and Alves Garre, S. and Ambrosone, A. and Ameli, F. and Andre, M. and Aphecetche, L. and others},
  title = {{Characterizing Candidate Blazar Counterparts of the Ultra-high-energy Event KM3-230213A}},
  journal = {Astron. J.},
  volume = {172},
  pages = {139},
  year = {2026},
  doi = {10.3847/1538-3881/ae7a4e},
}

@article{KM3NeT:2025vut,
  author = {Adriani, O. and Aiello, S. and Albert, A. and Alhebsi, A. R. and Alshamsi, M. and Alves Garre, S. and Ambrosone, A. and Ameli, F. and Andre, M. and Aphecetche, L. and others},
  title = {{On the Potential Cosmogenic Origin of the Ultra-high-energy Event KM3-230213A}},
  journal = {Astrophys. J. Lett.},
  volume = {984},
  pages = {L41},
  year = {2025},
  doi = {10.3847/2041-8213/adcc29},
}

@article{Long_2014,
  author = {Long, A. J. and Lunardini, C. and Sabancilar, E.},
  title = {{Detecting non-relativistic cosmic neutrinos by capture on tritium: phenomenology and physics potential}},
  journal = {J. Cosmol. Astropart. Phys.},
  volume = {2014},
  number = {08},
  pages = {038},
  year = {2014},
  doi = {10.1088/1475-7516/2014/08/038},
}

@article{Lesgourgues:2006nd,
  author = {Lesgourgues, J. and Pastor, S.},
  title = {{Massive neutrinos and cosmology}},
  journal = {Phys. Rept.},
  volume = {429},
  pages = {307--379},
  year = {2006},
  doi = {10.1016/j.physrep.2006.04.001},
}

@article{Abbasi_2025,
  author = {Abbasi, R. and Ackermann, M. and Adams, J. and Agarwalla, S. K. and Aguilar, J. A. and Ahlers, M. and Alameddine, J. M. and Amin, N. M. and Andeen, K. and Argüelles, C. and others},
  title = {{Search for Extremely-High-Energy Neutrinos and First Constraints on the Ultrahigh-Energy Cosmic-Ray Proton Fraction with IceCube}},
  journal = {Phys. Rev. Lett.},
  volume = {135},
  pages = {031001},
  year = {2025},
  doi = {10.1103/physrevlett.135.031001},
}

@misc{Kuznetsov:2025ehl,
  author = {Kuznetsov, M. Y. and Petrov, N. A. and Savchenko, Y. S.},
  title = {{Ultra-high energy event KM3-230213A as a cosmogenic neutrino in light of minimal UHECR flux models}},
  year = {2025},
  eprint = {2509.09590},
  archiveprefix = {arXiv},
}

@article{Planck:2018vyg,
  author = {Aghanim, N. and Akrami, Y. and Akrami, M. and Aumont, J. and Baccigalupi, C. and Ballardini, M. and Banday, A. J. and Barreiro, R. B. and Bartolo, N. and Basak, S. and others},
  title = {{Planck 2018 results. VI. Cosmological parameters}},
  journal = {Astron. Astrophys.},
  volume = {641},
  pages = {A6},
  year = {2020},
  doi = {10.1051/0004-6361/201833910},
}

@misc{Blum:2014ewa,
  author = {Blum, K. and Hook, A. and Murase, K.},
  title = {{High energy neutrino telescopes as a probe of the neutrino mass mechanism}},
  year = {2014},
  eprint = {1408.3799},
  archiveprefix = {arXiv},
}

@article{Herrera:2023nww,
  author = {Herrera, G. and Murase, K.},
  title = {{Probing light dark matter through cosmic-ray cooling in active galactic nuclei}},
  journal = {Phys. Rev. D},
  volume = {110},
  pages = {L011701},
  year = {2024},
  doi = {10.1103/PhysRevD.110.L011701},
}

@misc{Arguelles:2026rjr,
  author = {Arg{\"u}elles, C. A. and Dev, P. S. B. and Dutta, B. and Herrera, G. and Kamp, N. and Kumar, J. and Meighen-Berger, S. A. and Rai, M. and Shoemaker, I. M.},
  title = {{Astrophysical Neutrino Sources as Colliders}},
  year = {2026},
  eprint = {2607.19254},
  archiveprefix = {arXiv},
}

@article{Farzan_2008,
  author = {Farzan, Y. and Smirnov, A. Y.},
  title = {{Coherence and oscillations of cosmic neutrinos}},
  journal = {Nucl. Phys. B},
  volume = {805},
  pages = {356--376},
  year = {2008},
  doi = {10.1016/j.nuclphysb.2008.07.028},
}

@article{Elbers:2025DESI,
  author = {Elbers, W. and Avilés, A. and Noriega, H. E. and Chebat, D. and Menegas, A. and Frenk, C. S. and Garcia-Quintero, C. and Gonzalez, D. and Ishak, M. and Lahav, O. and others},
  title = {{Constraints on Neutrino Physics from DESI DR2 BAO and DR1 Full Shape}},
  journal = {Phys. Rev. D},
  volume = {112},
  pages = {083513},
  year = {2025},
  doi = {10.1103/w9pk-xsk7},
}

@article{Murase:2019xqi,
  author = {Murase, K. and Shoemaker, I. M.},
  title = {{Neutrino Echoes from Multimessenger Transient Sources}},
  journal = {Phys. Rev. Lett.},
  volume = {123},
  pages = {241102},
  year = {2019},
  doi = {10.1103/PhysRevLett.123.241102},
}

@article{Cline:2022qld,
  author = {Cline, J. M. and Gao, S. and Guo, F. and Lin, Z. and Liu, S. and Puel, M. and Todd, P. and Xiao, T.},
  title = {{Blazar Constraints on Neutrino-Dark Matter Scattering}},
  journal = {Phys. Rev. Lett.},
  volume = {130},
  pages = {091402},
  year = {2023},
  doi = {10.1103/PhysRevLett.130.091402},
}

@article{IceCube:2013dkx,
  author = {Aartsen, M. G. and Abbasi, R. and Ackermann, M. and Adams, J. and Aguilar, J. A. and Ahlers, M. and Altmann, D. and Arguelles, C. and Auffenberg, J. and Bai, X. and others},
  title = {{Energy Reconstruction Methods in the IceCube Neutrino Telescope}},
  journal = {JINST},
  volume = {9},
  number = {03},
  pages = {P03009},
  year = {2014},
  doi = {10.1088/1748-0221/9/03/P03009},
}

@article{Dror:2017nsg,
  author = {Dror, J. A. and Lasenby, R. and Pospelov, M.},
  title = {{Dark forces coupled to nonconserved currents}},
  journal = {Phys. Rev. D},
  volume = {96},
  pages = {075036},
  year = {2017},
  doi = {10.1103/PhysRevD.96.075036},
}

\end{document}